\documentstyle[11pt,amssymb]{article}

\makeatletter
\@addtoreset{equation}{section}
\makeatother

\def\dalemb#1#2{{\vbox{\hrule height .#2pt
        \hbox{\vrule width.#2pt height#1pt \kern#1pt
                \vrule width.#2pt}
        \hrule height.#2pt}}}

\def\hF{\hat F}
\def\tA{\widetilde A}
\def\tcA{{\widetilde{\cal A}}}
\def\tcF{{\widetilde{\cal F}}}
\def\hA{\hat{\cal A}}

\def\0{{\sst{(0)}}}
\def\1{{\sst{(1)}}}
\def\2{{\sst{(2)}}}
\def\3{{\sst{(3)}}}
\def\4{{\sst{(4)}}}
\def\5{{\sst{(5)}}}
\def\6{{\sst{(6)}}}
\def\7{{\sst{(7)}}}
\def\8{{\sst{(8)}}}

\def\tV{\widetilde V}
\def\tW{\widetilde W}
\def\tH{\widetilde H}
\def\tE{\widetilde E}
\def\tF{\widetilde F}
\def\tA{\widetilde A}
\def\tP{\widetilde P}
\def\tD{\widetilde D}
\def\bA{\bar{\cal A}}
\def\bF{\bar{\cal F}}
\def\Z{\rlap{\sf Z}\mkern3mu{\sf Z}}
\def\R{\rlap{\rm I}\mkern3mu{\rm R}}
\def\G{{\cal G}}
\def\gg{\bf g}
\def\CS{{\cal S}}
\def\S{{\cal S}}
\def\P{{\cal P}}
\def\ep{\epsilon}
\def\td{\tilde}
\def\wtd{\widetilde}

\let\a=\alpha \let\b=\beta \let\g=\gamma \let\d=\delta \let\e=\epsilon
    
\let\l=\lambda     
\let\s=\sigma

\def\nn{\nonumber} \def\bd{\begin{document}} \def\ed{\end{document}}
\def\ds{\documentstyle} \let\fr=\frac \let\bl=\bigl \let\br=\bigr
\let\Br=\Bigr \let\Bl=\Bigl 
\let\bm=\bibitem
\let\na=\nabla
\let\pa=\partial \let\ov=\overline 
\newcommand{\be}{\begin{equation}} 
\newcommand{\ee}{\end{equation}} 
\def\ba{\begin{array}}
\def\ea{\end{array}}
\def\ft#1#2{{\textstyle{{\scriptstyle #1}\over {\scriptstyle #2}}}}
\def\fft#1#2{{#1 \over #2}}
\def\del{\partial}
\def\sst#1{{\scriptscriptstyle #1}}
\def\oneone{\rlap 1\mkern4mu{\rm l}}
\def\ie{{\it i.e.\ }}
\def\via{{\it via}}
\def\semi{{\ltimes}}
\def\v{{\cal V}}
\def\str{{\rm str}}

\newcommand{\ho}[1]{$\, ^{#1}$}
\newcommand{\hoch}[1]{$\, ^{#1}$}
\newcommand{\bea}{\begin{eqnarray}} 
\newcommand{\eea}{\end{eqnarray}} 
\newcommand{\ra}{\rightarrow}
\newcommand{\lra}{\longrightarrow}
\newcommand{\Lra}{\Leftrightarrow}
\newcommand{\ap}{\alpha^\prime}
\newcommand{\bp}{\tilde \beta^\prime}
\newcommand{\tr}{{\rm tr} }
\newcommand{\Tr}{{\rm Tr} } 
\newcommand{\NP}{Nucl. Phys. }
\newcommand{\tamphys}{\it Center for Theoretical Physics,
Texas A\&M University, College Station, Texas 77843\\
and SISSA, Via Beirut No. 2-4, 34013 Trieste, Italy\hoch{2}}
\newcommand{\ens}{\it Laboratoire de Physique Th\'eorique de l'\'Ecole
Normale Sup\'erieure\hoch{2,3}\\
24 Rue Lhomond - 75231 Paris CEDEX 05}

\newcommand{\auth}{E. Cremmer\hoch{\dagger}, B. Julia\hoch{\dagger}, 
H. L\"u\hoch{\dagger} and C.N. Pope\hoch{\ddagger1}}

\begin{document}
\begin{flushright}
\hfill{CTP TAMU-24/98}\\
\hfill{LPTENS-98/05}\\
\hfill{SISSARef-48/98/EP}\\
\hfill{hep-th/9806106}\\
\hfill{June 1998}\\
\end{flushright}


\begin{center}
{ \large {\bf Dualisation of Dualities II:}}

{\large {\bf Twisted self-duality of doubled fields and superdualities}}

\vspace{10pt}
\auth

\vspace{10pt}

{\hoch{\dagger}\ens}

\vspace{10pt}
{\hoch{\ddagger}\tamphys}

\vspace{10pt}

\underline{ABSTRACT}
\end{center}

    We introduce a doubled formalism for the bosonic sector of the
maximal supergravities, in which a Hodge dual potential is introduced
for each bosonic field (except for the metric).  The equations of
motion can then be formulated as a twisted self-duality condition on
the total field strength $\G$, which takes its values in a Lie
superalgebra.  This doubling is invariant under dualisations; it
allows a unification of the gauge symmetries of all degrees, including
the usual U-dualities that have degree zero.  These ``superdualities''
encompass the dualities for all choices of polarisation ({\it i.e.}\
the choices between fields and their duals).  All gauge symmetries
appear as subgroups of finite-dimensional supergroups, with Grassmann
coefficients in the differential algebra of the spacetime manifold.

{\vfill\leftline{}\vfill
\footnoterule
{\footnotesize	\hoch{1} Research supported in part by DOE 
Grant DE-FG03-95ER40917 \vskip	-12pt} \vskip 14pt
{\footnotesize \hoch{2} Research supported in part by EC under TMR
contract ERBFMRX-CT96-0045 \vskip -12pt} \vskip 14pt
{\footnotesize
        \hoch{3} Unit\'e Propre du Centre National de la Recherche
Scientifique, associ\'ee \`a l'\'Ecole Normale Sup\'erieure \vskip -12pt}
                       \vskip 10pt
{\footnotesize \hoch{\phantom{3}} et \`a l'Universit\'e de Paris-Sud 
\vskip -12pt}} 

\pagebreak
\setcounter{page}{1}

\section{Introduction}

    The study of the rigid (global) symmetry groups of the various
supergravities has provided many insights into the understanding of
the structure of the theories \cite{CJ,cjgroups}.  In recent years the
global symmetries have acquired a new significance, in the context of
the conjectured non-perturbative U-duality symmetries of string
theories and M-theory \cite{HT,W}.

     It is useful to develop a universal framework within which the
rigid symmetries can be studied, in which, for example, all the
maximal supergravities can be discussed in an essentially
dimension-independent way.  Some steps in that direction were taken in
\cite{lpsol}, where a formalism for describing the bosonic sector of
all the maximal supergravities was developed.  Then, in \cite{cjlp},
the global symmetries of the scalar sector were discussed in full
generality, but the higher-rank fields were treated on more or less a
``case by case'' basis.  In this paper we shall present a new
description of the bosonic equations of motion in maximal
supergravities, including all the higher-rank fields. It has been
realised long ago that spacetime symmetries become internal symmetries
upon dimensional reduction, in \cite{cjlp} the dualisation of
dualities has been understood to exchange internal symmetries with
gauge symmetries. It is a natural idea, in order to achieve a deeper
understanding of duality symmetries that would be immune to
dualisation, to treat more uniformly gauge and internal symmetries:
this can be largely achieved.
 
   It was clear for many years that the twisted self-duality structure
of supergravities had to be more general than the few cases already
known in the 1980's. For instance the two-dimensional case with its
affine symmetry enlarged by the Moebius subgroup of the circle
diffeomorphisms was a confirmation of that hope \cite{N,B} in the
Moebius sector.  Furthermore the central extension of the affine
symmetry originates in the gravity sector of the theory and suggests
that gravity and matter could be unified by the magic of these
theories even without invoking any supersymmetry.  Our approach can be
motivated also by considering the situation in even spacetime
dimensions, where, as is well known, the rigid symmetries can usually
be realised only in terms of local field transformations of solutions
of the equations of motion, where they act on the field strengths
themselves rather than on the potentials.  This feature is especially
starkly illustrated by the consideration of a field strength of degree
$n$ in $D=2n$ dimensions, where the field and its Hodge dual are
members of an irreducible multiplet under the rigid symmetry.  For
example, in $D=8$ the 4-form field strength $F_\4$ and its Hodge dual
form a doublet under the $SL(2,\R)$ factor in the $SL(2,\R)\times
SL(3,\R)$ rigid U-duality group.  Obviously, therefore, it is not
possible to realise the $SL(2,\R)$ symmetry in terms of local
transformations on the potential $A_\3$ for $F_\4$, and so in
particular the $SL(2,\R)$ cannot be realised at the level of the
standard action.  However, as we showed in \cite{cjlp}, it {\it is}
possible to introduce a formalism in which the symmetry is realised on
potentials, by introducing a second 3-form potential $\tA_\3$, with
field strength $\tF_4$.  The ensuing doubling of the degrees of
freedom is counterbalanced by imposing, after varying the ```doubled''
Lagrangian, a constraint that the original and the doubled fields
strengths are related by Hodge duality.  (Actually, because of the
presence of a dilaton $\phi$ in the eight-dimensional theory, the
constraint takes the form $\tF_\4= e^{-\phi} \, {*F_\4}$ \cite{cjlp}.)
In fact although it can be useful to consider the Lagrangian for the
doubled system it is in some sense a ``gilding of the lilly,'' since
the constraint itself already implies all the equations for $F_\4$.
In other words, the Bianchi identities for $F_\4$ and $\tF_\4$,
together with the constraint, imply the equations of motion for
$\tF_\4$ and $F_\4$ respectively.

    In this paper we pursue the idea of introducing ``doubled'' fields
for all fields including the dilatonic scalars. This can be also
motivated by arguing that U-duality is not invariant under dualisation
\cite{lptdual,cjlp} but is transmuted partly into gauge symmetries so
F-duality \cite{JJ} must include all the latter so as to
deserve the name of Full duality.\footnote{We do, however, postpone
the extension to the gravity sector.  Also, we are considering only
the bosonic sectors here.}  Thus every bosonic field equation, with
the exception of the Einstein equation, can be expressed as the
statement that each field strength is equal to the dual of its double.
We shall show that by introducing generators for each field and its
double, we can write a combined single field $\G$, such that the
equations of motion read simply
\be
     {*\G} = \CS \, \G\ .\label{geq}
\ee
Here $*$ denotes the Hodge dual, and $\CS$ is an involution or a map
of square minus one (let us say a pseudo-involution) that exchanges
the generators for fields and those for their partners under doubling.
The field $\G$ can itself be written in terms of the exponential $\v$
of linear combinations of the generators mentioned above, with
potentials (including the doubled potentials) as coefficients, as $\G
= d\v\, \v^{-1}$.  This is a generalisation of the parameterisation of
scalar group manifolds in the Borel gauge, discussed in \cite{cjlp}
and known as the Iwasawa decomposition in Mathematics. In this
viewpoint, the Cartan-Maurer equation
\be
d\G - \G\wedge \G =0\label{cmeq}
\ee
follows as an identity.  One can also take another viewpoint and
instead view (\ref{geq}) as the {\it definition} of the doubled field
strengths such that the Cartan-Maurer equation (\ref{cmeq}) gives the
equations of motion for the fields.  In this alternative viewpoint the
ability to write $\G=d\v\, \v^{-1}$ is viewed as a consequence of
(\ref{cmeq}). Equation (\ref{cmeq}) is a zero curvature equation but
where the field strengths play the role of (generalised) Yang-Mills
potentials. Note that in (\ref{geq}) also, the potentials do not
appear.  Previous attempts to use differential algebras in Physics,
see for instance \cite{DAF}, have imposed extra restrictions requiring
freeness, disallowing Hodge duals or considering potentials and not
field strengths as the basic objects.  The equations we just presented
in (\ref{cmeq}) can be interpreted as the defining equations of a
minimal (defined here as quadratically nonlinear) differential algebra
in the sense of Sullivan \cite{Su}. Actually (\ref{geq}) spoils the
``freeness" by imposing relations beyond those of graded commutativity
and most essential, the basic fields are the field strengths not the
potentials.  Lagrangian realisations of these theories require the
choice of independent potentials, solutions of (\ref{geq}), and make
use of half as many potentials as one starts with.
  
    The paper is organised as follows.  In section 2 we present a
detailed discussion for $D=11$ supergravity, showing how the equation
of motion for the 4-form field can be re-expressed in the doubled
formalism.  This illustrates many of the basic ideas that will recur
in the later sections, including the fact that the generator
associated with the original 3-form potential is an odd (fermionic)
one, and thus the extended algebra of the doubled formalism is a
superalgebra.  In section 3 we extend the discussion to the important
case of ten-dimensional type IIA supergravity, and then in section 4
we generalise to cover all the $D$-dimensional maximal supergravities
that come from $D=11$.  We show that the underlying algebras in these
cases are deformations of $G\semi G^*$, where $G$ itself is the
semi-direct product of the Borel subalgebra of the superalgebra
$SL(11-D|1)$ and a rank-3 tensor representation, and $G^*$ is the
co-adjoint representation of $G$.  In section 5, we apply the doubled
formalism to scalar coset manifolds and group manifolds, beginning
with a detailed study of the symmetries for the $O(2) \backslash
SL(2,\R)$ coset, and finishing with the general case, there one can
verify the full $G$ invariance of the doubled formalism.  In section
6, we obtain the doubled formalism for type IIB supergravity in
$D=10$.  Interestingly, this is the only example among the maximal
supergravities where the generators are all exclusively bosonic.

\section{$D=11$ Supergravity}

     Our first example of a theory that can be expressed in terms of a
doubled field equation is the bosonic sector of eleven-dimensional
supergravity.  The Lagrangian is given by \cite{cjs}
\be
{\cal L}_{11} = R\, {*\oneone} - \ft12 {*F_\4} \wedge F_\4 - \ft16
F_\4\wedge F_\4 \wedge A_\3\ ,\label{d11lag}
\ee
where $F_\4=dA_\3$, and the bracketed suffices denote the degrees of the
differential forms.  Varying with respect to $A_\3$, we obtain the
equation of motion
\be
d{*F_\4} + \ft12 F_\4\wedge F_\4 = 0\ .\label{d11eom}
\ee
Note that the action, and hence the equations of motion, are invariant
under the abelian gauge transformation $\delta A_\3=d\lambda_\2$.
Eqn.~(\ref{d11eom}) can be written as $d({*F_\4} + \ft12 A_\3\wedge
F_\4)=0$, and so we can write the field equation in the first-order
form
\be
{*F_\4} = \tF_\7\equiv d\tA_\6 - \ft12 A_\3\wedge F_\4\ ,\label{d11fo}
\ee
where we have introduced a dual potential $\tA_\6$.  Taking the
exterior derivative of this equation gives rise to the second-order
equation of motion (\ref{d11eom}).  Note that it is not possible to
eliminate the 3-form potential $A_\3$ and write the equation of motion
purely in terms of the dual potential $\tA_\6$; nevertheless the
equation of motion could still be rewritten as the closure of a form.
(After we obtained these results a $D=11$ Lagrangian involving both a
3-form and a 6-form potential, together with further auxiliary fields,
was proposed in \cite{BBS}.)
  
     It is easily checked that the first-order equation (\ref{d11fo}) is
invariant under the following infinitesimal gauge transformations:
\be
\delta A_\3  =  \Lambda_\3\ , \qquad\quad
\delta \tA_\6= \wtd \Lambda_\6 - 
\ft12 \Lambda_\3 \wedge A_\3\ \label{d11gauge}
\ee
where $\Lambda_\3$ and $\wtd \Lambda_\6$ are 3-form and 6-form gauge
parameters, satisfying $d\Lambda_\3=0$ and $d\wtd \Lambda_6=0$.  (Note
that we work with ``gauge parameters'' that are closed forms of
degrees equal to the associated potentials, see \cite{bj80}.  This
leads to a more uniform treatment when we discuss the global
symmetries of 0-form potentials later.)

The commutators of infinitesimal gauge transformations are
therefore given by
\bea
{[}\delta_{\Lambda_\3}, \delta_{\Lambda_{\3}'}{]} &=&
\delta_{\wtd \Lambda_\6''}\ ,
\qquad \wtd \Lambda_\6'' = \Lambda_\3 \wedge \Lambda_\3'\ ,\nn\\
{[}\delta_{\Lambda_\3},\delta_{\wtd\Lambda_\6}{]} &=&0\ ,\qquad
{[} \delta_{\wtd\Lambda_\6},\delta_{\wtd\Lambda_\6'}{]} =0\ .
\label{d11comvar}
\eea
Note that the introduction of the dual potential $\tA_\6$, and the use
of the parameter $\Lambda_\3$, has the
consequence that the realisation of the originally abelian gauge symmetry 
of the potential
$A_\3$ has now become non-abelian as acting on the dual potential
$\tA_\6$.\footnote{If instead we use unconstrained parameters $\mu_\2$
and $\mu_\5$, where $\Lambda_\3= d\mu_\2$ and $\Lambda_\6=d\mu_\5$,
then eqn.\ (\ref{d11gauge}) can be rewritten as $\delta A_\6 = d\mu'_\5
+\ft12 \mu_\2\wedge dA_\3$, \ $\delta A_\3 = d\mu_\2$, where $\mu'_\5
= \mu_\5 -\ft12 \mu_\2 \wedge A_\3$, and then all
the gauge transformations commute.   This abelianisation by means of a
field-dependent redefinition of the gauge parameters does not appear
to be possible unless the $\Lambda$ parameters are exact forms.
This shows that the
non-commutativity is associated with the global parts of the gauge 
invariances.} However it is not a Yang-Mills type 
non-linearity but rather an odd one involving anticommutation rather 
than commutators, which can be traced back to the Chern-Simons term in
$D=11$ supergravity. 
As we mentioned above, the field strengths are now quadratically coupled as
if they were Yang-Mills connections.  
If we introduce generators $V$ and $\tV$ for the $\Lambda_\3$ and
$\tilde\Lambda_\6$ transformations respectively, we see that the
commutation relations (\ref{d11comvar}) translate into the (Lie)
superalgebra
\be
\{V,V\}= -\tV\ ,\qquad {[}V,\tV {]}=0\ ,\qquad {[} \tV, \tV {]} =0\ .
\label{d11com}
\ee
Note that the generators are even or odd according to whether the
degrees of the associated field strengths are odd or even.  In other
words, the product $A_\3\, V$ and the product $\tA_\6\, \tV$ are both
even elements, but $V$ is an odd generator.  Also, when the exterior
derivative passes over a generator, the latter acquires a minus if it
is odd.  Thus $d(V\, X)=-V\, dX$, while $d(\tV\, X) = \tV\, dX$, for
any $X$. We shall return to the general structure of the superalgebras
corresponding to (\ref{d11com}) and identify them in section 4, but
let us right away discuss the simplest one given here. $\tV$ is even
and in the centre so it can be diagonalised. For each of its
eigenvalue one has a Clifford algebra in one generator. It can be
viewed as a deformation (quantisation) of the Grassmann superalgebra
on one generator. This deformation is precisely the result of adding
the Chern-Simons term into the Lagrangian. It will of course operate
in any dimension of spacetime.

     Moving a stage further, we can combine the doubled set of fields
that describe the non-gravitational degrees of freedom of the extended
$D=11$ supergravity equations:
\be
\v = e^{A_\3\, V}\, e^{\tA_\6\, \tV}\ .\label{d11coset}
\ee
This parameterisation is suggested by (\ref{d11fo}). 
By an elementary calculation one checks that the field strength $\G=d\v\,
\v^{-1}$ following from (\ref{d11coset}) is given by
\bea
\G &=& dA_\3\, V + (d\tA_\6 - \ft12 A_\3 \wedge dA_\3) \, \tV\ ,\nn\\
  &=& F_\4 \, V + \tF_\7\, \tV\ .\label{gcalc}
\eea  
The gauge transformations (\ref{d11gauge}) can be re-expressed in the
simple form
\be
\v' = \v\,e^{\Lambda_\3\, V}\, e^{\wtd\Lambda_\6\, \tV}\ .
\label{cosetgauge}
\ee
It is straightforward to see that $\G$ is invariant under these gauge
transformations, since $\Lambda_\3$ and $\wtd \Lambda_\6$ are closed
forms.  The gauge transformations act on the right; they are analogous
to the rigid $G$ action on the right of the scalar coset space
$K\backslash G$, and $\G$ is the analogue of the $K$-tensor $dg
g^{-1}$.  Let us notice that in the scalar sector the local $K$ action
on the left will be fixed in the Borel gauge so it does not count as a
gauge invariance in the new sense of differential algebras.  The
first-order field equation (\ref{d11fo}) can now be compactly written
as a twisted self-duality condition:
\be
{*\G} = \CS \, \G\ ,\label{d11master}
\ee
where $\CS$ is a pseudo-involution that maps between the
generators $V$ and $\tV$:
\be
\CS \, V = \tV\ ,\qquad \CS \,\tV = -V\ .\label{d11om}
\ee
Note that here we have $\CS^2\, V = -V$ and $\CS^2\, \tV = -\tV$, so $\CS
^2 = - {\rm id}$.  In general, the eigenvalue of $\CS^2$ on a given
generator is the same as the eigenvalue of $*^2$ on the associated
field strength.  In the more general examples in lower dimensions, we
shall see that $\CS^2$ acts sometimes as an involution, and sometimes
as a pseudo-involution. Let us insist however that $\CS$ does not
preserve the commutation relations (\ref{d11com}) but it is 
analogous to the scalar case situation where $\CS$ is a 
$K$-tensor \cite{JJ}.  

    There is another view of the above construction.  Since the
doubled field strength $\G$ is written as $\G= d\v\, \v^{-1}$, it
follows by taking an exterior derivative that we have the
Cartan-Maurer equation $d\G= -d\v\, d\v^{-1} = d\v\, \v^{-1}\, d\v\,
\v^{-1}$, and hence
\be
   d\G - \G\wedge \G = 0\ .\label{cm}
\ee
Now, substituting (\ref{d11fo}) into (\ref{gcalc}), we can write the 
doubled field as
\be
 \G= F_\4\, V + {*F_\4}\, \tV\ .\label{d11mf}
\ee
It follows from this that
\bea
\G\wedge \G &=& F_\4\, V\, F_\4 \, V + F_\4\, V\, {*F_\4}\, \tV
           + {* F_\4}\, \tV\, F_\4\, V \ ,\nn\\
&=& \ft12 \{F_\4\, V , F_\4 \, V\} + \{F_\4\, V , {*F_\4}\, \tV\} \ ,\nn\\
 &=& \ft12 F_\4 \wedge F_\4\, \{V,V\} - F_\4\wedge {*F_\4}\, {[} V,\tV {]}
       \nn\\
 &=& - \ft12 F_\4\wedge F_\4\, \tV  \ ,\label{ggcomp}
\eea
and so the original second-order equation of motion (\ref{d11eom}) can
be obtained simply by substituting (\ref{d11mf}) into the
Cartan-Maurer equation (\ref{cm}).  Note that in (\ref{ggcomp}), we
temporarily suspended the writing of the wedge-product symbols
$\wedge$.  In getting from the first line to the second, we used that
for {\it any} $X$, we can write $X\, X$ as $\ft12 \{X,X\}$.  Passing
to the third line, we used that $V$ is odd (\ie it behaves like an
odd-degree differential form), while $\tV$ is even.  Thus in
particular, we acquired a minus sign in turning $V\, {*F_\4}$ into
$-{*F_\4}\, V$.  Finally, to reach the last line, we used that $V$ and
$\tV$ satisfy the (anti)-commutation relations given in
(\ref{d11com}).
  
\section{Type IIA Supergravity}

     The formalism developed above may be extended straightforwardly
to the maximal supergravities obtained by dimensional reduction from
eleven-dimensional supergravity.  We shall give the general
$D$-dimensional results in the next section.  Here, we consider the
important special case of type IIA supergravity.  The Lagrangian for
the bosonic fields can be written as
\bea
{\cal L}_{10} &=& R {*\oneone} -\ft12 {*d\phi}\wedge d\phi - \ft12
e^{-\fft32\phi}\, {*{\cal F}_\2}\wedge {\cal F}_\2 -\ft12 e^{\phi}\,
{*F_\3}\wedge F_\3\nn\\
&& - \ft12 e^{-\fft12\phi}\, {*F_\4}\wedge F_\4 -\ft12
dA_\3\wedge dA_\3\wedge A_\2\ ,
\eea
where $F_\4=dA_\3 -dA_\2\wedge {\cal A}_\1$, $F_\3=dA_\2$ and ${\cal
F}_\2= d{\cal A}_\1$. From this, it follows that the equations of motion
for the antisymmetric tensor and scalar fields are:
\bea
d(e^{-\fft12\phi} \, {*F_\4}) &=& -F_\4\wedge F_\3\ .\nn\\
d(e^\phi\, {*F_\3}) &=&-{\cal F}_2\wedge (e^{-\fft12\phi}{*F_\4}) -
\ft12 F_\4\wedge F_\4\ ,\label{2aeom}\\
d(e^{-\fft32\phi}\, {*{\cal F}_\2}) &=& -F_\3\wedge (e^{-\fft12\phi}\, 
{*F_\4}) \ ,\nn\\
d{*d\phi} &=& \ft14 F_\4\wedge ( e^{-\fft12\phi}\,{*F_\4}) + \ft12
F_\3\wedge (e^\phi\,{*F_\3}) +\ft34 {\cal
F}_\2\wedge (e^{-\fft32\phi}\, {*{\cal F}_\2}) \ .\nn
\eea

     It is not hard to re-write these second-order field equations in
first-order form, by extracting an overall exterior derivative from
each equation. This means that all the equations of motion are
(generalised) conservation laws \cite{bj80,JS}.  (This is best done by
starting with the equation for $F_\4$, and working down through the
degrees of the fields, ending with the dilaton.) Thus, introducing a
``doubled'' set of potentials $\{\psi,\wtd{\cal A}_\7, \tA_\6,
\tA_\5\}$ dual to $\{\phi, {\cal A}_\1, A_\2,A_\3\}$ respectively, we
can write (at least locally) the following first-order equations:
\bea
e^{-\fft12\phi}\, {*F_\4} &\equiv & \tF_\6 = d\tA_\5 - A_\2\wedge
dA_\3\ ,\nn \\ 
e^\phi\, {*F_\3} &\equiv& \tF_\7 = d\tA_\6 -\ft12 A_\3\wedge dA_\3 
-{\cal A}_\1
\wedge (d\tA_\5 -A_\2 \wedge dA_\3)\ ,\nn\\
e^{-\fft32\phi}\, {*{\cal F}_\2} &\equiv & \wtd{\cal F}_\8 = 
d\wtd{\cal A}_\7 -A_\2\wedge (d\tA_\5 -\ft12 A_\2\wedge dA_\3)
\ ,\label{2alin}\\ 
{*d\phi} & \equiv & \wtd P = d\psi + \ft12 A_\2\wedge d\tA_\6 
+\ft14 A_\3\wedge (d\tA_\5  -A_\2\wedge dA_\3)\nn\\
 && + \ft34 {\cal A}_\1\wedge
(d\wtd{\cal A}_\7 -A_\2\wedge (d\tA_\5-\ft12 A_\2\wedge dA_\3))\ .\nn
\eea
It is straightforward to check that by taking the exterior derivatives
of these equations, and substituting the first-order equations back
in where appropriate, we recover precisely the equations of motion 
(\ref{2aeom}).  

    In principle, we could now follow the same strategy that we
described for $D=11$ supergravity in section 2, and derive the
enlarged set of gauge transformations for all the potentials,
including the doubled potentials for the dual field strengths.  From
the commutators of these gauge transformations we could then derive a
superalgebra of generators associated with the potentials, which would
be the analogue of (\ref{d11comvar}).  In practice, this is a
cumbersome procedure, and it is easier to derive the superalgebra by
instead looking at the field equations.  We first note that these can
be written, using the tilded field strengths that are defined in a
natural fashion in terms of the duals of the untilded ones in
(\ref{2alin}), as
\bea
d\wtd F_\6 &=& - F_\4 \wedge F_3\ ,\nn\\
d\wtd F_\7 &=& - {\cal F}_\2 \wedge {\wtd F}_\6 - \ft12 F_\4 \wedge
F_\4\ ,\nn\\
d \wtd {\cal F}_8 &=& -F_\3 \wedge {\wtd F_6}\ ,\label{2abilinear}\\
d{\wtd P} &=&  \ft14 F_\4 \wedge \wtd F_\6 + \ft12 F_\3 \wedge \wtd 
F_\7 + \ft34 {\cal F}_\2 \wedge \wtd{\cal F}_\8\ .\nn
\eea
The fact that the right-hand sides are all simply bilinear in field
strengths suggests that it should be possible again to write the
equations in the Cartan-Maurer form $d\G-\G\wedge \G=0$,  as in the
case of eleven-dimensional supergravity treated in section 2.  Thus  
let us introduce the doubled field strength $\G$:
\bea
\G&=&\ft12d\phi\, H + e^{-\fft34\phi}\, {\cal F}_\2\, W_1 + 
e^{\fft12\phi}\, F_\3\, V^1  +e^{-\fft14\phi}\, F_\4\, V \nn\\
&& + e^{\fft14\phi}\, \tF_\6 \, \tV 
+ e^{-\fft12\phi}\, \tF_\7\, \tV_1 
+ e^{\fft34\phi}\, \wtd{\cal F}_\8\, \tW^1
+\ft12 \wtd P\, \tH\ .\label{2ag}
\eea
Note that $\G$ is defined so as to be invariant under the gauge
symmetries of the original Lagrangian, including the constant shift
symmetry of the dilaton $\phi$ together with the corresponding
constant scalings of the other gauge potentials.  (It is the
requirement that $\G$ be invariant under this shift symmetry that
determines the exponential factors in the various terms in
(\ref{2ag}).)  The generators $H$, $V^1$, $\tV_1$, $\tH$, being
associated with field strengths whose potentials are of even degree,
are themselves even.  On the other hand the generators $W_1$, $V$,
$\tV$ and $\tW^1$ are associated with potentials of odd degrees, and
they will therefore be odd.  (The notation here will be generalised in
section 4, when we discuss the dimensional reduction to $D$
dimensions.  The ``1'' suffices and superscripts on generators
indicate that they are associated with potentials arising in the first
step of the reduction from $D=11$.)

    We may again introduce the (pseudo)-involution operator $\CS$,
which is defined to act on an untilded generator $X$ to give the
corresponding tilded generator $\widetilde X$ associated with the dual
potential; $\CS \, X =\widetilde X$.  Acting on $\widetilde X$, we
have $\CS\, \widetilde X = \pm X$; the operator $\CS^2$ has eigenvalue
$+1$ or $-1$ in accordance with the eigenvalue of $*^2$ on the
corresponding field strength. Thus the field $\G$ defined in
(\ref{2ag}) automatically satisfies the twisted self-duality equation
${*\G}= \CS \, \G$.

    We now find that the equations of motion (\ref{2abilinear}) can
indeed be written simply as the ``curvature-free" condition
$d\G-\G\wedge \G=0$, where the generators satisfy the following
commutation and anti-commutation relations.  Firstly the commutators
with the Cartan generator $H$, governed by the weights of the various
fields appearing in (\ref{2ag}), are
\bea
&&{[} H, W_1 {]} = -\ft32 W_1\ ,\qquad
{[} H, V^1 {]} = V^1\ ,\qquad
{[} H, V {]} = -\ft12 V\ ,\nn\\
&&{[} H, \wtd W_1 {]} = \ft32 \wtd W_1\ ,\qquad
{[} H, \wtd V^1 {]} = -\wtd V^1\ ,\qquad
{[} H, \wtd V {]} = \ft12 \wtd V\ .\label{2ahcom}
\eea
Next, the commutators and anti-commutators associated with the bilinear
structures on the right-hand sides of the Bianchi identity for $F_\4$,
and those for $\tF_\6$, $\tF_\7$ and $\wtd{\cal F}_\8$ in 
(\ref{2abilinear}), are
\bea
&&{[}W_1, V^1{]} = -V\ ,\qquad \{W_1,\tV\} = -\tV_1\ ,\qquad
{[}V^1,V{]} = - \tV\ ,\nn\\
&&{[} V^1,\tV{]} = -\tW^1\ ,\qquad \{V,V\} = -\tV_1\ .\label{2acom}
\eea
Finally, those associated with the right-hand side in the equation for
$\tP$ in (\ref{2abilinear}) are 
\be
\{W_1, \tW^1\}= \ft38 \tH\ ,\qquad 
{[}V^1, \tV_1{]} =\ft28 \tH\ ,\qquad 
\{ V,\tV\} = \ft18 \tH\ .\label{2athcom}
\ee
Note that here and in the sequel, all commutators and anti-commutators
that are not listed do vanish.  We shall discuss the structure of this
superalgebra in the next section, where superduality algebras for
$D$-dimensional maximal supergravities are obtained.

  As in the eleven-dimensional example of the previous section, the
Cartan-Maurer equation $d\G-\G\wedge \G=0$ for the doubled field $\G$ can
be solved by writing $\G=d\v\,\v^{-1}$, with
$\v$ most conveniently given by
\be
\v = e^{\fft12\phi H}\, e^{{\cal A}_\1 W_1}\, e^{A_\2 V^1}\, e^{A_\3 V}\,
e^{\tA_\5 \tV}\, e^{\tA_\6 \tV_1}\, e^{\wtd{\cal A}_\7 \tW^1}\,
e^{\fft12\psi \tH}\ .
\ee
A detailed calculation of $d\v \v^{-1}$ gives precisely (\ref{2ag}),
where the tilded field strengths are now given by the right-hand sides
of the first-order equations in (\ref{2alin}).  From this viewpoint,
where $\G$ is defined to be $d\v\, \v^{-1}$, the equation
$d\G-\G\wedge \G=0$ is trivially satisfied, and the field equations,
in the first-order form (\ref{2alin}), arise from the twisted
self-duality equation ${*\G}=\CS \, \G$.

        The type IIA supergravity has a classical global $\R$
symmetry, which corresponds to continuous shifts of the dilaton and
rescalings of the higher-degree potentials.  It is straightforward to
see that this symmetry is preserved in the doubled equation ${*\G}=\CS
\G$, since $\G$, given by (\ref{2ag}), is invariant under this global
symmetry provided that the dual fields rescale accordingly.  In fact
the doubled-equation formalism puts the local gauge symmetries of the
higher-degree fields and the constant shift symmetry of the dilaton on
an equal footing.  The transformation rules for these symmetries can
be expressed as
\be
\v' = \v\, e^{\fft12\Lambda_\0 H}\, e^{\Lambda_\1 W_1}\, e^{\Lambda\2
V^1}\, e^{\Lambda_\3 V}\,
e^{\wtd \Lambda_\5 \tV}\, e^{\wtd\Lambda_\6 \tV_1}\, 
e^{\wtd\Lambda_\7 \tW^1}\,
e^{\fft12\Lambda_\8 \tH}\ ,
\ee
where the gauge parameters $\Lambda_{(i)}$ and $\wtd \Lambda_{(i)}$
are all closed forms.  The commutators of these transformations
generate the algebra presented in (\ref{2ahcom}), (\ref{2acom}) and
(\ref{2athcom}).  The superalgebra of gauge symmetries of this bosonic
theory seems to grow out of control, nevertheless it can be
reorganised into a manageable form (that is by human beings).  Again
we may contract away the trilinear Chern-Simons term and its
associated commutators, namely those producing tilded generators out
of untilded ones. Then the untilded generators form a subalgebra $G$,
and its dual space $G^*$ transforms under it as its contragredient or
dual representation. In other words we are actually considering a
deformation (by the Chern-Simons term) of the Lie (super-)algebra $G
\semi G^*$ where the semi-direct product is a standard construction
for any linear representation of $G$ to be treated as an abelian
algebra. We may note in passing that the coadjoint representation is
equivalent to the adjoint one for a semisimple algebra, or more
generally for an algebra admitting a non degenerate invariant
quadratic form in the adjoint representation; these are sometimes
called contragredient (super-)algebras \cite{Ka,Sc}. Let us note also
that the commutators (\ref{2athcom}) represent a central extension by
$\tH$.  It too can be contracted away; we shall return to this
deformation later.  If one were to contract away both $\tH$ and
$\tW^1$ one would find a $\Z /3\Z$ grading.

\section{$D$-dimensional Maximal Supergravity}

     In this section, we consider the general case of maximal
supergravity in $3\le D\le 9$ dimensions, obtained by spacelike
toroidal dimensional reduction from either $D=11$ supergravity or type
IIB supergravity.  We shall adopt the notation and conventions of
\cite{lpsol,cjlp}, where the dimensional reductions from $D=11$ are
discussed. (But note that the sign of the Chern-Simons term in
(\ref{d11lag}) is taken to be the opposite of the one chosen in those
references.)  Since the general case is quite complicated, we divide
the analysis into four subsections and three appendices.  First, we
obtain the first-order equations of motion for the doubled systems of
fields in each dimension.  Then, we discuss the associated coset
constructions.  Next is the introduction of an unexpected twelfth
fermionic dimension and finally the discussion of the deformation
theory.  Certain dimension-dependent details of the constructions as
well as a more general discussion of the fermionic dimension are
relegated to the appendices.

\subsection{First-order equations for $D$-dimensional supergravity}

  The $D$-dimensional Lagrangian, in the language of differential
forms, is given by \cite{lpsol}
\bea
{\cal L}&=& R\, {*\oneone} -\ft12{*d\vec\phi}\wedge d\vec\phi-  \ft12
e^{\vec a\cdot\vec\phi}\,  {*F_\4}\wedge F_\4 -
\ft12\sum_i  e^{\vec a_i\cdot\vec\phi}\, {*F_{\3 i}}\wedge F_{\3 i} \nn\\
&&-
\ft12\sum_{i<j}  e^{\vec a_{ij}\cdot\vec\phi}\, {*F_{\2 ij}}\wedge F_{\2 ij}
-\ft12\sum_i  e^{\vec b_i\cdot\vec\phi}\, {*{\cal F}_\2^i}\wedge
{\cal F}_\2^i -
\ft12\sum_{i<j<k}  e^{\vec a_{ijk}\cdot\vec\phi}\, {*F_{\1 ijk}}
\wedge F_{\1 ijk}\nn\\
&& -\ft12\sum_{i<j}  e^{\vec b_{ij}\cdot\vec\phi}\, {*{\cal F}^i_{\1
j}}\wedge {\cal F}^i_{\1 j} +{\cal L}_{FFA}\ .\label{dgenlag}
\eea
Let us recall that the vector notation represents vectors in the root
space of $GL(n,\R)$, with $n=(11-D)$.  The Chern-Simons terms ${\cal
L}_{FFA}$ are given for each dimension in \cite{lpsol,cjlp}.  An
important property of these terms is that their variation with respect
to the various potentials $A_\3$, $A_{\2 i}$, $A_{\1 ij}$ and $A_{\0
ijk}$ takes, up to a total derivative, the form
\be
-\delta{\cal L}_{FFA} =  dX\wedge \delta A_\3+ 
 dX^i\wedge \delta A_{\2 i} + \ft12 dX^{ij}\wedge \delta A_{\1 ij}
+ \ft16 dX^{ijk}\, \delta A_{\0 ijk}\ ,\label{Xeq}
\ee
where the quantities $X$, $X^i$, $X^{ij}$ and $X^{ijk}$ can be
determined easily in each dimension.  They are given in appendix A.
Again the existence of the $X$'s reflects the abelian gauge invariance
and source-freeness of the Chern-Simons integral and the ensuing
possibility to rewrite the would be equations of motion of the
Lagrangian ${\cal L}_{FFA}$ as total derivatives \cite{bj80,JS}. Here
source-freeness means that a choice of action can be made such that
any given potential can appear always differentiated.
  
 We shall work with the hatted $\hat{\cal A}_1^i=\gamma^i{}_j\,
{\cal A}_1^j$ Kaluza-Klein potentials, introduced in (A.18) of 
ref.~\cite{cjlp}.
Thus the various field strengths are given by
\bea
{\cal F}_\2^{i} &=& \tilde\gamma^i{}_j\, \hat{\cal F}_\2^{j}\ ,\qquad
{\cal F}_{\1 j}^i = \gamma^k{}_j\, d{\cal A}_{\0 k}^i\ ,\nn\\
F_{\2ij} &=&  \gamma^k{}_i\, \gamma^\ell{}_j\, \hat F_{\2k\ell}\ ,\qquad
F_{\3i} = \gamma^j{}_i\, \hat F_{\3j}\ ,\qquad F_\4 = \hat F_\4\ ,
\label{hatfieldstr}
\eea
where these (Kaluza-Klein modified) field strengths read
\bea
\hat {\cal F}_\2^{i} &=& d \hat{\cal A}_\1^{i}\ ,\qquad
\hat F_{\2ij} = dA_{\1ij} - dA_{\0ijk}
\wedge\hat {\cal A}_\1^{k}\ ,\nn\\
\hat F_{\3i} &=& dA_{\2i} +
dA_{\1ij}\wedge \hat {\cal A}_\1^{j}
+\ft12 dA_{\0ijk}\wedge  \hat {\cal A}_\1^{j}\wedge
\hat {\cal A}_\1^{k}\ ,\label{interm}\\
\hat F_\4 &=& dA_\3 - dA_{\2i} \wedge\hat {\cal A}_\1^{i} +\ft12
dA_{\1ij}\wedge  \hat {\cal A}_\1^{i}\wedge  \hat {\cal A}_\1^{j}
-\ft16 dA_{\0ijk}\wedge  \hat {\cal A}_\1^{i}\wedge
\hat{\cal A}_\1^{j}
\wedge  \hat {\cal A}_\1^{k}\ .\nn
\eea
(Here $\tilde\g^i{}_j = \delta^i{}_j + {\cal A}^i{}_{\0 j}$, and
$\g^i{}_j$ is its inverse. See (A.19) and (A.29) in ref.~\cite{cjlp}.)

 We are now in a position to start constructing the first-order
equations, by writing down the second-order equations following from
(\ref{dgenlag}), and then stripping off a derivative in each case.  If
we handle the various field equations in the appropriate order, this
turns out to be a fairly straightforward deductive process.  The order
to follow is first to look at the equation of motion coming from
varying $A_\3$ in (\ref{dgenlag}), then $A_{\2 i}$, then $A_{\1 ij}$,
then $A_{\0 ijk}$, then $\hat{\cal A}_1^i$, then ${\cal A}^i_{\0 j}$,
and finally $\vec\phi$.  We shall look explicitly here at the first
two of these, and then present only the results for the others.  Thus
varying (\ref{dgenlag}) with respect to $A_3$ we get
\be
- \delta{\cal L} = e^{\vec a\cdot\vec\phi}\, {*F_4}\wedge d\delta A_\3
+ dX\wedge \delta A_\3\ .
\ee
Integrating by parts, this gives\footnote{When the degree of the
field strength becomes larger than or equal to D we take the dual to
vanish in this paper.}
\be
-(-1)^D\, d( e^{\vec a\cdot\vec\phi}\, {*F_4}) + dX=0\ .
\ee
Thus we can immediately strip off the derivative, and write the
first-order equation
\be
e^{\vec a\cdot\vec\phi}\, {*F_4}\equiv \tF_{(D-4)} =
d\tA_{(D-5)} + (-1)^D\, X \ .
\label{df4f}
\ee
Varying (\ref{dgenlag}) with respect to $A_{\2 j}$, and integrating by
parts, we get the field equation
\be
(-1)^D\, \sum_i d(e^{\vec a_i\cdot\vec\phi}\, {*F_{\3 i}}\,
\gamma^j{}_i) + (-1)^D\, d(e^{\vec a\cdot\vec\phi}\, {*F_4} \wedge
\hat{\cal A}_1^j) + dX^j=0\ .
\ee
We can now strip off the derivative, and then use the previous result
(\ref{df4f}), to give the first-order equation
\be
e^{\vec a_i\cdot\vec\phi}\, {*F_{\3 i}}\, = \td\gamma^i{}_j\, \tF_{(D-3)}^j
\ ,\label{df3f}
\ee
where
\be
\tF_{(D-3)}^j=
d\tA_{(D-4)}^j - d\tA_{(D-5)} \wedge \hat{\cal A}_1^j - (-1)^D\,
(X^j + X\wedge  \hat{\cal A}_1^j ) \ .\label{df3fdef}
\ee
(We have also multiplied by a $\td\gamma$ here, which has allowed us to
obtain equations for each $i$ value separately.)

 Proceeding in a similar vein, we obtain the first-order equations
\bea
e^{\vec a_{ij}\cdot\vec\phi}\, {*F_{\2 ij}} &=& \td\gamma^i{}_k\,
\td\gamma^j{}_\ell \, \tF_{(D-2)}^{k\ell}\ ,\label{df2f}  \\
e^{\vec a_{ijk}\cdot\vec\phi}\, {*F_{\1 ijk}} &=& \td\gamma^i{}_\ell\,
\td\gamma^j{}_m\, \td\gamma^k{}_n\, \tF_{(D-1)}^{\ell mn}\ ,\label{df1f}
\eea
where
\bea
\tF_{(D-2)}^{k\ell} &=&d\tA_{(D-3)}^{k\ell} -d\tA_{(D-4)}^{k}\,
\hA_\1^\ell + d\tA_{(D-4)}^{\ell}\,\hA_\1^k + d\tA_{(D-5)}\, \hA_\1^k \,
\hA_\1^\ell \nn\\
&&+(-1)^D\, (X^{k\ell} + X^k\, \hA_\1^\ell - X^\ell\, \hA_\1^k +
X\, \hA_\1^k\, \hA_\1^\ell)\ ,\label{df2fdef}\\
\tF_{(D-1)}^{\ell mn} &=&
d\tA_{(D-2)}^{\ell mn} -3d\tA_{(D-3)}^{[\ell m}\,
\hA_\1^{n]} + 3d\tA_{(D-4)}^{[\ell}\,\hA_\1^m\, \hA_\1^{n]} -
d\tA_{(D-5)}\, \hA_\1^\ell\, \hA_\1^m\, \hA_\1^n \nn\\
&&-(-1)^D\, (X^{\ell mn} + 3 X^{[\ell m}\, \hA_\1^{n]} +
3 X^{[\ell}\, \hA_\1^m\, \hA_\1^{n]} + X\, \hA_\1^\ell\, \hA_\1^m\,
\hA_\1^n) \ .\label{df1fdef}
\eea
(As usual, we drop the $\wedge$ symbols when the going gets tough.)

    Now we turn to the equation of motion coming from varying
$\hA_\1^k$ in (\ref{dgenlag}).  For this, we again make use of the
first-order equations obtained previously in order to simplify the
result.  We then arrive at the equation
\bea
(-1)^D\, \sum_i d(e^{\vec b_i\cdot\vec\phi}\, {*{\cal F}_\2^i}\,
\td\gamma^i{}_k ) = -\ft12 \tF_{(D-2)}^{ij}\, dA_{\0 ijk} + \tF^i_{(D-3)} \,
\hat F_{\2 ik} -\tF_{(D-4)} \, \hat F_{\3 k}\ . 
\label{da111}
\eea
This is still a bit messy-looking, since the terms here involve a lot
of $\hA_\1^i$ Kaluza-Klein potentials.  But, remarkably, if we
substitute the definitions of the $\tF$ and $\hat F$ fields in terms
of the potentials, we find that all the $\hA_\1^i$ potentials cancel
out in (\ref{da111}), and we are left simply with:
\bea
\sum_i d(e^{\vec b_i\cdot\vec\phi}\, {*{\cal F}_\2^i}\,
\td\gamma^i{}_k ) &=& (-1)^D\, (-\ft12 d\tA_{(D-3)}^{ij}\, dA_{\0 ijk} +
d\tA_{(D-4)}^i\, dA_{\1 ik} - d\tA_{(D-5)}\, dA_{\2 k}) \nn\\
&&-\ft12 X^{ij}\,  dA_{\0 ijk} - X^i\, dA_{\1 ik} - X\, dA_{\2 k}\ .
\label{intres}
\eea
We still have the task of stripping the derivative off this equation.
This is clearly straightforward for the upper line on the right-hand
side.  For the lower line, it requires some more detailed analysis of
the Chern-Simons terms in each dimension, which give rise to the
quantities $X$, $X^i$ and $X^{ij}$, defined in (\ref{Xeq}).  We find
that the lower line is always a closed form, and so a derivative can
indeed be stripped off.  In other words, there exists a $Y_k$ such
that
\be
X\, dA_{\2 k} + X^i\, dA_{\1 ik} + \ft12 X^{ij}\, dA_{\0 ijk} = dY_k
\ ,\label{ydef0}
\ee
where the quantities $Y_k$ in each dimension $D$ are given in appendix A.
Thus we can now strip off the derivative in (\ref{intres}), to get the
first-order equation
\be
e^{\vec b_i\cdot\vec\phi}\, {*{\cal F}_2^i} = \gamma^j{}_i \,
\tcF_{(D-2)j}\ ,\label{dcf2f}
\ee
where
\be
\tcF_{(D-2)j} = d\tcA_{(D-3)j} -A_{\2 j}\, d\tA_{(D-5)} + (-1)^D\,
A_{\1 kj}\, d\tA_{(D-4)}^k -\ft12 A_{\0 jk\ell}\, d\tA_{(D-3)}^{k\ell}
-Y_j\ .\label{cfdef}
\ee

     Now we turn to the equation of motion coming from varying ${\cal
A}_{\0 k}^i$.  This will receive various contributions coming from the
fact that the field strengths $F_{\3 i}$, etc., involve $\gamma$.
After a little calculation, and substitution of the previous results
for first-order equations, the field equation can be put in the form
\be
(-1)^D\,  \sum_j d(e^{\vec b_{ij}\cdot\vec\phi}\, {*{\cal F}^i_{\1
j}}\, \gamma^k{}_j) -\sum_{\ell <j} e^{\vec b_{\ell j}\cdot\vec\phi} \,
{*{\cal F}^\ell_{\1 j}}\, \gamma^k{}_j\, {\cal F}^\ell_{\1 i} =
\gamma^j{}_i\, {\cal B}^k{}_j\ ,\label{xxx2}
\ee
where
\be
{\cal B}^k{}_j = \tF^k_{(D-3)}\, \hat F_{\3 j} + \tF_{(D-2)}^{k
\ell}\, \hat F_{\2 j\ell} +\ft12 \tF_{(D-1)}^{k\ell m}\, \hat F_{\1
j\ell m} -\tcF_{(D-2)j}\, d\hA_\1^k\ .
\ee
When re-expressed in terms of the potentials, this is again an
expression that undergoes ``miraculous'' simplifications, giving
\bea
{\cal B}^k{}_j &=& d\tA_{(D-5)}\, d(\hA_\1^k\, A_{\2 j}) +
d\tA_{(D-4)}^k\, dA_{\2 j} -d\tA_{(D-4)}^\ell\, d(\hA_\1^k\, A_{\1
j\ell})-\ft12 d\hA_\1^k\, d\tcA_{(D-3)j}\nn\\
&&+d\tA_{(D-3)}^{k\ell}\, d A_{\1 j\ell} +\ft12 d\tA_{(D-3)}^{\ell m}\,
d(A_{\0 j\ell m}\, \hA_\1^k) +\ft12 d\tA_{(D-2)}^{k\ell m}\, dA_{\0
j\ell m}\nn\\
&&+(-1)^D\, \Big( X\, dA_{\2 j} + X^\ell\, dA_{\2 \ell j} +
            \ft12 X^{\ell m}\,  dA_{\0 j\ell m}\Big)\,  \hA_\1^k
                + Y_j\, d\hA_\1^k\nn\\
&&+(-1)^D\, \Big( - X^k\, dA_{\2 j} + X^{k\ell}\, dA_{\1 j\ell} - \ft12
X^{k\ell m}\, dA_{\0 j\ell m} \Big)\ .\label{xxx1}
\eea
It is manifest that the first two lines on the right-hand side can be
written as exact differentials. It is also clear that we can do this for
the third line, after recognising that the three terms in the bracket
are nothing but the exact form $dY_j$ defined in (\ref{ydef0}).  For the
final line, we can write it as an exact form if we can find quantities
$Q^k{}_j$, such that
\be
X^k\, dA_{\2 j} +X^{k\ell}\, dA_{\1 \ell j}
+\ft12 X^{k\ell m}\, dA_{\0 j\ell m} = -dQ^k{}_j \ .\label{xxx6}
\ee
We find that this can indeed be done, and the results are presented
for each dimension in appendix A. In fact the structure of
(\ref{xxx6}) is quite analogous to that of (\ref{ydef0}), the closure
of their left-hand sides is equivalent upon integration by parts to
that of expressions of the form $ \sum dX^B\, R_B{}^C \, A_C$ with $B$
and $C$ collective indices but we have seen that $dX^B$ is the term in
the equation of motion of $A_B$ that comes from ${\cal L}_{FFA}$, and
hence these mysterious equations reflect nothing but invariances of
the pure Chern-Simons action under various diffeomorphisms of the
compactified coordinates.

   Thus we have that ${\cal B}^k{}_j = dW^k{}_j$, where
\bea
W^k{}_j &=&Q^k{}_j+ A_{\2 j}\, \hA_\1^k\, d\tA_{(D-5)} - (-1)^D\, A_{\2
j}\, d\tA_{(D-4)}^k  -(-1)^D\, A_{\1 j\ell}\, \hA_\1^k\, d\tA_{(D-4)}^\ell
\nn\\
&& +A_{\1 j\ell}\, d\tA_{(D-3)}^{k\ell} + \ft12 A_{\0 j\ell m}\, \hA_\1^k\,
d\tA_{(D-3)}^{\ell m} -\ft12 (-1)^D \, A_{\0 j\ell m}\,
d\tA_{(D-2)}^{k\ell m}\nn\\
&&-\hA_\1^k\, d\tcA_{(D-3)j} + Y_j\, \hA_\1^k\ .\label{xxx3}
\eea
We have found the non-trivial result that it is possible to strip off
a derivative in the second-order equations (\ref{xxx2}), by writing:
\be
e^{\vec b_{ij}\cdot\vec\phi}\, {*{\cal F}^i_{\1 j}} =
\gamma^\ell{}_i\, \td\gamma^j{}_k\, \tcF^k_{(D-1)\ell}\ ,
\ee
where
\be
 \tcF^k_{(D-1)\ell}= d\tcA^k_{(D-2)\ell} + (-1)^D\, W^k{}_\ell\ .
\ee
This gives us the first-order equation for the ${\cal F}^i_{\1 j}$
fields.

     This completes the derivation of first-order equations for all
the field strengths (including non-dilatonic 1-form field strengths)
in $D$-dimensional supergravity.  It now remains to obtain the
first-order equations for the dilatonic scalars $\vec\phi$.  After
varying (\ref{dgenlag}) with respect to $\vec\phi$, and using the
various first-order equations already obtained, we can write the
second-order equations of motion as
\bea
-(-1)^D\, d{*d\vec\phi} &=& \ft12 \vec a\, \tF_{(D-4)}\, F_\4 +
\ft12 \sum_i
\vec a_i\, \td\gamma^i{}_j\, \tF_{(D-3)}^j\, F_{\3 i} + \ft12 \sum_{i<j}
\vec a_{ij}\, \td\gamma^i{}_k \, \td\gamma^j{}_\ell \,
\tF_{(D-2)}^{k\ell}\,F_{\2 ij} \nn\\
&&+\ft12\sum_{i<j<k} \vec a_{ijk}\, \td\gamma^i{}_\ell\, \td\gamma^j{}_m
\, \td\gamma^k{}_n\, \tF_{(D-1)}^{\ell mn} \, F_{\1 ijk}
+\ft12 \sum_i \vec b_i\, \gamma^j{}_i\, \tcF_{(D-2)j} \, {\cal F}_2^i
\nn\\
&&
+\ft12 \sum_{i<j} \vec b_{ij}\, \gamma^\ell{}_i\, \td\gamma^j{}_k\,
\tcF^k_{(D-1)\ell}\,  {\cal F}^i_{\1 j}\ .\label{phi2o}
\eea
This is the trickiest equation to turn into first-order form.

      We must consider the structure of the second-order equation
(\ref{phi2o}), after replacing the field strengths by their
expressions in terms of potentials.  Firstly, we find that all terms
of bilinear or higher order in $\hA_\1^i$ cancel out.  Secondly, we
note that the only occurrences of a differentiated $\hA_\1^i$ are from
the ${\cal F}_\2^i$ field strength in the penultimate term in
(\ref{phi2o}).  This means that if we are to be able to strip off a
derivative, the set of terms linear in an undifferentiated $\hA_\1^i$
must themselves assemble into the form $Z\, \hA_\1^i$, where $Z$
itself is the total derivative of the factor multiplying $d\hA_\1^i$.
It is easiest first to consider the case when we temporarily set the
axions ${\cal A}_{\0 j}^i$ to zero.  It then becomes clear that in
each dimension there must exist a vector of $(D-1)$-forms $\vec Q$,
whose exterior derivatives satisfy
\be
\vec a\, X\, dA_\3 -\sum_i \vec a_i  X^i\, d A_{\2 i}
+\ft12 \sum_{ij} \vec a_{ij}\,  X^{ij}\, d A_{\1 ij}
-\ft16 \sum_{ijk}\vec a_{ijk}\,  X^{ijk}\, d A_{\0 ijk} = d\vec Q \ .
\label{qid}
\ee
One can indeed find such quantities $\vec Q$ in each dimension.  (In
doing this, and in proving the other previously-mentioned results for
stripping off a derivative from (\ref{phi2o}), it is necessary to make
extensive use of the various ``sum rules'' satisfied by the dilaton
vectors.  These are of the form $\vec b_{ij}=\vec b_i -\vec b_j$,
$\vec a_i=\vec a-\vec b_i$, {\it etc}.  See \cite{lpsol,cjlp}.)  It
may become less surprising if we make the same integration by parts as
we did after (\ref{xxx6}); the corresponding sum with the ${\cal
A}_{\0 j}^i$ reinserted is now of the form $ \sum dX^B \, \vec R_B{}^B
A_B$ and now vanishes by (Weyl) dimensional analysis again of the
internal coordinates.

  We now allow the axions ${\cal A}_{\0 j}^i$ to be non-zero again.
It is not hard to see that the expressions for $\vec Q$ remain
identical in structure, but now all indexed quantities are ``dressed''
with $\gamma$ matrices (for downstairs indices) or $\td\gamma$
matrices (for upstairs indices).  The underlying reason for this
``dressing'' phenomenon is a silver rule of supergravity, namely that
twisted self-duality holds for the ``flattened" field strengths that
transform under the subgroup $K$, whereas the natural potentials
transform under the full group $G$ (see, for instance,
\cite{JJ}). This will become obvious when we study the coset
construction of the doubled field in the next subsection.  It may be
useful to recognise that the matrices $\gamma$ and $\td\gamma$ are
actually moving frames intertwining between analogues of curved and
flat indices, which unfortunately have not been distinguished here.

The expressions for $\vec Q$ that we find in each dimension are given
in appendix A.  We can then strip off a derivative from
(\ref{phi2o}), giving the first-order equation
\bea
{*d\vec\phi} &=& -\ft12 (-1)^D\, \vec a\, A_\3\, d\tA_{(D-5)} +
\ft12 \sum_i \vec a_i\, \gamma^j{}_i\, \td\gamma^i{}_k\,
A_{\2 j}\, d\tA_{(D-4)}^k \nn\\
&&-\ft12 (-1)^D\, \sum_{i<j} \vec a_{ij}\,
\gamma^\ell{}_i\,
\gamma^m{}_j\, \td\gamma^i{}_p\, \td\gamma^j{}_q\,
A_{\1 \ell m}\, d\tA_{(D-3)}^{pq}\nn\\
&&+\ft12 \sum_{i<j<k} \vec a_{ijk}\, \gamma^\ell{}_i\, \gamma^m{}_j\,
\gamma^n{}_k\, \td\gamma^i{}_p\, \td\gamma^j{}_q\, \td\gamma^k{}_r\,
A_{\0\ell mn}\, d\tA_{(D-2)}^{pqr}\nn\\
&&-\ft12(-1)^D\, \sum_i \vec b_i\, \td\gamma^i{}_j\, \gamma^k{}_i\,
\hA_\1^j\, \Big(d\tcA_{(D-3)k} - A_{\2 k}\, d\tA_{(D-5)} + (-1)^D\,
A_{\1\ell k}\, d\tA_{(D-4)}^\ell\nn\\
&&\qquad\qquad\qquad\qquad\qquad\qquad
-\ft12 A_{\0 k\ell m}\, d\tA_{(D-3)}^{\ell m} -Y_k\Big)\nn\\
&&+\ft12 \sum_{i<j} \vec b_{ij}\, \gamma^k{}_i\, {\cal A}_{\0 j}^i\, d
\tcA_{(D-2)}^j{}_k -\ft12 \vec Q\ .\label{phi1o}
\eea
Detailed calculation shows that the exterior derivative of this equation
indeed gives (\ref{phi2o}).

\subsection{``Coset" construction for $D$-dimensional supergravity}

     Having obtained the first-order equations in the previous
section, we can now write down the doubled field $\G$, and then look
for a parameterisation for $\v$, such that $\G=d\v\, \v^{-1}$.
Thus as before, the doubled field is written as
\bea
\G&=& \ft12 d\vec\phi\cdot\vec H  +\sum_{i<j} e^{\fft12 \vec
b_{ij}\cdot \vec\phi}\, {\cal F}^i{}_{\1 j}\, E_i{}^j +\sum_i 
e^{\fft12 \vec
b_i \cdot\vec\phi}\, {\cal F}_\2^i\, W_i + \sum_{i<j<k} e^{\fft12 \vec
a_{ijk}\cdot\vec\phi}\, F_{\1 ijk}\, E^{ijk} \nn\\
&&+\sum_{i<j} e^{\fft12 \vec
a_{ij}\cdot\vec\phi}\, F_{\2 ij}\, V^{ij} +\sum_{i} e^{\fft12 \vec
a_{i}\cdot\vec\phi}\, F_{\3 i}\, V^{i} + e^{\fft12 \vec a\cdot\vec\phi}\,
F_\4\, V +e^{-\fft12 \vec a\cdot\vec\phi}\,\tF_{(D-4)}\, \tV \nn\\
&&+ \sum_{i} e^{-\fft12 \vec a_{i}\cdot\vec\phi}\, \hF_{(D-3)}^i\, \tV_{i}
+\sum_{i<j} e^{-\fft12 \vec
a_{ij}\cdot\vec\phi}\, \hF_{(D-2)}^{ij}\, \tV_{ij}+ 
\sum_{i<j<k} e^{-\fft12 \vec
a_{ijk}\cdot\vec\phi}\, \hF_{(D-1)}^{ijk}\, \tE_{ijk}\nn\\
&&+\sum_i e^{-\fft12 \vec
b_i \cdot\vec\phi}\, \hat{\cal F}_{(D-2)i}\, \tW^i +\sum_{i<j} e^{-\fft12
\vec b_{ij}\cdot \vec\phi}\, \hat{\cal F}^j{}_{(D-1)i}\, \tE^i{}_j
+\ft12 F_{\psi}\, \vec{\tH}\ ,\label{master}
\eea
where the hatted dual field strengths are dressed (one could say 
flattened on all their internal  indices) with $\gamma$
and $\td\gamma$ matrices in the systematic way:
\bea
\hF_{(D-3)}^i &=& \td\gamma^i{}_j\, \tF_{(D-3)}^j\ ,\quad
\hF_{(D-2)}^{ij} = \td\gamma^i{}_k\, \td\gamma^j{}_\ell\,
\tF_{(D-2)}^{k\ell}\ ,\quad
\hF_{(D-1)}^{ijk} = \td\gamma^i{}_\ell\, \td\gamma^j{}_m\, \td\gamma^k{}_n
\, \tF_{(D-1)}^{\ell mn}\ ,\nn\\
\hat{\cal F}_{(D-2)i} &=& \gamma^j{}_i\, \tF_{(D-2)j} \ ,\qquad
\hat{\cal F}^j_{(D-1)i} = \gamma^k{}_i\,\td\gamma^j{}_\ell \,
\tF^\ell_{(D-1)k}\ ,
\eea
and $F_{\vec\psi}$ is defined to be the right-hand side of the
first-order dilaton equation (\ref{phi1o}).

     We shall parameterise $\v$ as
\bea
\v &=& e^{\fft12\vec\phi\cdot\vec H}\, h\, e^{\hA_\1^i\, W_i}\,
e^{\fft16 A_{\0 ijk}\, E^{ijk}}\, e^{\fft12 A_{\1 ij}\, V^{ij}}\,
e^{A_{\2 i}\, V^i}\, e^{A_\3\, V}\times\label{gencoset}\\
&& \times e^{\tA_{(D-5)}\, \tV}\,
e^{\tA_{(D-4)}^i\, \tV_i}\, e^{\fft12 \tA_{(D-3)}^{ij}\, \tV_{ij}}\,
e^{\fft16 A_{(D-2)}^{ijk}\, \tE_{ijk}}\,
e^{\tcA_{(D-3)i}\, \tW^i}\, e^{\tcA_{(D-2)i}^j\, \tE^i{}_j}\,
e^{\fft12\vec\psi\cdot\vec{\tH}}\ ,\nn
\eea
where $h$ is defined as the product
\be
h= \prod_{i<j} e^{{\cal A}^i_{\0 j}\, E_i{}^j}\ ,\label{hbor}
\ee
with the terms arranged in anti-lexical order, namely
\be
(i,j)=\cdots  (3,4), (2,4), (1,4), (2,3), (1,3), (1,2)\ .
\ee
Note that only in the terms involving ${\cal A}_{\0 j}^i$ is it
necessary to separate the individual fields of an $SL(11-D,\R)$
multiplet into separate exponential factors.

     By comparing the terms in the doubled field (\ref{master}) that
are bilinear in fields with the bilinear terms coming from $d\v\,
\v^{-1}$, we can read off all the commutation and anti-commutation
relations for the generators.  Note that a generator is {\it even} if
it is associated in (\ref{gencoset}) with a potential of {\it even}
degree, and it is {\it odd} if it is associated with a potential of
{\it odd} degree.  Two odd generators satisfy an anticommutation
relation, while all other combinations satisfy commutation relations.
We have seen in the previous subsection that the bosonic Lie derivatives
become partly fermionic when we change the statistics of the internal 
coordinates by treating them separately from the remaining spacetime 
coordinates.
 
    The (anti)-commutators divide into two sets. There are those that
are independent of the dimension $D$; we shall present these first.
Then, there are additional (anti)-commutators that are specific to the
dimension; these are all associated with terms in (\ref{master})
coming from the Chern-Simons terms ${\cal L}_{FFA}$, and consequently
they all involve the epsilon tensor.  These are given dimension by
dimension in appendix B.  We find that the dimension-independent
commutators are as follows.  Firstly, we have the commutators of all
generators with $E_i{}^j$, which characterise their
$SL(11-D,\R)$-covariance properties:
\bea
&&{[} E_i{}^j, E_k{}^\ell {]} = \delta^j_k\, E_i{}^\ell -
\delta^\ell_i\, E_k{}^j\ ,\qquad
{[}E_i{}^j, E^{k\ell m} {]} = - 3\delta^{[k}_i\, E^{\ell m]j}\ ,\nn\\
&&{[} E_i{}^j, V^k{]} = -\delta^k_i\, V^j\ ,\qquad
  {[} E_i{}^j, V^{k\ell}{]} = 2\delta^{[k}_i\, V^{\ell] j}\ ,\qquad
   {[}E_i{}^j, W_k{]} = \delta^j_k\, W_i\ ,\nn\\
&&{[}E_i{}^j, \tE_{k\ell m} {]} = 3\delta^j_{[k}\, \tE_{\ell m]i} \ ,\qquad
{[}E_i{}^j, \tW^k {]} = -\delta^k_i\, \tW^j\ ,\nn\\
&&{[}E_i{}^j, \tV_k{]} = \delta^j_k\, \tV_i\ ,\qquad
   {[}E_i{}^j, \tV_{k\ell}{]} = -2\delta^j_{[k}\, \tV_{\ell]i}\ ,\nn\\
&&{[}E_i{}^j, \tE^i{}_\ell{]} = -\tE^j{}_\ell\ ,\hbox{\ no\ sum\ on\ $i$,
\ $j\ne\ell$}\ ,\nn\\
&&{[} E_i{}^j, \tE^k{}_j {]} = \tE^k{}_i\ , \hbox{\ no\ sum\ on\ $j$,
\ $i\ne k$}\ .\label{slnr}
\eea
Next, we have the commutators of all generators with $\vec H$, which
 are expressed in terms of the roots under the chosen Cartan subalgebra:
\bea
&&{[}\vec H, E_i{}^j {]} = \vec b_{ij}\, E_i{}^j\ ,\qquad
  {[}\vec H, E^{ijk} {]} = \vec a_{ijk}\, E^{ijk}\ ,\qquad
  {[}\vec H, V^{ij} {]} = \vec a_{ij}\, V^{ij}\ ,\nn\\
&&{[}\vec H, V^i {]} = \vec a_i\, V^i\ ,\qquad
  {[}\vec H, V {]} = \vec a\, V\ ,\qquad
  {[}\vec H, W_i {]} = \vec b_i\, W_i\ , \nn\\
&&{[}\vec H, \tE^i{}_j {]} = -\vec b_{ij}\, \tE^i{}_j\ ,\qquad
  {[}\vec H, \tE_{ijk} {]} = -\vec a_{ijk}\, \tE_{ijk}\ ,\qquad
  {[}\vec H, \tV_{ij} {]} = -\vec a_{ij}\, \tV_{ij}\ ,\nn\\
&&{[}\vec H, \tV_i {]} = -\vec a_i\, \tV_i\ ,\qquad
  {[}\vec H, \tV {]} = -\vec a\, \tV\ ,\qquad
  {[}\vec H, \tW^i {]} = -\vec b_i\, \tW^i\ . \label{weights}
\eea
The rest of the dimension-independent commutators are:
\bea
&&{[} W_i, E^{jk\ell}{]} = - 3\delta_i^{[j}\, V^{k\ell]}\ ,\qquad
\{ W_i, V^{jk} \} = -2\delta_i^{[j}\, V^{k]}\ ,\qquad
{[} W_i, V^j{]} = -\delta_i^j\, V\ ,\nn\\
&&{[} W_i, \tV \}= -\tV_i\ ,\qquad
{[} W_i, \tV_j \} = \tV_{ij}\ ,\qquad
{[} W_i, \tV_{jk} \} = -\tE_{ijk}\ ,\nn\\
&&{[} V^i, \tV{]} = -\tW^i\ ,\qquad
   {[}V^{ij}, \tV_k \} = -2 \delta^{[i}_k\, \tW^{j]}\ ,\qquad
   {[} E^{ijk}, \tV_{\ell m}{]} = -6 \delta_\ell^{[i}\, \delta_m^j\,
    \tW^{k]}\ ,\nn\\
&&{[} V^i, \tV_j{]} = -\tE^i{}_j\ ,\qquad
   {[}V^{ij}, \tV_{k\ell} \} = 4 \delta^{[i}_{[k}\, \tE^{j ]}{}_{\ell]}\ ,
    \label{rest}\\
&&{[}E^{ijk}, \tE_{\ell mn}{]} = -18 \delta^{[i}_{[\ell}\,
     \delta^j_m\, \tE^{k]}{}_{n]}\ ,\qquad
   {[}W_i, \tW^j \} = - \tE^j{}_i\ , \nn\\
&&{[} V, \tV \} = -\ft14 \vec a\cdot\vec{\tH}\ ,\qquad
   {[}V^i, \tV_i{]} = \ft14 \vec a_i\cdot\vec{\tH}\ ,\qquad
   {[}V^{ij}, \tV_{ij} \} = -\ft14\vec a_{ij}\cdot\vec{\tH}\ ,\nn\\
&&{[}E^{ijk}, \tE_{ijk}{]} = \ft14\vec a_{ijk}\cdot\vec{\tH}\ ,\qquad
   {[} W_i, \tW^i \} = -\ft14 \vec b_{i}\cdot\vec{\tH}\ ,\qquad
   {[} E_i{}^j, \tE^i{}_j {]} = \ft14 \vec b_{ij}\cdot\vec{\tH}\ .\nn
\eea

    It is straightforward to verify, with the aid of a computer, that
these commutation and anti-commutation relations indeed satisfy the
Jacobi identities.  We have also verified that the augmented set of
commutation relations in each dimension $D$, where we include also the
dimension-dependent ones given in appendix B,  satisfy the Jacobi
identities.\footnote{In the next subsection, after rewriting the
(anti)-commutation relations in a more transparent form, we shall be
able to present a more digestible proof of the Jacobi identities.}

\subsection{A twelfth (fermionic) dimension}

    These algebras in $D$ dimensions can be written in a considerably
more elegant and transparent form.  To do this, we first extend the
range of the $i,\ldots$ indices to $\a=(i,0)$, where 0 will, for
convenience, be formally defined to be larger than any of the values
taken by $i$.  Thus we have an enlargement from $n=(11-D)$ bosonic
dimensions to $(n|1)$, with the extra dimension turning out to be
fermionic.  We define the extended generators $E_\a{}^\b$,
$\tE^\a{}_\b$, $V^{\a\b\g}$ and $\tV_{\a\b\g}$ as follows:
\bea
&&E_i{}^j = E_i{}^j\ ,\qquad E_i{}^0 =W_i \ ,\nn\\
&&V^{ijk} = E^{ijk}\ ,\qquad V^{ij0} = V^{ij}\ ,\qquad
  V^{i00} =V^i\ ,\qquad V^{000}= V\ ,\nn\\
&&\tE^i{}_j = \tE^i{}_j\ ,\qquad \tE^i{}_0 = -\tW^i \ ,\nn\\
&&\tV_{ijk} = \tE_{ijk}\ ,\qquad \tV^{ij0} = -\tV_{ij}\ ,\qquad
  \tV_{i00} = 2\tV_i\ ,\qquad \tV_{000}= -6 \tV\ . \label{extgen}
\eea
The generators $V^{\a\b\g}$ and $\tV_{\a\b\g}$ are graded
antisymmetric, {\it i.e.}\ $V^{\a\b\g}=V^{[\a\b\g\}}$ and
$\tV_{\a\b\g}=\tV_{[\a\b\g\}}$.  In other words, the index 0 is
symmetrised with itself but the other $i$ indices are antisymmetrised.

    In terms of these generators, the dimension-independent part
(\ref{slnr}), (\ref{weights}) and (\ref{rest}) of the algebras can be
re-written as
\bea
&&{[} E_\a{}^\b , E_\g{}^\d \} = \d_\g^\b\, E_\a{}^\d -
  \d_\a^\d\, E_\g{}^\b\ ,\qquad
  {[}\vec H, E_\a{}^\b {]} = \vec b_{\a\b}\, E_\a{}^\b\ ,\nn\\
&&{[}E_\a{}^\b, V^{\g\d\s} \} = -3\, \d_\a^{[\g}\, V^{\d\s\} \b}\ ,
\qquad {[} \vec H, V^{\a\b\g} {]} =\vec a_{\a\b\g} \,  V^{\a\b\g} 
\ ,\nn\\
&&{[}E_\a{}^\b, \tE^\g{}_\d \} = -\d_\a^\g\, \tE^\b{}_\d +
  \d_\d^\b\, \tE^\g{}_\a + \ft14 \d_\d^\b\, \d_\a^\g\, 
  \vec b_{\a\b}\cdot\vec{\tH} \ ,\quad
  {[} \vec H, \tE^\a{}_\b {]} = -\vec b_{\a\b}\, \tE^\a{}_\b
\ ,\nn\\
&&{[} E_\a{}^\b , \tV_{\g\d\s} \} = 3\, \d^\b_{[\g}\, 
  \tV_{\d\s\} \a}\ ,\qquad
  {[} \vec H, \tV_{\a\b\g} {]} = -\vec a_{\a\b\g}\, 
    \tV_{\a\b\g}\ ,\nn\\
&&{[} V^{\a\b\g}, \tV_{\d\s\l} \} = -18\, \d_{[\d}^{[\a}\, 
  \d_{\s}^{\b}\, \tE^{\g \}}{}_{\l \}} + \ft32\, 
  \d^{\a\b\g}_{\d\s\l}\, \vec a_{\a\b\g}\cdot\vec{\tH}\ ,\label{transp}
\eea
while the dimension-dependent (anti)-commutators, given in appendix B, can
all be expressed in the simple form
\be
{[} V^{\a_1\a_2\a_3}, V^{\b_1\b_2\b_3} \} = \ft16 \, (-1)^{11-D}\,
  \ep^{\b_1\b_2\b_3\a_1\a_2\a_3\g_1\g_2\g_3}\, \tV_{\g_1\g_2\g_3}
\ . \label{ddep}
\ee
Here, the 9-index $\ep$ tensor is graded antisymmetric with
$\e^{i_1\cdots i_n 0\cdots 0}$  simply equal to $\ep^{i_1\cdots
i_n}$ when the number of 0's is $D-2$
but vanishes otherwise.  In (\ref{transp}) we have defined
\be
\vec b_{i0} = \vec b_i\ ,\qquad \vec a_{ij0} = \vec a_{ij}\ ,\qquad
\vec a_{i00}= \vec a_i\ ,\qquad \vec a_{000} = \vec a\ .
\ee

    As we shall now explain, the algebra (\ref{transp}) contains the
superalgebra $SL_+(n|1)$ (the Borel subalgebra of $SL(n|1)$),
generated by $\vec H$ and $\tE_\a{}^\b$ with $\a<\b$ (subject, as
before, to the formal rule that the 0 value is regarded as being
greater than any other value $i$).  It is clear that the Cartan
generators $\vec H$ should be expressible in terms of the diagonal
generators $E_\a{}^\b$ with $\a=\b$, and likewise we would expect that
$\vec{\tH}$ should be expressible in terms of the diagonal generators
amongst the $\tE^\a{}_\b$.  The expressions for the Cartan generators
of $SL(n|1)$ in terms of the $E_\a{}^\a$ are presented in appendix C,
equation (\ref{sl1sncart}).  In our case, we may therefore write
\be
\vec H = \sum_{\a=0}^n (\vec b_\a+\vec c)\, E_\a{}^\a\ ,
\ee
where $\vec c= \ft{9}{n-1}\, \vec s$, and by definition $\vec
b_0\equiv 0$.  It is useful also to ``invert'' this expression, and
give the diagonal generators $E_\a{}^\b$ in terms of $\vec H$.  Of
course there will not be a unique result, since there are only $n$
Cartan generators $\vec H$ while there are $(n+1)$ diagonal
generators.  In fact the sum $\sum_\a E_\a{}^\a$ commutes with all
generators in $SL(n|1)$.  Rather than simply setting it to the
identity (or zero), however, it is useful instead to introduce one
additional generator ${\cal D}$, a ``dilatation,'' which commutes with
all the generators $E_\a{}^\b$ for all $\a$ and $\b$.  In terms of
this, it turns out that the solution for the diagonal generators is
\bea
E_0{}^0 &=& \ft12 (D-2)\, \vec s\cdot\vec H - {\cal D}\ ,\nn\\
E_i{}^i &=& \ft12(\vec b_i + \vec s)\cdot\vec H + {\cal D}\ .
\eea
Another useful expression is obtained by considering the simple-root
Cartan generators $K_\a$, defined by
\be
K_0 \equiv E_n{}^n + E_0{}^0\ ,\qquad K_i \equiv E_i{}^i
-E_{i+1}{}^{i+1}\ .
\ee
In terms of these, we find that 
\be
\vec H = \vec c\, K_0 +\sum_i \vec\beta_i\, K_i\ ,\qquad {\rm
where}\qquad \vec \beta_i = \sum_{j=1}^i (\vec b_j + \vec c)\ .
\ee
For the diagonal $\tE^\a{}_\a$ generators, we find that they are
related to the generators $\vec {\tH}$ by
\be
\tE^\a{}_\a = \ft14 (\vec b_\a + \vec c)\cdot \vec {\tH}\ .
\ee

    With the introduction of the additional generator ${\cal D}$, the
original $SL_+(n|1)$ algebra generated by $\vec H$ and $E_\a{}^\b$
with $\a<\b$ is in fact enlarged to $GL_+(n|1)$.  By doing this, it
turns out that the 3-tensors $V^{\a\b\g}$ and $\tV_{\a\b\g}$, which
were previously seen to be irreducible representations of the Borel
subalgebra $SL_+(n|1)$ of $SL(n|1)$, can be viewed also as irreducible
(tensor density) representations of the full algebra $GL(n|1)$.  To do
this, we assign dilatation weights to $V^{\a\b\g}$ and $\tV_{\a\b\g}$
as follows:
\be
{[} {\cal D}, V^{\a\b\g} {]} = \ft16(D-3)\, V^{\a\b\g}\ ,\qquad
{[} {\cal D}, \tV_{\a\b\g} {]} = -\ft16(D-3)\, \tV_{\a\b\g}\ .
\ee
We then find that the algebra (\ref{transp}) is a subalgebra of the
simpler superalgebra: 
\bea
&&{[} E_\a{}^\b , E_\g{}^\d \} = \d_\g^\b\, E_\a{}^\d -
  \d_\a^\d\, E_\g{}^\b\ ,\nn\\
&&{[}E_\a{}^\b, V^{\g\d\s} \} = -3\, \d_\a^{[\g}\, V^{\d\s\} \b} 
+ \ft16(D-1)\, \d_\a^\b\,  V^{\g\d\s} 
\ ,\nn\\
&&{[}E_\a{}^\b, \tE^\g{}_\d \} = -\d_\a^\g\, \tE^\b{}_\d +
  \d_\d^\b\, \tE^\g{}_\a \ ,\nn\\
&&{[} E_\a{}^\b , \tV_{\g\d\s} \} = 3\, \d^\b_{[\g}\, 
  \tV_{\d\s\} \a} - \ft16(D-1)\, \d_\a^\b\, \tV_{\g\d\s}\ ,\nn\\
&&{[} V^{\a\b\g}, \tV_{\d\s\l} \} = -18\, \d_{[\d}^{[\a}\, 
  \d_{\s}^{\b}\, \tE^{\g \}}{}_{\l \}}\ ,\label{transp2}
\eea
where now the index ranges on $E_\a{}^\b$ and $\tE^\a{}_\b$ are simply
restricted by the Borel subalgebra conditions $\a\le\b$ (again with
the understanding that 0 is larger than any $i$).

       The $D$-dimensional algebra (\ref{transp}) has the form
\bea
{[} T^{\sst{A}}, T^{\sst B} {\}} &=& f^{\sst{AB}}{}_{\sst C}\, T^{\sst
C} + g^{\sst {ABC}}\, \wtd T_{\sst C}\ ,\nn\\
{[} T^{\sst A}, \wtd T_{\sst B} {\}} &=&  f^{\sst{CA}}{}_{\sst B}\, 
\wtd T_{\sst C}\ ,\label{genalg}\\
{[} \wtd T_{\sst A}, \wtd T_{\sst B} {\}} &=& 0\ ,\nn
\eea
where $T^{\sst{A}}$ represents the untilded generators $\{ E_\a{}^\b,
\vec H, V^{\a\b\g} \}$, while $\wtd T_{\sst A}$ represents the tilded
(dual) generators $\{ \tE^\a{}_\b, \vec{\tH}, \tV_{\a\b\g} \}$.  This
dual structure is the central point of our discussion and we are
naturally led to consider a splitting (a quantum mechanics or
symplectic geometer would say a polarisation) of the doubled set of
potentials into two halves; on the one hand the fundamental fields to
be used in a Lagrangian, and on the other their dual potentials. This
splitting is not unique, as our previous paper has testified
\cite{cjlp}, but the most obvious choice is the one implied by
selecting precisely the untilded generators as the fundamental ones.
Let us call $\P$ the vector space of superalgebra generators of the
fundamental fields. The dual potentials live in the dual space
$\P^*$. Let us remark that the second line of eq. (\ref{genalg})
becomes just the transformation law of the super-coadjoint
representation \cite{Stens} (in other words super-contragredient to
the adjoint) after we set the $g^{\sst{ABC}}$ equal to zero.

    It is worth remarking that the terms appearing on the right-hand
side of ${[} T^{\sst A}, \wtd T_{\sst B} {\}}$ can be deduced very
easily from the right-hand side of ${[} T^{\sst{A}}, T^{\sst B} {\}}$,
by using the following ``Jade Rule.''  This rule states that if we
have untilded generators $X$, $Y$ and $Z$ where ${[}X,Y\} = Z$, then
it follows that we will necessarily also have ${[}X,\wtd Z \} =
(-1)^{XY+1}\, \wtd Y$.  One can easily see from (\ref{genalg}) that
the jade rule is equivalent to the statement that
$f^{\sst{AB}}{}_{\sst C}$ is graded (anti)-symmetric in its upper
indices.  In our algebras, the origins of the jade rule can be traced
back to the supergravity theories from which we first derived the
(anti)-commutation relations: If $A_{\sst X}$, $A_{\sst Y}$ and
$A_{\sst Z}$ are the three potentials associated with the generators
$X$, $Y$ and $Z$, then a ``Chern-Simons type" modification in a field
strength, of the form $F_{\sst Z} = dA_{\sst Z} + A_{\sst Y} \wedge
F_{\sst X}$, leads both to a ``Bianchi identity contribution''
$dF_{\sst Z} \sim F_{\sst Y} \wedge F_{\sst X}$ and a ``field equation
contribution'' $d\wtd F_{\wtd{\sst Y}} \sim \wtd F_{\wtd{\sst
Z}}\wedge F_{\sst X}$, where $\wtd F_{\wtd{\sst Y}}$ and $\wtd
F_{\wtd{\sst Z}}$ are the duals of the field strengths $F_{\sst Y}$
and $F_{\sst Z}$.

     It is easy to check the jade rule in examples.  For instance, we
see from (\ref{rest}) that ${[} W_i, V^j{]} = -\delta_i^j\, V$.  By
the jade rule, there should therefore also be an (anti)-commutator
${[} W_i, \wtd V \} = - \wtd V_i$, and indeed we see it too in
(\ref{rest}).  (On has to be careful to make the sign reversals
indicated in the bottom line of (\ref{extgen}) before applying the
jade rule, in order to get precise agreement.)

  Before discussing further specific details of our algebra, it is of
interest to consider the general class of algebras defined by
(\ref{genalg}).  We now turn to this in the next subsection.

\subsection{Deformation of ``cotangent algebras"}

It is easily seen that the Jacobi identities for (\ref{genalg}) impose the
following requirements on the structure constants:
\bea
f^{ {[} \sst{AB} }{}_{\sst D}\, f^{ {|} \sst{D} {|} \sst{C} \}}{}_{\sst E} 
&=&0\ ,\nn\\
f^{{[}\sst{AB} }{}_{\sst D}\, g^{\sst{ {|} D|C} \} \sst{E}} +
f^{\sst{E} {[} \sst{C}}{}_{\sst D}\, g^{\sst{AB} \} \sst{D}} &=& 0\ .
\label{jac1}
\eea
The first is just the usual requirement for the $T^{\sst A}$
generators to form a superalgebra after contracting away the tilded
generators.\footnote{Note that the $\wtd T_{\sst A}$ generators carry
the same indices as the $T^{\sst A}$ generators.  In odd dimensions,
$\wtd T_{\sst A}$ and $T^{\sst A}$ have opposite statistics, and so it
becomes important in general to distinguish the statistics of the
indices from that of the generators.  Any graded representation leads
to a BF exchanged representation upon shifting the gradation by one,
and the corresponding semi-direct products are different. The symmetry
property of the structure constants changes accordingly.  Our
convention will be that an index $\sst{A}$ {\it always} has the
statistics corresponding to $T^{\sst A}$, regardless of whether in a
particular commutator it is actually associated with $T^{\sst A}$ or
$\wtd T_{\sst A}$.  In other words, the statistics factor
$(-1)^{T^{\sst A}}$ is associated with the index $A$, where we use the
standard convention that an odd generator $X$ has $(-1)^X = -1$, while
if it is even it has $(-1)^X =+1$.  An example is the first Jacobi
identity in (\ref{jac1}), which requires that the symmetry of the
indices appearing on $f^{\sst CA}{}_{\sst B}$ in the second line in
(\ref{genalg}) is the same as in the first line in (\ref{jac1}), where
none of the indices on $f^{\sst{AB}}{}_{\sst C}$ is linked to a tilded
generator.}  There is no unique solution for the additional structure
constants $g^{\sst{ABC}}$.  One solution would be to take
$g^{\sst{ABC}}$ to be proportional to $f^{\sst{ABC}}$, where the lower
index of $f^{\sst{ AB}}{}_{\sst C}$ is raised using the Cartan-Killing
metric $h^{\sst{AB}} \equiv f^{\sst{AC}}{}_{\sst D}\,
f^{\sst{BD}}{}_{\sst C}$; $f^{\sst{ABC}} = h^{\sst{CD}}\,
f^{\sst{AB}}{}_{\sst D}$.  However, this is not the solution that
arises in our case, as we shall see in detail below.  In our algebras
the $g^{\sst{ABC}}$ are the structure constants associated with the
dimension-dependent commutation relations given by (\ref{ddep}).
These terms have their origin in the ${\cal L}_{FFA}$ terms in the
Lagrangian.

   In order to understand these algebras, we found it useful first to
contract them by setting $g^{\sst{ABC}}=0$. In this way the vector
space $\P$ acquires a Lie algebra structure $G$.  This can be achieved
by rescaling the $\wtd T_{\sst A}$ generators to zero, giving rise to
\be
{[} T^{\sst A}, T^{\sst B} {\}} = f^{\sst{AB}}{}_{\sst C}\, T^{\sst C}
\ ,\qquad
{[} T^{\sst A}, \wtd T_{\sst B} {\}} = f^{\sst{CA}}{}_{\sst B}\, 
\wtd T_{\sst C}\ ,\qquad
{[} \wtd T_{\sst A}, \wtd T_{\sst B} {\}} = 0\ .\label{genalg1}
\ee
We shall keep the same notation for the (rescaled) generators $\wtd
T_{\sst A}$ as there will be no ambiguity.  Then the full (contracted)
algebra becomes $G\semi {G^*}$, where $G$ is generated by $T^{\sst
A}$, and in the semi-direct product ${G^*}$, generated by $\wtd
T_{\sst A}$, denotes the co-adjoint representation of $G$.  We will
see in the next section that it has the same general structure as the
algebra (\ref{doubalg}) in the doubled formalism describing the sigma
model Lagrangians.

    Here $G$ is generated by $\vec H$, $E_\a{}^\b$, and $V^{\a\b\g}$.  In
particular $\vec H$ and $E_i{}^j$ form the Borel subalgebra of
$GL(n,\R)$, denoted by $GL_+(n,\R)$, where $n=11-D$.  The
$E_i{}^0=W_i$ generators are odd, and are associated with internal
diffeomorphisms.  Thus the generators $\vec H$, $E_\a{}^\b$, which are
all associated with Kaluza-Klein fields, generate the Borel subalgebra
$SL_+(n|1)$ of the superalgebra $SL(n|1)$ (see appendix C for details
and a more general and purely bosonic discussion).  The generators
$V^{\a\b\g}=\{V, V^i, V^{ij}, E^{ijk}\}$, associated with the fields
coming from the 3-form potential in $D=11$, form a linear graded
antisymmetric 3-tensor representation of $SL(n|1)$.  Thus the algebra
$G$ for $D$-dimensional supergravity can be denoted by
\be
G=SL_+(n|1) \semi (\wedge v)^3\ .
\label{Ggroup}
\ee
with $v$ the appropriate fundamental representation.  Note that $G$ is
associated with the complete set of ``gauge'' symmetries of the
$D$-dimensional supergravity coming from the dimensional reduction
from $D=11$ (comprising both local gauge symmetries for vector and
tensor potentials, and global symmetries for scalars and axionic
potentials). The contraction occurs by considering only the action on
the fundamental potentials obtained without any dualisation.

      Having understood the contracted algebra where the $g^{\sst{
ABC}}$ coefficients are set to zero, we may now study the deformation
of this algebra where the $g^{\sst{ABC}}$ are non-vanishing.  It is
useful first to note that the analysis of the algebra (\ref{genalg})
can be further refined.  We may divide the generators into two
subsets, labelled generically by $T^a$ to denote the subset $\{\vec H,
E_\a{}^\b\}$ which are the $SL_+(n|1)$ generators, and by $T^{\bar a}$
to denote the 3-tensor representation generators $V^{\a\b\g}$.  The
dual generators $\widetilde T_{\sst A}$ are correspondingly split as
$\widetilde T_a$ representing $\vec{\tH}$ and $\tE^\a{}_\b$, and
$\widetilde T_{\bar a}$ representing $\tV_{\a\b\g}$.  In the algebras
that we encounter in the dimensionally-reduced supergravities, certain
of the sets of structure constants in the refined form of
(\ref{genalg}) vanish, and we have
\bea
&&{[} T^a, T^b \} = f^{ab}{}_c\, T^c\ ,\qquad 
  {[} T^a, T^{\bar b} \} = f^{a\bar b}{}_{\bar c}\, T^{\bar c}\ ,\qquad 
  {[} T^{\bar a}, T^{\bar b} \} = g^{\bar a\bar b \bar c}\, \wtd
T_{\bar c}\ ,\nn\\
&&{[} T^a, \wtd T_b \} = f^{ca}{}_b\, \wtd T_c\ ,\qquad 
  {[} T^a, \wtd T_{\bar b} \} = f^{\bar c a}{}_{\bar b}\, 
\wtd T_{\bar c}\ ,\qquad 
  {[} T^{\bar a}, \wtd T_{\bar b} \} = f^{c\bar a}{}_{\bar b}\, \wtd
T_c\ .\label{genalg2}
\eea
In particular, we see that the deformation $g^{\sst{ABC}}$ arises 
in the sector $g^{\bar a \bar b \bar c}$, describing (anti)-commutators 
involving only the tensor-representation generators $V^{\a\b\g}$. 

    Before discussing the specific algebras of the doubled formalism
in more detail, it is again worthwhile to give the Jacobi identities
for general algebras of the form (\ref{genalg2}).  Aside from the
obvious Jacobi requirements for the structure constants $f^{ab}{}_c$
and $f^{a\bar b}{}_{\bar c}$ themselves, we find that the conditions
on the deformations $g^{\bar a \bar b \bar c}$ can be written as
\bea
&&\!\!\! f^{e [\bar a}{}_{\bar d}\, g^{\bar b \bar c \}\bar d} = 0\ ,
\label{jac2}\\
&&\!\!\! (-1)^{\bar c \bar d}\, f^{a\bar b}{}_{\bar d}\, 
   {[}g^{\bar c\bar d \bar
e} +(-1)^{\bar e(\bar c+\bar d)+\bar c\bar d}\, g^{\bar c\bar e \bar d}{]}
-(-1)^{a\bar b}\, f^{a\bar c}{}_{\bar d}\, {[}g^{\bar b \bar d\bar e} +
(-1)^{\bar e(\bar b+\bar d) + \bar b \bar d}\, g^{\bar b \bar e \bar
d}{]}= 0\ .\nn
\eea
The first condition is, modulo the second condition, the statement
that the deformation constants $g^{\bar a\bar b \bar c}$ form an
invariant tensor of $G$.  A simple way of satisfying the second
condition is by requiring that each of the two terms vanishes
separately, which is achieved by the single requirement
\be
g^{\bar a \bar b\bar c} = (-1)^{\bar a\bar b+\bar b\bar c + \bar a\bar c +
1}\, g^{\bar a\bar c\bar b}\ .\label{req}
\ee
The symmetry in the first two indices is of course dictated by
(\ref{genalg2}), {\it i.e.}\ $g^{\bar a\bar b\bar c} = (-1)^{\bar
a\bar b+1} g^{\bar b\bar a\bar c}$.  It should be emphasised again
that the statistics of the indices are determined by the untilded
generators $T^{\bar a}$, and not the tilded generators $\wtd T_{\bar
a}$. Using this graded antisymmetry it is now an easy exercise to
rewrite (\ref{req}) as graded antisymmetry of the $g^{\bar a\bar b\bar
c}$ in $\bar a$ and $\bar c$. Returning now to our concrete situation
we can check explicitly all the Jacobi identities.
 
    The structure constants $f^{ab}{}_c$ and $f^{a\bar b}{}_{\bar c}$
appear in the commutation relations given in (\ref{transp}), and, as
we discussed above, they are the structure constants of the
superalgebra $G\semi G^*$, where $G$ is given in (\ref{Ggroup}).  The
deformation coefficients $g^{\bar a\bar b\bar c}$ can be read off from
the dimension-dependent commutation relations given in (\ref{ddep});
it can now be easily verified that they satisfy the requirement
(\ref{req}), and satisfy the Jacobi conditions (\ref{jac2}).

     Having obtained all the commutators by comparing the bilinear
terms in the doubled field (\ref{master}) with the bilinear terms from
$d\v \, \v^{-1}$, it is now a matter of detailed computation to verify
that the complete calculation of $d\v\, \v^{-1}$ gives rise to the
complete expression (\ref{master}) for the doubled field.  In
particular, this depends crucially on the ordering of the various
factors in the expression (\ref{gencoset}) for $\v$.  This ordering is
dictated by the strategy that we followed when stripping off the
derivatives from the second-order equations to obtain first-order
equations.  Namely, whenever a derivative was to be extracted from an
expression such as $dB\wedge dB'$, where $B$ and $B'$ represent two
potentials in the theory, we always extracted the derivative so that
the potential {\it further to the left} in (\ref{gencoset}) lost its
derivative.  Thus if $B$ appears further to the left than $B'$, we
extract the derivative by writing $dB\wedge dB'$ as $d(B\wedge dB')$:
One can see by considering the detailed calculation of $d\v\, \v^{-1}$
that when the $d$ lands on a particular factor in (\ref{gencoset}),
the factors sitting further to the right cancel out, while the factors
sitting further to the left provide a ``dressing'' of undifferentiated
potentials, in a pattern governed by the details of the commutation
relations.

   Most of the ``dressings'' alluded to above come from the action of
the coset factor $h$ for the axions ${\cal A}_{\0 j}^i$, given in
(\ref{hbor}).  This has the effect of dressing all upstairs indices
with factors of $\td\gamma^i{}_j$, and all downstairs indices with
factors of $\gamma^i{}_j$.  This follows from the fact that the
generators $T_{i_1\cdots i_p}{}^{j_1\cdots j_q}$ associated with a
field $B^{i_1\cdots i_p}{}_{j_1\cdots j_q}$ satisfy the $SL(11-D,\R)$
commutation relations
\bea
{[} E_k{}^\ell, T_{i_1\cdots i_p}{}^{j_1\cdots j_q} {]} &=&
\delta^\ell_{i_1}\, T_{k i_2\cdots i_p}{}^{j_1\cdots j_q}
+\delta^\ell_{i_2}\, T_{i_1 k i_3\cdots i_p}{}^{j_1\cdots j_q}+\cdots
\nn\\
&&-\delta^{j_1}_k\, T_{i_1\cdots i_p}{}^{\ell j_2\cdots j_q}
-\delta^{j_2}_k\, T_{i_1\cdots i_p}{}^{j_1 \ell j_3\cdots j_q} -\cdots
\ .
\eea
Then $h$ acts on $T_{i_1\cdots i_p}{}^{j_1\cdots
j_q}$ valued fields as follows:
\be
h\, (B^{i_1\cdots i_p}{}_{j_1\cdots j_q}\,
T_{i_1\cdots i_p}{}^{j_1\cdots j_q})\, h^{-1}
=\gamma^{k_1}{}_{j_1}\cdots \gamma^{k_q}{}_{j_q}\,
\td\gamma^{i_1}{}_{\ell_1}\cdots \td\gamma^{i_p}{}_{\ell_p}\,
B^{\ell_1\cdots\ell_q}{}_{k_1\cdots k_p}\,
T_{i_1\cdots i_p}{}^{j_1\cdots j_q}\ .\label{hlemma}
\ee
Consequently, all fields associated with terms in $\v$ that get
sandwiched between $h$ and $h^{-1}$ will acquire a ``dressing'' of
$\gamma$ and $\td\gamma$ factors on their indices.

    Another dressing that frequently occurs involves the Kaluza-Klein
vectors $\hA_\1^i$. For example, the sequence of commutators in the
top line in (\ref{rest}) is responsible for generating the various
higher-order terms in the hatted field strengths (\ref{hatfieldstr}).

     These observations, together with the fact that the right-hand
sides of all commutation relations arise with ``unit strength,'' and
that all higher-order terms in the doubled field (\ref{master}) occur
with the ``expected'' combinatoric factors, enable us to see that the
all-orders computation of $d\v\, \v^{-1}$ should indeed give
(\ref{master}).

\section{Doubled Formalism for Scalar Cosets}

      In the previous sections, we studied the doubled-formalism for
eleven-dimensional supergravity and lower dimensional maximal
supergravities coming from dimensional reductions of
eleven-dimensional supergravity.  We showed that the complete set of
gauge symmetries of the doubled formalism (including the constant
shifts of dilatonic and axionic scalars) form a closed algebra that is
a deformation of $G\semi G^*$, where $G$ is the superalgebra given in
(\ref{Ggroup}).  The full set of gauge symmetries leaves the complete
set of generalised field strengths $\G$ invariant.

     The full symmetry of the doubled equations of motion is larger
than the symmetry described above, since there can also be
transformations that change $\G$ whilst nevertheless leaving the
equation ${*\G} = \S\, \G$ invariant.  In the case of the doubled
system of equations for maximal supergravity in $D=11-n$ dimensions,
the global part of the $\G$-preserving symmetry is in fact $E_n^+$,
the Borel subalgebra of the familiar maximally-noncompact $E_n$
symmetry.  We expect that the doubled formalism, however, is in fact
invariant under the full global $E_n$ algebra, with the anti-Borel
generators describing transformations under which $\G$ varies but
${*\G}$ is still equal to $\S\, \G$.
     
          Owing to the fact that the global symmetries are realised
non-linearly on the scalar manifold, the generalisation of the usual
scalar coset to the ``doubled'' formalism is more complicated than the
analogous generalisation of the discussion for higher-degree forms.
In this section, therefore, we shall present a discussion of the
global symmetries of the doubled system in the comparatively-simple
example of an $SL(2,\R)$-symmetric scalar theory.  We shall then
generalise the results to any symmetric space (in particular
principal) sigma model. This construction encompasses the classical
work of \cite{ft} and \cite{TM}.

\subsection{Doubled equations for $SL(2,R)$ coset}

    Let us consider the $SL(2,\R)$-invariant scalar Lagrangian 
\be
{\cal L}_{10} = -\ft12 e\,  (\del\phi)^2 -\ft12e\, e^{2\phi}\,
(\del\chi)^2 \label{sl2rlag}
\ee
in $D$ spacetime dimensions.
This can be written as ${\cal L}= \ft14 e\, \tr(\del {\cal M}^{-1}\,
\del{\cal M})$, where ${\cal M}=\v_0^{\rm T}\, \v_0$ and
\be
\v_0= \pmatrix{e^{\ft12\phi} & \chi\, e^{\ft12\phi}\cr
                    0 & e^{-\ft12\phi}}\ .\label{vdef}
\ee
In the language of differential forms, we have
\be
{\cal L}_{10}= -\ft12 {*d\phi}\wedge d\phi - \ft12 e^{2\phi}\,
{*d\chi}\wedge d\chi\ .
\ee
The resulting equations of motion are:
\bea
d {*d\phi} &=& e^{2\phi}\, d\chi\wedge {*d\chi}\ ,\nn\\
d(e^{2\phi}\, {*d\chi}) &=&0\ .\label{sl2eom}
\eea
This enables us to write down two first order equations
\bea
*d\phi &=& d\psi+\chi\, d\wtd\chi \ ,\nn\\
e^{\phi}\, {*d\chi} &=& e^{-\phi}\,  d\wtd\chi\ ,\label{sl2rfo}
\eea
where we have introduced dual $(D-2)$-forms $\psi$ and $\wtd\chi$ for the
dilaton $\phi$ and the axion $\chi$ respectively.  Taking the exterior
derivatives of these first order equations, we
obtain the second-order equations of motion (\ref{sl2eom}).

             It is well known that the Lagrangian (\ref{sl2rlag}) is
invariant under the $SL(2,\R)$ global symmetry $\tau\longrightarrow
\tau'=(a\tau + b)/(c\tau + d)$, where $\tau = \chi + {\rm i}
e^{-\phi}$.  We shall now check that this symmetry can be extended to
act on the dual forms and to preserve the first order equations
(\ref{sl2rfo}).  It is straightforward to see that (\ref{sl2rfo}) is
invariant under the Borel subgroup of the $SL(2,\R)$, corresponding to
the matrices
\be
\pmatrix{ 1/a& b\cr 0 & a}\nn
\ee
acting from the right on the matrix $\v_0$ given in (\ref{vdef}).
In fact the Borel subgroup is generated by the constant
shift symmetries of the dilaton $\phi$ and the axion $\chi$
\bea
\phi\longrightarrow \phi' = \phi -2\log a\ ,&&
\chi\longrightarrow \chi'= a^2\, \chi + a b\ ,\nn\\
\wtd \chi \longrightarrow \wtd \chi'= a^{-2}\,  \wtd \chi  \ ,&&
\psi \longrightarrow \psi'= \psi - b\, a^{-1}\, \wtd \chi\ .\label{boreltr}
\eea
The higher-degree potentials $\wtd\chi$ and $\psi$ have, in addition, 
the (diagonal) local gauge symmetries
\be
\wtd \chi \longrightarrow \wtd \chi' = \wtd \chi + \Lambda_{\wtd \chi}
\ ,\qquad
\psi \longrightarrow \psi'=\psi + \Lambda_{\psi}\ ,\label{sl2rgauge}
\ee
where $d\Lambda_{\wtd\chi} =0=d\Lambda_{\psi}$.  In fact the Borel
symmetries and the local gauge symmetries of $\wtd \chi$ and $\psi$
form a closed algebra.  As in the appendix of \cite{cjlp}, we can
introduce generators $H$ and $E_\pm$ for $SL(2,\R)$.  Then the Cartan
generator $H$ and the positive-root generator $E_{+}$ are associated
with the dilaton and axion respectively.  We also introduce the new
generators $\tH$ and $\tE_+$, associated with $\psi$ and $\wtd \chi$
respectively.  The commutators of the transformations (\ref{boreltr})
and (\ref{sl2rgauge}) implies that these generators satisfy the
algebra
\be
{[}H, E_+{]} = 2E_+\ ,\qquad {[}H, \tE_+{]} = -2\tE_+\ ,\qquad
{[} E_+, \tE_+ {]} = \ft12 \tH\ .\label{sl2rcom}
\ee
Note that the first commutator defines the Borel subalgebra of the
$SL(2,\R)$ symmetry, and that $\tH$ commutes with everything. 
Then we find that if we define
\be
\v = e^{\fft12\phi\, H}\, e^{\chi\, E_+}\, e^{\wtd\chi\, \tE_+}\,
e^{\fft12\psi\, \tH}\ ,
\ee
where $\wtd\chi$ is the $(D-2)$-form dual to the axion $\chi$, and 
$\psi$ is the
$(D-2)$-form dual to the dilaton $\phi$, then $\G=d\v\v^{-1}$ is given by
\be
\G = \ft12d\phi\, H + e^{\phi}\, d\chi\, E_+ + e^{-\phi}\, d\wtd\chi\, \tE_+
+ \ft12(d\psi+ \chi\, d\wtd\chi)\, \tH\ .\label{sl2rmaster}
\ee
Thus the doubled equation $*\G=\Omega \G$ gives precisely the two
first-order equations (\ref{sl2rfo}). The Borel and gauge
transformation rules (\ref{boreltr}) and (\ref{sl2rgauge}) can be
re-expressed as
\be
\v' = \v\, e^{\log a\, H}\, e^{b\, E_+}\, e^{\Lambda_{\wtd\chi}\, \tE_+}\,
e^{\fft12\Lambda_{\psi}\, \tH}\ ,\label{sl2rtrans}
\ee
which leaves $\G$ invariant. In particular each quantity coupled to
each generator in (\ref{sl2rmaster}) is independently invariant under
the transformations in (\ref{sl2rtrans}).

        Thus we see that as in the case of eleven-dimensional
supergravity, the first order equations (\ref{sl2rfo}) of the
$SL(2,\R)$ coset can be re-expressed as the doubled equation $*\G=\S
\G$, and that $\G$ itself is invariant under the Borel subgroup of the
global $SL(2,\R)$ scalar symmetry, and under the gauge symmetries of
the $(D-2)$-form dual fields.  Here, the involution $\S$ acts on the
generators by $\S\, H= \tH$, $\S\, E_+= \tE_+$.  Obviously, therefore,
the Borel transformations are also an invariance of the doubled
equation of motion $*\G=\S\, \G$.

This equation is in fact invariant under a larger global symmetry
group, namely the entire $SL(2,\R)$ global symmetry.  To show this, we
note that any $SL(2,\R)$ matrix can be decomposed as
\be
\pmatrix{ a& b\cr c & d} = B_{L}\, \pmatrix{ 0& 1\cr -1 & 0} B_{R}
\ee
where $B_{L}$ and $B_{R}$ are Borel matrices, given by
\be
B_{L}= \pmatrix{ 1& a/c\cr 0 & 1}\, \qquad
B_{R} = \pmatrix{ -c& -d\cr 0 & -1/c}\ .
\ee
Since $\G$ itself is invariant under the Borel transformations, it
remains only to verify that the equations are invariant under
transformations generated by the inversion group element
\be
\pmatrix{ 0& 1\cr -1 & 0}\ ,\label{nontriv}
\ee
corresponding to $\tau'=-1/\tau$.  Defining $P=d\psi$ and $Q=d\wtd
\chi$, it is
straightforward to verify that the equations (\ref{sl2rfo}) are invariant
under this transformation, provided that $P$ and $Q$ transform as 
\be
P'=-P\ ,\qquad Q'=-(\chi^2 + e^{-2\phi}) Q - 2\chi P\ .
\ee
(Note that the Bianchi identities $dP'=0$ and $dQ'=0$ are indeed
satisfied, modulo the equations (\ref{sl2rfo}).)  Thus we have
verified that the doubled equation ${*\G}=\S\, \G$ is also invariant
under the entire global $SL(2,\R)$, although $\G$ itself transforms
non-trivially under the inversion generated by (\ref{nontriv}).  The
entire $SL(2,\R)$ symmetry is realised on the scalars $\chi$ and
$\phi$, but, locally, only on the derivatives $P$ and $Q$ of the
potentials $\psi$ and $\wtd \chi$.  Note, incidentally, that the
natural ``field strengths'' for the potentials $\psi$ and $\wtd \chi$
are
\be
\bar P= P +\chi\, Q = d\psi+ \chi\, d\wtd\chi\ ,\qquad
\bar Q = e^{-\phi} d\wtd\chi\ .\label{dsfs}
\ee
In particular, it is these quantities that appear in the coefficients
of the generators $\tH$ and $\tE_+$ in (\ref{sl2rmaster}), and hence
they are invariant under the Borel subgroup of transformations
(\ref{boreltr}).      Under the inversion $\tau' =-1/\tau$, 
these field strengths transform as a $U(1)$ doublet
\be
\pmatrix {\bar P' \cr \bar Q'} =
\pmatrix{ \cos\theta & \sin\theta\cr
                   -\sin\theta & \cos\theta} 
\pmatrix {\bar P\cr \bar Q}\label{ppqtrans}
\ee
with field-dependent parameter $\tan\ft12\theta = \chi \, e^{\phi}$.

     As in the case of $D=11$ supergravity that we discussed in
section 2, we may present an alternative derivation of the algebra
(\ref{sl2rcom}) which does not require the introduction of the
potentials.  Let us first introduce a field strength 
\be
\G_0= d\v_0\, \v_0^{-1} 
=\ft12 d\phi\, H + e^\phi\, d\chi\, E_+\ ,\label{g0def}
\ee
where $\v_0$ is given by (\ref{vdef}),
and then define $\G=\G_0+\S\, {*\G_0}$.  Thus we have
\be
\G=\ft12 d\phi\, H + e^\phi\, d\chi\, E_+ + e^{\phi}\, {*d\chi}\, \tE_+
   +\ft12 {*d\phi} \, \tH\ .
\ee 
The doubled equation ${*\G}=\S\, \G$ is now trivially satisfied, and
instead the second-order equations of motion follow from the
Cartan-Maurer equation
\be
d\G - \G\wedge \G=0\ ,
\ee
provided that we take the generators to have the non-vanishing
commutation relations given in (\ref{sl2rcom}).

\subsection{Noether currents of the global symmetry}

       As we have mentioned earlier, the dual fields in the doubled
formalism are introduced to equate the duals of the (generalised)
Noether currents of the gauge symmetries.  Thus the transformation of
the dual fields of the doubled formalism under the full global
symmetry can be derived from the transformation rules of the Noether
currents.  To see this explicitly, note that the first-order equations
(\ref{sl2rfo}) can also be expressed as
\bea
d\psi &=& *(d\phi -e^{2\phi} \chi d\chi) \equiv *J_0\ ,\nn\\
d\wtd\chi &=& e^{2\phi} * d\chi \equiv * J_+\ ,\label{sl2rfo2}
\eea
where $J_0$ and $J_+$ are precisely the conserved Noether currents,
associated with the constant shift symmetries of the scalars, {\it
i.e.}\ the Borel symmetries.  Specifically, $J_0$ is the Noether
current associated with the Cartan generator, and $J_+$ is the Noether
current associated with the positive-root generator of $SL(2,\R)$.  In
the case of $SL(2,\R)$, there is a third Noether current $J_-$,
associated with the negative-root generator of $SL(2,\R)$, given by
\bea
J_- &=& d\chi + 2\chi d\phi - e^{2\phi} \chi^2 d\chi\nn\\
&=& 2\chi\, J_0 + (\chi^2 + e^{-2\phi})\, J_+\ .\label{j3}
\eea
The fact that $J_-$ is a linear combination of the two Borel currents
$J_0$ and $J_+$ is not surprising, since we have only two independent
fields, $\phi$ and $\chi$. This dependence is a general feature of
sigma models resulting from the local gauge invariance without
propagating gauge fields that can be restored there \cite{CJ}, and
results from the vanishing of the would-be Noether currents of the
gauge symmetry \cite{JS}.  The complete set of Noether currents $(J_0,
J_+, J_-)$ form the adjoint representation of
$SL(2,\R)$.\footnote{Here we are concerned with the Noether current
${\cal J}$ with $d*{\cal J}=0$ giving rise to equations of motion.
The dual currents associated with Bianchi identity $d^2\phi=0=d^2\chi$
are given by $d\chi$ and $d\phi$.  Unlike the Noether currents, which
form the adjoint representation, acting on the dual currents with
$SL(2,\R)$ generates an infinite number of currents, forming an
infinite-dimensional representation of $SL(2,\R)$ \cite{euclid}.}
They transform as
\be
X\longrightarrow X' = \Lambda X \Lambda^{-1}\ ,
\ee
where
\be
X=\pmatrix{ J_0 & -J_- \cr -J_+ & - J_0}
\ee
and $\Lambda$ is a constant $SL(2,\R)$ matrix. Since $J_-$ is a linear
combination of $J_0$ and $J_+$ with field-dependent coefficients, it
follows that the linear $SL(2,\R)$ transformation of the full set of three
Noether currents can be re-expressed as a non-linear transformation of
the Borel currents $J_0$ and $J_+$, namely
\be
\pmatrix{J_0\cr J_+} \longrightarrow \pmatrix {J_0\cr J_+}'
= \Lambda_{\phi,\chi} \pmatrix{J_0\cr J_+}\ ,
\ee
where $\Lambda_{\phi, \chi}$ is some specific field-dependent $2\times
2$ matrix.  Thus, the first-order equations (\ref{sl2rfo2}) are
invariant under $SL(2,\R)$, provided that $d\psi$ and $d\wtd\chi$
transform in the same way as $J_0$ and $J_+$ under $\Lambda_{\phi,
\chi}$.  Thus we may assign this transformation rule to $d\psi$ and
$d\wtd\chi$.  However, we must check that this is consistent with the
Bianchi identities for $d\psi$ and $d\wtd\chi$.  This is in fact
clearly the case, since, as we have already seen, the transformed
$J'_0$ $J'_+$ and $J'_-$ currents can be expressed as linear
combinations of the three original Noether currents $(J_0, J_+, J_-)$,
with constant coefficients.  Thus it is manifest, since the Noether
currents are conserved, that the transformed $J'_0$ and $J'_+$
currents are also conserved, even if we choose to express them in
terms of the field-dependent combinations of the original $J_0$ and
$J_+$ currents.  Since the calculations for checking the Bianchi
identities for the transformed $d\psi$ and $d\wtd\chi$ fields will be
identical, the conclusion will also be identical, namely that the
$SL(2,\R)$ transformations preserve the Bianchi identities.

\subsection{Generalisation to general cosets and principal $\sigma$-models}

      The above discussion can be easily generalised to an arbitrary
coset $K \backslash G$ with maximally non-compact (ie split) group $G$
and $K$ its maximal compact subgroup.  The coset can be parameterised
by the Borel subgroup elements, as in the $SL(2,\R)$ example. The
dilatons $\phi_i$ couple to the Cartan generators $H_i$, and the
axions $\chi_m$ couple to the positive-root generators $E_{+}^m$, so
that the coset representative can be written as \be
\v=\exp(\fft12\phi_i H_i)\, \exp( \chi_m E_{+}^m)\ .\label{borellp}
\ee The Lagrangian is of the form (\ref{cosetlag}), and is invariant
under the full global symmetry $G$.  This can be seen from the Iwasawa
decomposition, which asserts that any group element $\gg$ in $G$ can
be written in the form $ {\bf k} \times \gg_B$, where ${\bf k}$ is an
element of the maximal compact subgroup and $\gg_B$ is an element of
the Borel subgroup.  Thus for any group element $\gg$, there is a
(field-dependent) compensating transformation ${\bf k}$ such that
\be
\v\longrightarrow \v' = {\bf k} \, \v\, \gg\label{comp} 
\ee 
is back in the
Borel gauge.  The scalar Lagrangian (\ref{cosetlag}) can also be
written as 
\be 
{\cal L}= \ft14 {\rm tr}\, (\del{\cal M}^{-1}\, \del{\cal M})\ ,
\label{mlag}
\ee
using \cite{cjlp, JJ} the Cartan involution and $K$-invariant metric
$\eta$ with ${\cal M}=\v^\# \, \eta \, \v$, and so it is evident that
it is invariant under the transformation (\ref{comp}) for any element
$\gg$ in $G$.

    We can calculate the Noether currents for the global symmetries
$G$, by the standard procedure of replacing the global parameters by
spacetime-dependent ones, and collecting the terms in the variation of
the Lagrangian where derivatives fall on the parameters.
Infinitesimally, we have $\delta{\cal M} =\epsilon^{\#}\, {\cal M} +
{\cal M}\, \epsilon$, where $\gg=1+\epsilon$, and hence we find
$\delta{\cal L} = -{\rm tr}\, (\del\epsilon\, {\cal M}^{-1}\,
\del{\cal M})$, implying that the Noether currents ${\cal J}$ are
given by
\be
{\cal J} = {\cal M}^{-1}\, d{\cal M}\ .\label{gennoether}
\ee
Under the global $G$ transformations, they therefore transform
linearly:
\be
{\cal J}\longrightarrow {\cal J}' =  \gg^{-1}\, 
{\cal J}\, \gg\ .\label{noetherlin}
\ee
The Noether currents are not all linearly independent, since the
number of currents exceeds the number of scalar fields.  In fact they
satisfy the relations
\be
{\rm tr}\, ({\cal J}\, \v^{-1}\, h_i\, \v) =0\ ,\label{lindep}
\ee
where $h_i$ denotes the generators of the maximal compact subgroup of
the global symmetry group.  This can be seen by substituting
(\ref{gennoether}) into (\ref{lindep}), and writing ${\cal M}$ as
$\v^\# \, \v$, giving
\be
\tr\, \Big( (d\v\, \v^{-1} + \eta (d\v\, \v^{-1})^{\#} \eta ) \, h_i\Big)=0\ .
\label{nccon}
\ee
In this form, the relation is manifestly true since for any generator
${\cal T}$ of $G$, it is the case that ${\cal T} + \eta {\cal T}^{\#}
\eta $ is non-compact, and hence orthogonal to $h_i$.  (Note that
(\ref{nccon}) can be shown to follow from the statement that the
Noether currents for transformations associated with the denominator
gauge group vanish \cite{JS}.)  Thus the total number of relations in
(\ref{lindep}) on the dim($G$) Noether currents is equal to the
dimension of the maximal compact subgroup of $G$.  The total number of
linearly-independent Noether currents is therefore equal to the
dimension of the scalar coset manifold.

   In the Borel parameterisation (\ref{borellp}) of the scalar coset,
the transformations in the Borel subgroup of $G$ are generated by
constant shifts of the dilatons and axions, implying that the
equations of motion can be expressed as $d *J_{0}^i = 0 = d *J_{+}^m$,
where $J_0^{i}$ and $J_{+}^m$ are the Noether currents associated with
these shift symmetries.  The explicit forms for these currents for the
$E_{n(n)}$ global symmetry groups of the maximal supergravities are
given in section 4.  Thus the first-order equations can be expressed
as
\be
d\psi^i = {*J_{0}^i}\ ,\qquad d\wtd\chi^m = {*J_{+}^m}\ ,\label{groupfo}
\ee
where $\psi^i$ and $\wtd\chi^m$ are the associated dual potentials.
The Noether currents $J_{-}^m$ of the transformations generated by the
negative-root generators can be expressed as linear combinations of
$J_{0}^i$ and $J_{+}^m$, with scalar-dependent coefficients.  Thus the
linear transformation of the complete set of the Noether currents
becomes a non-linear transformation when acting purely on the Borel
currents.  It follows that the first-order equation is invariant under
the full group $G$, provided that $d\psi^i$ and $d\wtd\chi^m$
transform covariantly, with the same non-linear transformation rules
as the Borel currents.  Under these transformations, the Bianchi
identities for the transformed dual fields are guaranteed, for the
same reason that we discussed in the $SL(2,\R)$ example.

         The full dualisation of the general coset model can be easily
understood.  In the Borel parameterisation of the coset, the scalars
appear in the Lagrangian (\ref{cosetlag}) through $\G_{0}$, which
satisfies the Bianchi identity $d\G_0 - \G_0\wedge \G_0 = 0$.  Thus we
can introduce Lagrange multipliers for this Bianchi identity.  The
resulting fully-dualised theory has no global symmetry.  The
disappearance of the global Borel subgroup can be easily understood,
since the doubled field $\G$ is invariant under the Borel subgroup,
and hence so is its Lagrange multiplier.  The rest of the
transformations involve undifferentiated scalars as coefficients, and
hence cannot be expressed in terms of the dual fields, the Noether
currents have to be exchanged with topological charges and rigid
invariance with gauge invariance.

         The above analysis has so far concentrated on symmetric
spaces for maximally non-compact groups.  The situation is analogous
for a general non-compact symmetric space.  In such a case, the coset
$\v$ can be parameterised by using the Iwasawa decomposition and the
so-called solvable subalgebra \cite{trigiante}.  In other words, it
can be parameterised by using a subset of the positive-root generators
and the Cartan generators.  Again the Noether currents for the
symmetries associated with these generators can be used to construct
first-order equations, and they are invariant under the full global
symmetry group.  This can be seen by means of the same argument as
that discussed in the case of maximally non-compact groups.  An
example of a coset with a non-compact group that is not maximally
non-compact is provided by the toroidal dimensional reduction of the
heterotic string, which, in $D$ dimensions has a
$O(26-D,10-D)/((O(26-D)\times O(10-D))$ scalar manifold.

          Another interesting example is that of non-linear sigma
models with (compact or non-compact) group-manifold target spaces.  An
important difference in this case is that some subgroup of global
symmetry is linearly realised on the 1-form field strengths, although
it is still non-linearly realised on the scalars, {\it i.e.}\ the
0-form potentials.  Thus it is manifest that the first-order equations
should have this global symmetry.  In fact, owing to the linearity of
the realisation of the global symmetry on the 1-forms, it follows that
after fully dualising the sigma model fields to higher-degree forms,
the Lagrangian will maintain this global symmetry. The specific
example of such a dualisation in $D=4$ can be found in \cite{ft}.

       The algebras of the doubled formalism for the scalar Lagrangians
all have the form
\be
{[} T^a, T^b {]} = f^{ab}{}_{c}\,  T^c\ ,\qquad
{[} T^a, \wtd T_b {]} = -f^{ac}{}_b \, \wtd T_c\ ,\qquad
{[} \wtd T_a, \wtd T_b {\}}=0\ ,\label{doubalg}
\ee
where the generators $T^a$ are associated with the scalars, while the
generators $\wtd T_a$ are associated with the duals of the
scalars. Similar Lie algebras seem to be interesting from the point of
view of integrable systems \cite{ALV}.  The algebra can be denoted
again by $G\semi {G^*}$, where $G$ is generated by $T^\a$ and ${G^*}$
is the co-adjoint of $G$.

In the case of a principal sigma model (\ie one with a group manifold
$G$ as its target space), $T^\a$ generates the group $G$ of the sigma
model, whilst in the case of a coset $G/H$ in the Borel gauge, $T^\a$
generates the Borel subalgebra $G_+$ of $G$.  When $G=SL(2,\R)$, it is
easy to verify that $G_{+} \semi {G_{+}^*}$ is a subalgebra of $G\semi
{G^*}$.  However, this statement is not true for generic groups $G$.
The gauge invariant treatment of symmetric space sigma models
including the principal models is sketched in Appendix D.

\section{Type IIB Supergravity}

    There is no covariant Lagrangian for type IIB supergravity, since
it includes a self-dual 5-form field strength.  However one can write
down covariant equations of motion \cite{sch}.  In order to make
manifest their global $SL(2,\R)$ symmetry, it is useful first to
assemble the dilaton $\phi$ and axion $\chi$ into a $2\times 2$
matrix:
\be
{\cal M} = \pmatrix{ e^\phi & \chi\, e^\phi\cr
                     \chi\, e^\phi & e^{-\phi} + \chi^2\, e^\phi}
\ee
Also, define the $SL(2,\R)$-invariant matrix
\be
\Xi = \pmatrix{ 0 & 1\cr
                   -1 & 0}\ ,
\ee
and the two-component column vector of 2-form potentials
\be
A_\2 = \pmatrix{A_\2^1 \cr A_\2^2}\ .
\ee
Here $A_\2^1$ is the R-R potential, and $A_\2^2$ is the NS-NS
potential.  The bosonic matter equations of motion can then be written as
\cite{trombone}
\bea
d{* H_\5} &=& -\ft12 \ep_{ij}\, F_\3^i\wedge F_\3^j\ ,\nn\\
d({\cal M} {* H_\3}) &=& H_\5 \wedge \Xi\, H_\3\ ,\nn\\
d(e^{2\phi}\, {*d\chi}) &=& - e^{\phi}\, F_\3^2\wedge
{* F_\3^1}\ ,\nn\\
d{*d\phi} &=& e^{2\phi}\, d\chi\wedge {*d\chi} +\ft12 e^{\phi}\,
F_\3^1\wedge {*F_\3^1} -\ft12 e^{-\phi}\, F_\3^2\wedge {*F_\3^2}\ ,
\label{2beom}
\eea
where $F_\3^1 = dA_\2^1 -\chi\, dA_\2^2$, $F_\3^2=dA_\2^2$,
$H_\3=dA_\2$, and $H_\5 = dB_\4 - \ft12 \ep_{ij}\, A_\2^i\wedge dA_\2^j$.  

     Introducing a ``doubled'' set of potentials $\{
\tA_\6,\psi,\wtd\chi\}$ for the $SL(2,\R)$ doublet $A_\2$, and for
$\phi$ and $\chi$, respectively, we find that one may write the equations
of motion (\ref{2beom}) in first-order form, as
\bea
{*H_\5} &=& dB_\4 - \ft12 \ep_{ij}\, A_\2^i\wedge dA_\2^j\ ,\nn\\
{\cal M} {*dA_\2} &=& d\tA_\6 - \ft12 \Xi\, A_\2\wedge (dB_\4 - \ft16
\ep_{ij}\, A_\2^i\wedge dA_\2^j)\ ,\nn\\
i\, e^{\phi}\, {*d\tau} &=& P + \tau\, Q\ ,\label{2bx}
\eea
where $\tau=\chi + i\, e^{-\phi}$, and the quantities $P$ and $Q$ are
defined by
\bea
P &=& d\psi +\ft12 A_\2^1\, d\tA_\6^1 - \ft12 A_\2^2\, d\tA_\6^2 -
\ft14 A_\2^1\, A_\2^2\, dB_\4 -\ft1{24} A_\2^2\, A_\2^2\, 
A_\2^1\, dA_\2^1\ ,\nn\\
Q &=& d\wtd\chi +A_\2^2 \, d\tA_\6^1 - \ft14 A_\2^2\, A_\2^2\, dB_\4 -
\ft1{36} A_\2^2\, A_\2^2\, A_\2^2\, dA_\2^1\ .\label{dpq}
\eea
Note that we do not need to introduce a ``double'' potential for
$B_\4$; the ``doubling'' in this case is automatically achieved by the
fact that $H_\5$, until the imposition of the self-duality constraint,
already has twice the physical degrees of freedom.  As usual, taking
the exterior derivatives of the equations in (\ref{2bx}) gives the
second-order field equations (\ref{2beom}).

   The first-order equations (\ref{2bx}) can, if desired, be written
also in the form
\bea
{*H_\5} &=& H_\5 = dB_\4 - \ft12 \ep_{ij}\, A_\2^i \wedge dA_\2^j \ ,\nn\\
e^{\phi}\, {*F_\3^1} &\equiv& \tF_\7^1 =
d\tA_\6^1 - \ft12 A_\2^2(dB_\4 -\ft16\, \ep_{ij}\,  A_\2^i\, 
dA_\2^j)\ ,\nn\\
e^{-\phi}\, {*F_\3^2} & \equiv & \tF_\7^2 =
d\tA_\6^2 + \ft12 A_\2^1(dB_\4 -\ft16 \, \ep_{ij}\, A_\2^i\, dA_\2^j ) \nn\\
&&\qquad -\chi\,
(d\tA_\6^1 - \ft12 A_\2^2(dB_\4 -\ft16\, \ep_{ij}\, A_\2^i\, dA_\2^j ))\ ,\nn\\
e^{2\phi}\, {*d\chi} &\equiv &Q\ ,\nn\\
{*d\phi} &\equiv & \tP = P+ \chi\, Q\ ,\label{2by}
\eea
where $P$ and $Q$ are defined in (\ref{dpq}).  

   The second-order equations of motion can be written in the bilinear form
\bea
dH_\5 &=& -F_\3^1\wedge F_\3^2\ ,\nn\\
d\tF_\7^1 &=& H_\5\wedge F_\3^2\ ,\nn\\
d\tF_\7^2 &=& -H_\5\wedge F_\3^1 - d\chi\wedge \tF_7^1\ ,\nn\\
dQ &=& -F_\3^2\wedge \tF_\7^1\ ,\label{2bbilin}\\
d\tP &=& d\chi\wedge Q + \ft12 F_\3^1\wedge \tF_\7^1 -\ft12
F_\3^2\wedge \tF_\7^2\ .\nn
\eea
We may then define a doubled field strength $\G$, given by
\bea
\G&=&\ft12 d\phi\, H + e^\phi\,  d\chi\, E_+ + e^{\fft12\phi}\, F_\3^1\, V_+
+ e^{-\fft12\phi}\, F_\3^2\, V_- + H_\5\, U\nn\\
&&+e^{-\fft12\phi}\, \tF_\7^1 \, \tV_+ + e^{\fft12\phi}\,
\tF_\7^2 \, \tV_-  + e^{-\phi}\, Q \, \tE_+
 + \ft12 \wtd P\, \tH\ .\label{2bmaster}
\eea
As usual, the various untilded generators are mapped into their
associated tilded ``dual'' generators by an (anti)-involution
$\S$, and so $\G$ automatically satisfies the equation 
${*\G}=\S\, \G$.  The equations of motion (\ref{2bbilin}) can then
be expressed in the Cartan-Maurer form 
 $d\G -\G\wedge \G=0$, provided that the generators
have the non-vanishing commutation relations
\bea
&&{[}H, E_+{]} = 2E_+\ ,\qquad {[}H, V_+{]} = V_+\ ,\qquad
{[}H, V_-{]} = -V_-\ ,\nn\\
&&{[}H, \tE_+{]} = -2\tE_+\ ,\qquad {[}H, \tV_+{]} = -\tV_+\ ,\qquad
{[}H, \tV_-{]} = \tV_-\ ,\nn\\
&&{[}E_+, V_-{]}= V_+\ ,\qquad {[}E_+, \tV_+{]}= -\tV_-\ ,\qquad
{[}V_+, V_-{]} = - U\ ,\nn\\
&& {[}V_+, U{]} =\ft1{2} \tV_-\ ,\qquad
   {[}V_-, U{]} =-\ft1{2} \tV_+\ ,\qquad
   {[}V_-, \tV_+{]}=\tE_+\ ,\nn\\ 
&& {[}E_+, \tE_+{]} = \ft12\tH\ ,\qquad
   {[} V_+, \tV_+{]} =\ft14\tH\ ,\qquad
   {[}V_-, \tV_-{]} = -\ft14\tH\ .\label{2bcom}
\eea
The commutators in the first two lines are determined by the weights
of the various fields, which are evident from (\ref{2bmaster}).  The
next two lines give the commutators associated with the terms on the
right-hand sides of various Bianchi identities given above.  The last
line gives the commutators associated with the equation for $d\tP$.
Note that in the type IIB theory all the generators are even, since
all the potentials are of even degree.

    We can solve the Cartan-Maurer equation by writing $\G=d\v\, \v^{-1}$,
where
\be
\v = e^{\fft12\phi\, H}\, e^{\chi\, E_+}\, e^{(A_\2^1\, V_+ + 
A_\2^2\, V_-)}\, e^{B_\4\, U}\, e^{(\tA_\6^1\, \tV_+ + \tA_\6^2\,
\tV_-)}\, e^{\wtd\chi\, \tE_+}\, e^{\fft12\psi\, \tH}\ .\label{2bcoset}
\ee
It is straightforward to check that all terms in (\ref{2bmaster}) are
correctly produced by (\ref{2bcoset}) and (\ref{2bcom}).

    In addition to the local gauge invariances of the higher-rank
fields, the type IIB theory also has a global $SL(2,\R)$ symmetry,
under the transformations
\bea
\tau &\longrightarrow& {a\tau +b \over c\tau +d}\ ,\qquad\qquad B_\4
\longrightarrow B_\4\ ,\nn\\
A_\2 &\longrightarrow& \pmatrix{ a& -b\cr -c & d}\, A_\2\ ,\qquad\qquad
\tA_\6\longrightarrow \pmatrix{ d& c\cr b & a}\, \tA_\6\ .\label{sl2}
\eea
The $SL(2,\R)$ transformations of the field strengths $\tP$ and $Q$
are exactly the same as the ones given in the previous section.

    As they stand, the commutation relations (\ref{2bcom}) for the
type IIB theory do not quite fit the pattern of the general algebras
we discussed in sections (4.3) and (4.4).  Specifically, it is easily
seen that they do not satisfy the jade rule that we discussed at the
end of section (4.3).  For example, given the commutator ${[} V_+, V_-
{]} = -U$ in (\ref{2bcom}), the jade rule would lead us to expect a
non-vanishing commutator ${[} V_+, \wtd U {]} = \tV_-$, while from
(\ref{2bcom}) we see that this does not occur.  Similarly, from ${[}
V_+, U {]} = \ft12 \tV_-$, we would expect from the jade rule that
${[} V_+, V_- {]} \sim -\ft12 \wtd U$, whereas in fact we have ${[}
V_+, V_- {]}= -U$.  In fact, of course, we do not even have a
generator $\wtd U$ in our theory.  The reason for this is that in our
expression (\ref{2bmaster}) for the total field strength $\G$, we
already made use of the fact that in the type IIB theory $H_\5$ is
self dual, and so we did not have to introduce a ``doubled'' field for
$H_\5$.  Related to this self-duality constraint is the fact that
there exists no Lagrangian for the type IIB theory.  In \cite{bbo}, it
was shown that one could derive the type IIB equations of motion from
a Lagrangian, if one initially relaxes the self-duality condition and
allows the 5-form field strength to be unconstrained in the
Lagrangian.  After varying to obtain the equations of motion one can
then impose self-duality as a consistent solution of the enlarged
equations, thereby recovering the equations of motion for type IIB
supergravity.

     We can perform a similar enlargement of the system of fields in
our description, and initially treat $H_\5$ as an unrestricted field
with no self-duality constraint.  We then have to introduce a doubled
field, say $\wtd H_\5$, and its associated generator $\wtd U$, in the
same way as we do for all other field strengths.  The resulting
changes in the equations of motion will imply, for example, that the
second line of (\ref{2beom}) will become
\be
d({\cal M} {*H_\3}) = \ft12(H_\5 + {*H_\5})\wedge \Xi\, H_\3\ .
\ee
Following through the consequences, we eventually find that the three
commutation relations in (\ref{2bcom}) that involve $U$ will be
modified, so that they will be replaced by
\bea
&&{[} V_+, V_- {]} =-\ft12 (U + \wtd U) \ ,\qquad 
{[} V_\pm, U- \wtd U {]} = 0\ ,\nn\\
&&{[} V_+, U + \wtd U {]} = \ft12 \tV_-\ ,\qquad
{[} V_-, U + \wtd U {]} = -\ft12 \tV_+ \ .
\eea

   With these replacements of commutation relations, the algebra
(\ref{2bcom}) does now have the form (\ref{genalg}).  The original
algebra (\ref{2bcom}) can be viewed as a subalgebra, obtained by
taking $U-\wtd U$, which commutes with everything, to be vanishing.
This choosing of a subalgebra is the analogue, at the level of the
algebra, of the imposition of the $H_\5={*H_\5}$ constraint at the
level of the field theory.

     It is interesting to note that whereas the type IIB theory has
an $SL(2,\R)$ global symmetry, the type IIA theory has an $SL(1|1)$
global symmetry.  This observation emphasises the similarity between
the symmetries of type IIA and type IIB, which both suggest a
twelve-dimensional origin (intriguingly, possibly fermionic).

\section{Conclusions}

     In this paper, we have studied the bosonic sectors of the various
$D$-dimensional maximal supergravities, in a unified formalism in
which every field, with the exception of the metric itself, is
augmented by the introduction of a ``doubled'' field, related to the
original one by Hodge dualisation.  This is done not only for the
various antisymmetric tensor gauge fields but also for the dilatonic
scalars, for which dual $(D-2)$-form potentials are also introduced.
The equations of motion for the various fields are then all
expressible in the form of the simple twisted self-duality equation
${*\G} = \CS\, \G$, where $\G$ is the total field strength, written as
a sum $\G= \sum_{i} \G_i\, X^i$ of each individual field strength
$\G_i$ (including those for the doubled potentials) times an
associated generator $X^i$.  The equations of motion can equivalently
be expressed as the zero-curvature condition $d\G -\G\wedge \G =0$,
provided that appropriate (anti)-commutation relations are imposed on
the generators $X^i$.  This condition can be interpreted as the
Cartan-Maurer equation allowing $\G$ to be written as $\G=d\v\,
\v^{-1}$, where $\v$ is expressed in the general form $\v=\exp( {\cal
A}_i\, X^i)$.  (The precise details of the parameterisation of $\v$
depend upon the choice of field variables.)

     In the simplest example, of $D=11$ supergravity, the two
generators $V$ and $\tV$ associated with the field strengths $F_\4$
and $\tF_\7$ satisfy a Clifford algebra, with $V$ being an odd
element, and $\tV$ a central element.  More generally, the maximal
supergravity in $D=11-n$ has an algebra which is a deformation of
$G\semi G^*$, where $G$ itself is the superalgebra $SL_+(n|1)\semi
(\wedge V)^3$, as described in section 4.4.  An exception is the
ten-dimensional type IIB theory, for which all the generators are
even, and hence the algebra is purely bosonic.

     The symmetries discussed above leave the total field strength
$\G=d\v\, \v^{-1}$ invariant, since they act on $\v$ as $\v\rightarrow
\v\, \exp(\Lambda_i\, X^i)$, where $\Lambda_i$ are the gauge
parameters, satisfying $d\Lambda_i=0$.  The 0-form parameters are
associated with constant shifts of the various dilatonic and axionic
scalar fields, and hence describe global (\ie rigid) symmetries, while
the higher-degree parameters are associated with local gauge
transformations of the higher-degree potentials.  This somewhat
unusual circumstance of having an algebra with both global and local
transformations arises because the total field strength $\G$ is the
sum of various antisymmetric tensors of differing degrees including,
in particular, degree 1 and degrees greater than 1.

     The full symmetry of the doubled equations of motion is larger
than the symmetry described above, since there can also be
transformations that change $\G$ whilst nevertheless leaving the
equation ${*\G} = \CS\, \G$ invariant.  In the case of the doubled
system of equations for maximal supergravity in $D=11-n$ dimensions,
the global part of the $\G$-preserving symmetry turns out to be
$E_n^+$, the Borel subalgebra of the familiar maximally-noncompact
$E_n$ symmetry.\footnote{There is no necessity of dualising
higher-degree field strengths here, since in the doubled formalism all
possible dualised field strengths are automatically present.  The
$E_n^+$ symmetry is realised as the global shift transformations of
the full set of 0-form potentials of the doubled formalism.}  We
believe that the doubled formalism, however, is also invariant under
the full global $E_n$ algebra, with the anti-Borel generators
describing transformations under which $\G$ varies but ${*\G}$ is
still equal to $\CS\, \G$.  Indeed we have shown that the doubled
formalism for the type IIB in $D=10$ is invariant under the full
$SL(2,\R)$ global symmetry.  We also showed in general that, at
least in the scalar sectors, the introduction of the doubled formalism 
preserves the original full $E_n$ global symmetry groups.
     
    The global symmetries of the doubled formalism for the scalar
sectors can be studied in detail in more general situations, for any
principal sigma model.  We discussed this in section 5.  In both cases
the doubled formalism retains the full global symmetries of the
original formulation, with the scalars themselves transforming in an
unaltered manner. For a group manifold $G$, the dual potentials are
invariant under the right action $G_R$ of the group $G$, but transform
covariantly under the left action $G_L$.  For a covariant formulation
of a coset target space $K\backslash G$ the situation is similar, with
the dual potentials being invariant under the global right action of
the symmetry $G$, and transforming covariantly under the local left
action of the subgroup $K$.  On the other hand, in a gauge-fixed
formulation in terms of the Borel generators of $G$, the dual
potentials are invariant only under the Borel subgroup of $G$, while
the anti-Borel generators describe symmetries that act covariantly
only on the doubled field strengths, but not on the doubled
potentials.  Thus the doubled gauge-fixed scalar coset theory retains
a local action of the entire global symmetry $G$ of the original
formulation only at the level of the doubled field strengths, and not
the doubled potentials.  In appendix D, we studied the symmetries that
remain when certain of the ``redundant'' fields of the doubled
formalism are eliminated.

     Other aspects of the symmetry algebras of the supergravity
theories can also be abstracted and studied in their own right.  In
appendix C, we considered the Kaluza-Klein reduction of $\wtd
D$-dimensional pure gravity to $D$ dimensions on a torus of $n=\wtd D
-D$ dimensions.  It is well known that the resulting theory has a
global $GL(n,\R)$ symmetry, with the Kaluza-Klein vectors transforming
linearly under the $SL(n,\R)$ subgroup.  We showed that the gauge
symmetry of the Kaluza-Klein vectors, together with the Borel
subalgebra $GL_+(n,\R)$ of $GL(n,\R)$, form the superalgebra
$SL_+(n|1)$.  In the special case of $D=3$ the Kaluza-Klein vectors
can be dualised to give additional axionic scalars, and the full
scalar Lagrangian then has a global $SL(n+1,\R)$ symmetry.


\appendix

\section{Dimension-dependent terms in first-order equations}

     In section 4, we gave a general derivation of the first-order
equations of motion for $D$-dimensional maximal supergravity.  The
contributions coming from the Chern-Simons terms in the Lagrangians are
dimension-dependent, and here we present their detailed forms in each
dimension. In each dimension, the variation of the appropriate
Chern-Simons term with respect to the various potentials $A_\3$, $A_{\2
i}$, $A_{\1 ij}$ and $A_{\0 ijk}$ takes the form
\be
-\delta{\cal L}_{FFA} = dX\wedge \delta A_\3 +
 dX^i\wedge \delta A_{\2 i} + \ft12 dX^{ij}\wedge \delta A_{\1 ij}
+ \ft16 dX^{ijk}\, \delta A_{\0 ijk}\ ,\label{Xeqapp}
\ee
where the quantities $X$, $X^i$, $X^{ij}$ and $X^{ijk}$ can be
determined easily in each dimension.  First, we list the results for
these quantities dimension by dimension:
\bea
D=11: \qquad && X = \ft12 A_\3\, dA_\3 \nn\\
&&\nn\\
D=10: \qquad && X=-A_{\2 1}\, dA_\3\ ,\qquad X^1 = \ft12 A_\3\, dA_\3
\nn\\
&&\nn\\
D=9:\qquad && X=\ft12 \e^{ij}\, (A_{\1 ij} \, dA_\3 - A_{\2 i}\, dA_{\2 j} )
\nn\\
&& X^i= -\e^{ij}\, A_{\2 j}\, dA_\3\nn\\
&& X^{ij} = \ft12 \e^{ij}\, A_\3\, dA_\3 \nn\\
&&\nn\\
D=8: \qquad && X= \e^{ijk}\,(-\ft16 A_{\0 ijk}\, dA_\3 -\ft12 A_{\1
ij}\, dA_{\2 k}) \nn\\
&& X^i = \ft12 \e^{ijk}\, (A_{\1 jk}\, dA_\3 - A_{\2 j}\, dA_{\2 k})
\nn\\
&& X^{ij} = -\e^{ijk}\, A_{\2 k}\, dA_\3\nn\\
&& X^{ijk} = \ft12 \e^{ijk}\, A_\3 \, dA_\3\nn\\
&&\nn\\
D=7:\qquad && X= \e^{ijk\ell}\, (\ft18 A_{\1 ij}\, dA_{\1 k\ell}
-\ft16 A_{\0 ijk}\, dA_{\2 \ell})\nn\\
&& X^i = -\e^{ijk\ell}\, (\ft12 A_{\1
jk}\, dA_{\2 \ell} +\ft16 A_{\0 jk\ell}\, dA_\3 )\nn\\
&& X^{ij} = \ft12\e^{ijk\ell}\, (A_{\1 k\ell}\, dA_\3 - A_{\2 k}\,
dA_{\2 \ell})\nn\\
&&X^{ijk} = -\e^{ijk\ell}\, A_{\2\ell}\, dA_\3\nn\\
&&\nn\\
D=6:\qquad && X=-\ft1{12} \e^{ijk\ell m}\, A_{\0 ijk}\, dA_{\1 \ell
m}\nn\\
&& X^i = \e^{ijk\ell m}\, (\ft18 A_{\1 jk}\, dA_{\1 \ell m} -\ft16
A_{\0 jk\ell}\, dA_{\2 m})\nn\\
&&X^{ij}= -\e^{ijk\ell m}\, (\ft12 A_{\1 k\ell}\, dA_{\2 m} +\ft16
A_{\0 k\ell m}\, dA_\3)\nn\\
&&X^{ijk}= \ft12\e^{ijk\ell m}\, (A_{\1 \ell m}\, dA_\3 - A_{\2\ell}\,
dA_{\2 m})\nn\\
&&\nn\\
D=5:\qquad && X=-\ft1{72} \e^{ijk\ell mn}\, A_{\0 ijk}\, dA_{\0 \ell
mn}\nn\\
&&X^i = -\ft1{12} \e^{ijk\ell mn}\, A_{\0 jk\ell}\, dA_{\1 mn}\nn\\
&&X^{ij}= \e^{ijk\ell mn}\, (\ft18 A_{\1 k\ell}\, dA_{\1 mn} -\ft16
A_{\0 k\ell m}\, dA_{\2 n})\nn\\
&&X^{ijk}= -\e^{ijk\ell mn}\, (\ft12 A_{\1 \ell m}\, dA_{\2 n}
+\ft1{6} A_{\0 \ell mn}\, dA_\3)\nn\\
&&\nn\\
D=4:\qquad&& X^i=-\ft1{72} \e^{ijk\ell mnp}\, A_{\0 jk\ell}\, dA_{\0
mnp} \nn\\
&&X^{ij} = -\ft1{12} \e^{ijk\ell mnp}\, A_{\0 k\ell m}\, dA_{\1
np}\nn\\
&&X^{ijk} = \e^{ijk\ell mnp}\, (\ft1{8} A_{\1 \ell m}\, dA_{\1 np} -
\ft1{6} A_{\0 \ell mn}\, dA_{\2 p})\nn\\
&&\nn\\
D=3:\qquad && X^{ij}= -\ft1{72} \e^{ijk\ell mnpq}\, A_{\0 k\ell m}\,
dA_{\0 npq}\nn\\
&& X^{ijk}=-\ft1{12} \e^{ijk\ell mnpq}\, A_{\0 \ell mn}\, dA_{\1
pq}\nn\\
&&\nn\\
D=2:\qquad&& X^{ijk}= -\ft1{72} \e^{ijk\ell mnpqr}\, A_{\0 \ell mn}\,
dA_{\0 pqr}\nn
\eea

     The task now is to show that the lower line on the right-hand
side of (\ref{intres}) can be written as the exterior derivative of
something, in each dimension.  In other words, we want to show that we
can write
\be
X\, dA_{\2 k} + X^i\, dA_{\1 ik} + \ft12 X^{ij}\, dA_{\0 ijk} = dY_k
\ ,\label{ydef}
\ee
and to find $Y_k$ explicitly for each dimension $D$.  The results are:
\bea
D=10: && Y_1 = -\ft12 A_{\2 1}\, A_{\2 1}\, dA_\3 \,\nn\\
&&\nn\\
D=9: && Y_k = \e^{ij}\, (\ft16 A_{\2 k}\, A_{\2 i}\, dA_{\2 j}
  -A_{\1 ik}\, A_{\2 j}\, dA_\3)\ ,\nn\\
&&\nn\\
D=8: && Y_k= \e^{ij\ell }\, (-\ft12 A_{\0 j\ell k}\, A_{\2 i}\,
dA_\3 -\ft14 A_{\1 j\ell}\, A_{\1 ik}\, dA_\3 -\ft12 A_{\1 ik}\, A_{\2
j}\, dA_{\2 \ell})\nn\\
&&\nn\\
D=7: && Y_k = \ft14 \e^{ij\ell m}\, (A_{\0 ijk}\, A_{\2 \ell}\,
dA_{\2 m} -A_{\0 ijk}\, A_{\1 \ell m}\, dA_\3 - A_{\1 ik}\, A_{\1
j\ell}\, dA_{\2 m}) \nn\\
&&\nn\\
D=6: && Y_k=\e^{ij\ell mn}\, (-\ft14 A_{\0 ijk}\, A_{\1 \ell m}\,
dA_{\2 n} +\ft1{24} A_{\1 ik}\, A_{\1 j\ell}\, dA_{\1 mn}\nn\\
&&\qquad\qquad  -\ft1{12} A_{\0 ijk}\, A_{\0 \ell mn}\, dA_\3)\nn\\
&&\nn\\
D=5: && Y_k=\e^{ij\ell mnp}\, (\ft1{24} A_{\0 ijk}\, A_{\0 \ell
mn}\, dA_{\2 p} -\ft1{16} A_{\0 ijk}\, A_{\1 \ell m}\, dA_{\1
np})\nn\\
&&\nn\\
D=4: && Y_k =-\ft1{48} \e^{ij\ell mnpq}\, A_{\0 ijk}\, A_{\0
\ell mn}\, dA_{\1 pq}\nn\\
&&\nn\\
D=3: && Y_k=\ft1{432} \e^{ij\ell mnpqr}\, A_{\0 ijk}\, A_{\0
\ell mn}\, dA_{\0 pqr}\nn\\
&&\nn\\
D=2: && Y_k=0\nn
\eea

     Now we turn to the quantities $Q^k{}_j$, which were introduced in
the derivation of the first-order equations for the axions ${\cal
A}^i_{\0 j}$.  They were defined in (\ref{xxx6}), as the forms whose
exterior derivatives would give
\be
dQ^k{}_j= -X^k\, dA_{\2 j} +X^{k\ell}\, dA_{\1 j\ell}
-\ft12 X^{k\ell m}\, dA_{\0 j\ell m} \ .\label{aaa1}
\ee
We find that indeed such forms exist, and are given by
\bea
D=9:&& Q^2{}_1 = -\ft12 A_{\2 1}\, A_{\2 1}\, dA_\3 \nn\\
&&\nn\\
D=8: && Q^k{}_j= \e^{k\ell m}\, (-A_{\1 j\ell}\, A_{\2 m}\, dA_\3
-\ft16 A_{\2 j}\, A_{\2 \ell}\, dA_{\2 m})\nn\\
&&\nn\\
D=7:&& Q^k{}_j = \e^{k\ell mn}\, (\ft12 A_{\0 j\ell m}\, A_{\2 n}\,
dA_\3 -\ft14 A_{\1 mn}\, A_{\1 j\ell}\, dA_\3 +\ft12 A_{\1 nj}\, A_{\2
\ell}\, dA_{\2 m})\nn\\
&&\nn\\
D=6: && Q^k{}_j =\ft14 \e^{k\ell mnp}\, ( A_{\0 j\ell m}\, A_{\1
np}\, dA_\3 -A_{\0 j\ell m}\, A_{\2 n}\, dA_{\2 p} -A_{\1 j\ell}\, A_{\1
np}\, dA_{\2 m})\nn\\
&&\nn\\
D=5: && Q^k{}_j =\e^{k\ell mnpq}\, (\ft14 A_{\0 j\ell m}\, A_{\1
np}\, dA_{\2 q} \nn\\
&&\qquad + \ft1{24} A_{\0 j\ell m}\, A_{\0 npq}\, dA_\3 +\ft1{24}
A_{\1 j\ell}\, A_{\1 mn}\, dA_{\1 pq})\nn\\
&&\nn\\
D=4: && Q^k{}_j =\e^{k\ell mnpqr}\, (\ft1{16} A_{\0 j\ell m}\, A_{\1
np}\, dA_{\1 qr} -\ft1{24} A_{\0 j\ell m}\, A_{\0 npq}\, dA_{\2 r})\nn\\
&&\nn\\
D=3: && Q^k{}_j = \ft1{48} \e^{k\ell mnpqrs}\, A_{\0 j\ell m}\,
A_{\0 npq}\,  dA_{\1 rs}\nn\\
&&\nn\\
D=2: &&  Q^k{}_j = -\ft1{432}\e^{k\ell mnpqrst}\, A_{\0 j\ell
m}\, A_{\0 npq}\, dA_{\0 rst}\nn
\eea
Note that they are defined only for $k>j$, and thus arise only in
$D\le9$.  Proving the above identities involves the use of Schoutens'
``over-antisymmetrisation'' identities on the lower indices.  These
can be used here, even though the number of lower indices is the same
as $11-D$, because all the lower indices are necessarily different
from $k$.

     Finally, we turn to the dimension-dependent quantities $\vec Q$,
which must satisfy equation (\ref{qid}).  As discussed at the end of
section 4.1, the expressions for $\vec Q$ are written in terms of
``dressed'' fields, where all downstairs indices are dressed with
$\gamma$, and all upstairs indices are dressed with $\td \gamma$.  It
is useful first to make the following definitions of dressed
quantities:
\be
\hat A_{\2 i} = \gamma^j{}_i \, A_{\2 j}\ ,\qquad \hat A_{\1 ij} =
\gamma^k{}_i\,\gamma^\ell{}_j\, A_{\1 k\ell}\ ,\qquad
\hat A_{\0 ijk} = \gamma^\ell{}_i\, \gamma^m{}_j\, \gamma^n{}_k
\, A_{\0 \ell mn}\ .\label{hatdef}
\ee
In terms of these, the required results for the $\vec Q$ are found to be:
\bea
D=10: && \vec Q = \ft12\, \vec a\, A_{\2 1}\, A_\3\, dA_\3\ ,\nn\\
&&\nn\\
D=9: && \vec Q= -\ft12 \vec a\, \e^{ij}\, \hat A_{\1 ij}\, A_\3\,
dA_\3+\ft12 \sum_i \vec a_i\, \hat A_{\2 i}\, \hat A_{\2 j}\, dA_\3 \,
\e^{ij}\ ,\nn\\ &&\nn\\
D=8: && \vec Q = -\ft16 \vec a\, \hat A_{\0 ijk}\, A_\3\, dA_\3\,
\e^{ijk} - \ft12 \sum_{ij} \vec a_{ij}\, \hat A_{\1 ij}\, \hat A_{\2 k}\,
dA_\3\, \e^{ijk} \nn\\
&& \qquad-\ft13 \sum_i \vec a_i\, \hat A_{\2 i}\, \hat A_{\2 j}\, d\hat
A_{\2 k}\, \e^{ijk}\ ,\nn\\
&&\nn\\
D=7: && \vec Q = \sum_{ij}\vec a_{ij}\, \Big( \ft18 \hat A_{\1 ij}\,
\hat A_{\1 k\ell}\, dA_\3 - \ft14 \hat A_{\1 ij}\, \hat A_{\2 k}\,
\gamma^m{}_\ell\, dA_{\2 m}\Big)\, \e^{ijk\ell} \nn\\
&& \qquad+ \ft16 \sum_{ijk} \vec a_{ijk}\, \hat A_{\0 ijk}\, \hat A_{\2
\ell}\, dA_\3\, \e^{ijk\ell}\ ,\nn\\
&&\nn\\
D=6: && \vec Q = -\ft18 \sum_{ij} \vec a_{ij}\, \hat A_{\1 ij}\,
\hat A_{\1 k\ell}\,\gamma^n{}_m\, d A_{\2 n}\, \e^{ijk\ell m} \nn\\
&&\qquad +\ft1{12} \sum_{ijk} \vec a_{ijk}\, \Big(\hat A_{\0 ijk}\, \hat
A_{\1
\ell m}\, dA_\3 - \hat A_{\0 ijk}\, \hat A_{\2 \ell}\,
\gamma^n{}_m\,  dA_{\2 n} \Big) \,
\e^{ijk\ell m}\ ,\nn\\
&&\nn\\
D=5: && \vec Q = \ft1{48} \sum_{ij} \vec a_{ij}\, \hat A_{\1 ij}\,
\hat A_{\1 k\ell}\, \gamma^p{}_m\, \gamma^q{}_n\, d A_{\1 pq}\,
\e^{ijk\ell mn}\nn\\
&&\qquad +\sum_{ijk} \vec a_{ijk}\, \Big( \ft1{72} \hat A_{\0 ijk}\,
\hat A_{\0 \ell mn}\, dA_\3 +\ft1{12} \hat A_{\0 ijk}\, \hat A_{\1 \ell
m}\, \gamma^p{}_n\, d A_{\2 p} \Big)\, \e^{ijk\ell mn}\ ,\nn\\
&&\nn\\
D=4: && \vec Q = \sum_{ijk}\vec a_{ijk}\, \Big( \ft1{24} \hat A_{\0
ijk}\, \hat A_{\1 \ell m}\,\gamma^q{}_n\, \gamma^r{}_p\, d A_{\1 qr}\,
\nn\\
&&\qquad\qquad\qquad\qquad -\ft1{72}
\hat A_{\0 ijk}\,
\hat A_{\0 \ell mn}\, \gamma^q{}_p \,dA_{\2 q} \Big)\,
\e^{ijk\ell mnp}\ ,\nn\\ &&\nn\\
D=3: && \vec Q = \ft1{144} \sum_{ijk} \vec a_{ijk}\, \hat A_{\0
ijk}\, \hat A_{\0 \ell mn}\, \gamma^r{}_p\, \gamma^s{}_q\, d A_{\1 rs}\,
\e^{ijk\ell mnpq}\ .\label{qres}
\eea

\section{Dimension-dependent commutators}

     In section 4.2 we derived the commutation and anticommutation
relations for the various generators appearing in the construction of
the doubled field $\G$ in each dimension.  Those associated with the 
contributions from the Chern-Simons terms in the Lagrangian are
dimension dependent, and here we present the detailed results for
each dimension $D$.  

     The commutation relations can be read off from the bilinear terms
in the doubled field $\G$ that involve the contributions from the
Chern-Simons terms ${\cal L}_{FFA}$, and which therefore all involve
the epsilon tensor.  We find that they are as follows:
\bea
D=11:&& \{ V, V \} = -\tV \ ,\nn\\
&&\nn\\
D=10:&& \{ V, V \} = -\tV_1\ ,\qquad {[} V, V^1{]} = \tV\ ,\nn\\
&&\nn\\
D=9: &&\{ V^{ij}, V \} = -\e^{ij}\, \tV\ ,\quad
   {[} V^i, V^j{]} = \e^{ij}\, \tV\ ,\quad
   {[} V^i, V{]} = \e^{ij}\, \tV_j\ ,\quad
   \{ V, V \} = -\tV_{12}\ ,\nn\\
&&\nn\\
D=8: && {[} E^{ijk}, V{]} = -\e^{ijk}\, \tV\ ,\qquad
    {[} V^{ij}, V^k{]} = \e^{ijk}\, \tV\ ,\qquad
    \{ V^{ij}, V \} = -\e^{ijk}\, \tV_k\ ,\nn\\
 &&{[} V^i, V^j{]} = \e^{ijk}\, \tV_k\ ,\qquad
   {[} V^i, V {]} = -\ft12\e^{ijk}\, \tV_{jk}\ ,\qquad
   \{ V, V \} = -\tE_{123}\ ,\nn\\
&&\nn\\
D=7: && {[} E^{ijk}, V{]} = \e^{ijk\ell}\,\tV_\ell\ ,\qquad
     {[} E^{ijk}, V^\ell{]} = \e^{ijk\ell}\, \tV\ ,\nn\\
&& \{ V^{ij}, V^{k\ell} \} = -\e^{ijk\ell}\, \tV\ ,\qquad
   {[} V^{ij}, V^k {]} =-\e^{ijk\ell}\, \tV_\ell \ ,\qquad
   \{ V^{ij}, V \} = -\ft12 \e^{ijk\ell}\, \tV_{k\ell}\ , \nn\\
&& {[} V^i, V^j{]} = \ft12\e^{ijk\ell}\, \tV_{k\ell}\ ,\qquad
   {[} V^i, V{]} = \ft16 \e^{ijk\ell}\, \tE_{jk\ell}\ ,\nn\\
&&\nn\\
D=6: && {[} E^{ijk}, V^{\ell m} {]} = -\e^{ijk\ell m}\, \tV\ ,\quad
  {[}E^{ijk}, V^\ell {]} = \e^{ijk\ell m}\, \tV_m\ ,\quad
  {[} E^{ijk}, V{]} = -\ft12\e^{ijk\ell m}\, \tV_{\ell m}\ ,\nn\\
&& \{ V^{ij}, V^{k\ell} \} = -\e^{ijk\ell m}\, \tV_m\ ,\qquad
   {[} V^{ij}, V^k{]} = \ft12\e^{ijk\ell m}\, \tV_{\ell m}\ ,\nn\\
&& \{ V^{ij}, V \} = -\ft16 \e^{ijk\ell m}\, \tE_{k\ell m} \ ,\qquad
   {[} V^i, V^j {]} = \ft16\e^{ijk\ell m}\, \tE_{k\ell m}\ ,\nn\\
&&\nn\\
D=5: && {[} E^{ijk}, E^{\ell mn} {]} = \e^{ijk\ell mn}\, \tV\ ,
\qquad
{[} E^{ijk}, V^{\ell m}{]} = \e^{ijk\ell mn}\, \tV_n\ ,\nn\\
&& {[} E^{ijk}, V^\ell {]} = \ft12 \e^{ijk\ell mn}\, \tV_{mn}\ ,\qquad
   {[} E_{ijk}, V {]} = \ft16 \e^{ijk\ell mn}\, \tE^{\ell mn}\ ,\nn\\
&& \{ V^{ij}, V^{k\ell} \} = -\ft12 \e^{ijk\ell mn}\, \tV_{mn}\ ,\qquad
   {[} V^{ij}, V^k {]} = -\ft16 \e^{ijk\ell mn}\, \tE_{\ell mn}\ ,\nn\\
&&\nn\\
D=4: && {[} E^{ijk}, E^{\ell mn} {]} = \e^{ijk\ell mnp}\, \tV_p
\ ,\qquad
{[} E^{ijk}, V^{\ell m} {]} =-\ft12 \e^{ijk\ell mnp}\, \tV_{np}\ ,\nn\\
&& {[} E^{ijk}, V^\ell {]} = \ft16\e^{ijk\ell mnp}\, \tE_{mnp}\ ,\qquad
\{ V^{ij}, V^{k\ell} \} = -\ft16 \e^{ijk\ell mnp}\, \tE_{mnp}\ ,\nn\\
&&\nn\\
D=3: &&
{[} E^{ijk}, E^{\ell mn}{]} = \ft12 \e^{ijk\ell mnpq}\, \tV^{pq}\ ,
\qquad
{[} E^{ijk}, V^{\ell m} {]} = \ft16 \e^{ijk\ell mnpq}\, \tE_{npq}\ .
\label{ddcom}
\eea

    We may observe that these commutation relations can all be
summarised in the single expression
\bea
{[} T^{\bar a}, T^{\bar b} \} = -(-1)^{{[} \bar b {]}}\, 
\epsilon^{\bar c \bar a\bar b}\, \wtd T_{\bar c}\ ,
\label{sign}
\eea
where, as in the notation of section 4.4, we use generic indices $\bar
a,\bar b,\ldots$ to represent antisymmetrised sets of $i,j,\ldots$
indices.  The symbol ${[}\bar a{]}$ denotes the number of such
$i,j,\ldots$ indices.  Appropriate $1/{[}\bar a{]}!$ combinatoric
factors are understood in summations over repeated generic indices.
Also, we have $T=V, T^i=V^i, T^{ij}=V^{ij}, T^{ijk} =E^{ijk}$, with a
similar set of definitions for $\wtd T_{\bar a}$.  It is useful also
to define generators $\wtd U^{\bar a}$, by
\be
\wtd U^{\bar a} = \epsilon^{\bar b\bar a}\, \wtd T_{\bar b}\ .
\ee
(In explicit notation, this means $\wtd U^{i_1\cdots i_p} = 
1/q!\, \epsilon^{j_1\cdots j_q i_1\cdots i_p}\, \wtd T_{j_1\cdots
j_q}$, where $p={[}\bar a{]}$, $q={[}\bar b {]}$, and $p+q=11-D$.)  In
terms of $\wtd U^{\bar a}$, the commutators (\ref{ddcom}) can all be 
written in the form
\be
{[} T^{\bar a}, T^{\bar b} \} = -(-1)^{{[}\bar b{]}}\, \wtd U^{\bar a\bar
b}\ .
\ee

    The above algebras can be understood directly through dimensional
reduction from the $D=11$ algebra $\{V,V\} = -\wtd V$.  To see this,
we define
\be
T^{\bar a} = d^{\bar a}z\, V\ ,\qquad \wtd U^{\bar a} = d^{\bar a}z
\, \wtd V\ ,
\ee
where $d^{\bar a}z$ denotes $1,dz^i,dz^i\wedge dz^j, dz^i\wedge
dz^j\wedge dz^k$ corresponding to ${[}\bar a{]}=0,1,2,3$.  Thus we
will have
\bea
{[} T^{\bar a}, T^{\bar b} \} &=& {[} d^{\bar a}z\, V, d^{\bar b}z\, V
\} = (-1)^{{[}\bar b{]}}\, d^{\bar a}z\, d^{\bar b}z\, \{V,V\} \nn\\
&=& -(-1)^{{[}\bar b{]}}\,  d^{\bar a}z\, d^{\bar b}z\, \wtd V = 
-(-1)^{{[}\bar b{]}}\, \wtd U^{\bar a\bar b}\ .
\eea
The peculiar sign in front of (\ref{sign}) follows from the fermionic 
character of $V$.

\section{Kaluza-Klein reduction and $SL_+(n|1)$}

    Let us consider the dimensional reduction of the pure gravity
Lagrangian $\hat{\cal L}= \hat e\, \hat R$ in $\tD$ dimensions on an
$n$-torus to $D=\tD-n$ dimensions.  This will give the $D$-dimensional
Lagrangian \cite{lpsol}
\be
{\cal L} = e\, R -\ft12 e\, (\del\vec\phi)^2 -\ft14 e\, \sum_i e^{\vec
b_i\cdot\vec\phi}\, ({\cal F}_\2^i)^2 -\ft12 e\, 
\sum_{i<j} e^{\vec b_{ij}\cdot\vec\phi}\, ({\cal F}_{\1 j}^i)^2\ ,
\ee
where the dilaton vectors are given by 
\be
\vec b_i= -\vec f_i\ ,\qquad b_{ij} = -\vec f_i + \vec f_j \ .
\ee
Here, $\vec f_i$ is given by
\be
\vec f_i = \Big(\underbrace{0,0,\ldots, 0}_{i-1}, (\tD-1-i) s_i, s_{i+1},
s_{i+2}, \ldots, s_n\Big)\ ,
\ee
where
\be 
s_i = \sqrt{\ft{2}{(\tD -1 -i)(\tD-2-i)}}\ .
\ee
It is also convenient to define
\be
\vec s = (s_1,s_2,\ldots, s_n)\ .
\ee
We note that $\vec f_i$ satisfies the sum rule
\be
\sum_i \vec f_i = (\tD -2)\, \vec s\ .
\ee
It is also easily established that $\vec f_i$ and $\vec s$ satisfy the
relations
\be
\vec s\cdot\vec s= \ft{2n}{(\tD -2)(D-2)}\ ,\qquad
\vec s\cdot\vec f_i = \ft{2}{D-2}\, \qquad
\vec f_i\cdot\vec f_j = 2\, \delta_{ij} + \ft2{D-2}\ .\label{dotp}
\ee
 From these, the following lemma can also be derived:
\be
\sum_i (\vec f_i\cdot\vec x)^2 = 2 \vec x\cdot\vec x + (\tD-2)\, 
(\vec s \cdot\vec x)^2\ ,\label{cl}
\ee
where $\vec x$ is an arbitrary vector.

    It has been shown previously that the dilaton vectors $\vec
b_{ij}$ form the positive roots of the $SL(n,\R)$ global symmetry
algebra of gravity compactified on the $n$-torus \cite{lpsweyl,cjlp}.
In this appendix, we show that the extended system $\vec b_{\a\b}$,
with $\a =(i,0)$, comprising $\vec b_{ij}$ and $\vec b_{i0}\equiv \vec
b_i$, form the positive roots of the superalgebra $SL(n|1)$.  As in
section 4.3, we extend the definition of the generators $E_i{}^j$
(with $i<j$) for the positive roots $\vec b_{ij}$ to $E_\a{}^\b$ (with
$\a<\b$), where $E_i{}^0=W_i$, and $W_i$ are the odd generators
associated with the weights $\vec b_i$.  (As before, we find it
convenient to make the formal definition that 0 is larger than any of
the values $i$.)  We may take a representation where $E_\a{}^\b$ is
the $(n+1)\times (n+1)$ matrix which is zero everywhere except for a
``1'' at the $\a$'th row and $\b$'th column.  In this representation,
it follows from the commutation relations ${[}\vec H, E_\a{}^\b {]}=
\vec b_{\a\b}\, E_\a{}^\b$ that the Cartan generators $\vec H$ are the
$(n+1)\times (n+1)$ matrices
\be
\vec H = {\rm diag}\, (\vec b_1+\vec c, \vec b_2 +\vec c, \ldots,
                     \vec b_n +\vec c, \vec c)\ ,\label{sl1sncart}
\ee
where $\vec c$ is an as-yet arbitrary vector.  We can now show that $\vec
H$ and $E_\a{}^\b$ ($\a<\b$) form the Borel subalgebra $SL_+(n|1)$ 
of the superalgebra $SL(n|1)$, provided that we choose 
\be
\vec c = \ft{\tD-2}{n-1}\, \vec s\ ,
\ee
so as to make the supertrace of $\vec H$ vanish.  (The $SL(n|1)$ 
supertrace of the matrix $X_{\a\b}$ is given by $\str (X)=
\sum_{i=1}^n X_{ii} -
X_{00}$.)  It only remains to show that the vectors $\vec b_{\a\b}$
with $\a<\b$ indeed form the positive roots of $SL_+(n|1)$.  

    To do this, we must first construct the Cartan-Killing metric.  We
can get it up to a Casimir factor as 
\be
K_{ab} = \ft12 \, \str\, (H_a\, H_b)\ ,
\ee
where $H_a$ denotes the $a$'th component of $\vec H$.  We can then
show, by making use of (\ref{cl}), that for any $\vec x$ 
\be
x_a\, x_b\, K_{ab} = \vec x\cdot\vec x -\ft{(D-1)(\tD-2)}{2(n-1)}\,
(\vec s\cdot\vec x)^2\ ,
\ee
and hence $K_{ab}= \delta_{ab} -\ft12 (D-1)(\tD -2)\, s_a\, s_b/(n-1)$.
The inverse metric is then easily seen to be
\be
K^{ab} = \delta_{ab} -\ft12{\scriptstyle{(D-1)(D-2)}}\, s_a\, s_b\ .
\ee
Thus the $SL(n|1)$ inner product between weight vectors $\vec x$ and
$\vec y$ is given by
\be
K(\vec x, \vec y) \equiv K^{ab}\, x_a \, y_b =
\vec x\cdot\vec y - \ft12{\scriptstyle{(D-1)(D-2)}}\, 
(\vec s\cdot\vec x)(\vec s\cdot\vec y)\ .\label{inner}
\ee
(Note that if we consider $n=1$, we find that the inverse metric for
$SL(n|1)$ vanishes, this is the special case where $SL(n|1)$ is not simple.)  
We can now deduce from (\ref{dotp}) that
\bea
K(\vec b_{ij}, \vec b_{k\ell}) &=& 2\,\d_{ik}+ 2\,\d_{j\ell} 
    -2\, \d_{i\ell} -2\, \d_{jk}\ ,\nn\\
K(\vec b_{ij}, \vec b_{k0}) &=& 2\, \d_{ik} -2\, \d_{jk}\ ,\nn\\
K(\vec b_{i0}, \vec b_{j0}) &=& 2\, \d_{ij} -2\ .
\eea
These are precisely the inner products of the positive-root vectors of
$SL(n|1)$ up to a normalisation factor that should follow from the Casimir
number alluded to above.  
In particular, we may augment the set of simple roots $\vec
b_{i,i+1}$ of $SL(n,\R)$ by including the null vector $\vec b_{n0}$.
These generate the Dynkin diagram for $SL(n|1)$, namely

\centerline{
\begin{tabular}{ccccccccccc}\\
 $\vec b_{12}$& &$\vec b_{23}$&&$\vec b_{34}$ & && &$\vec b_{n-1,n}$
& &$\vec b_{n0}$ \\
 $\scriptstyle{\bigcirc}$ &---& $\scriptstyle{\bigcirc}$ 
  &---& $\scriptstyle{\bigcirc}$&
---& $\cdots\cdots$&---&$\scriptstyle{\bigcirc}$&---& $\otimes$ \\
\end{tabular}}
\bigskip\bigskip

\centerline{Table 1: The dilaton vectors $\vec b_{i,i+1}$ and $\vec
b_{n0}$ generate the $SL(n|1)$ Dynkin diagram}
\bigskip\bigskip

     A new situation arises in the special case of a reduction to
$D=3$ dimensions.  If we leave the Kaluza-Klein 2-form field strengths
${\cal F}_\2^i$ undualised, then the theory will have the $SL_+(n|1)$
symmetry described above.  However, if we instead dualise the 2-form
field strengths, we will gain $n$ additional axionic scalars, while at
the same time losing all the vector potentials.  In this situation,
the obvious $GL(n,\R)$ symmetry from the $n$-torus compactification
can again be enlarged, but this time to the bosonic group $SL(n+1,\R)$
rather than the supergroup $SL_+(n|1)$.  In other words the
superalgebra $SL_+(1|1)$ is the other side of the Ehlers coin, namely
the still mysterious {\it a priori} $SL(2,\R)$ of gravity reduced to 3
dimensions; see \cite{JJ} for a recent review.  This can be foreseen by
noting that the Lagrangian for the undualised theory will be
\be
{\cal L} = e\, R -\ft12 e\, (\del\vec\phi)^2 -\ft14 e\, \sum_i e^{\vec
b_i\cdot\vec\phi}\, ({\cal F}_\2^i)^2 -\ft12 e\,
\sum_{i<j} e^{\vec b_{ij}\cdot\vec\phi}\, ({\cal F}_{\1 j}^i)^2\ .
\label{einstred}
\ee
After dualising ${\cal F}_\2^i$, the dilaton prefactors for the 
the dual fields $G_{\1 i}$ will be $e^{-\vec b_i\cdot\vec \phi}$.
It is now easily seen that the dilaton vectors $\{-\vec
b_i, \vec b_{ij}\}$ form the positive roots of $SL(n+1,\R)$, and that
the simple roots can be taken to be $-\vec b_1$, together with $\vec
b_{i,i+1}$ for $1\le i\le n-1$.  Thus we have the Dynkin diagram

\centerline{
\begin{tabular}{ccccccccccc}\\
 $-\vec b_{1}$& &$\vec b_{12}$&&$\vec b_{23}$ & && &$\vec b_{n-2,n-1}$
& &$\vec b_{n-1,n}$ \\
 $\scriptstyle{\bigcirc}$ &---& $\scriptstyle{\bigcirc}$
  &---& $\scriptstyle{\bigcirc}$&
---& $\cdots\cdots$&---&$\scriptstyle{\bigcirc}$&---&
   $\scriptstyle{\bigcirc}$ \\
\end{tabular}}
\bigskip\bigskip

\centerline{Table 2: In $D=3$, $-\vec b_1$ and
$\vec b_{i,i+1}$ and generate the $SL(n+1,\R)$ Dynkin diagram}
\bigskip\bigskip

   A detailed calculation confirms that indeed we have an $SL(n+1,\R)$
symmetry. First, we note that the Bianchi identity for the field
strengths ${\cal F}_\2^i$ can be written as $d(\gamma^i{}_j\, {\cal
F}_\2^j) =0$ \cite{cjlp}.  The fields ${\cal F}_\2^i$ can therefore be
dualised by introducing Lagrange multipliers $\chi_i$, and adding the
term $\chi_i\, d(\gamma^i{}_j\, {\cal F}_\2^j)$ to the Lagrangian
(\ref{einstred}).  We then treat ${\cal F}_\2^i$ as auxiliary fields,
and solve for them giving ${*{\cal F}_\2^i}\, e^{\vec
b_i\cdot\vec\phi} = G_{\1 i}\equiv \gamma^j{}_i\, d \chi_j$, and hence
the Lagrangian becomes
\be
{\cal L} = e\, R -\ft12 e\, (\del\vec\phi)^2 -\ft12 e\, \sum_i e^{-\vec
b_i\cdot\vec\phi}\, (G_{\1 i})^2 -\ft12 e\,
\sum_{i<j} e^{\vec b_{ij}\cdot\vec\phi}\, ({\cal F}_{\1 j}^i)^2\ .
\label{einstred2}
\ee
It is now evident that we can extend the range of the $i$ index to
$a=(0,i)$ (with $0<i$ here), and define axions $\bA^a_{\0 b}$ for
all $a<b$:
\be
\bA^0_{\0 i} = \chi_i\ ,\qquad \bA^i{}_{\0 j} = {\cal A}^i_{\0 j}\ .
\ee
(The bar over the potential indicates the extended set.)
Defining also $\bar\g^a{}_b$ as in \cite{cjlp}, but for the extended
set of axionic potentials $\bA^a{}_{\0 b}$, by $\bar\g^0{}_i= -\chi_i$ and
$\bar\g^i{}_j =\g^i{}_j$, we see that (\ref{einstred2}) assumes the
form
\be
{\cal L} = e\, R -\ft12 e\, (\del\vec\phi)^2 -\ft12 e\,
\sum_{a<b} e^{\vec b_{ab}\cdot\vec\phi}\, (\bF_{\1 b}^a)^2\ ,
\label{einstred3}
\ee
where $\bF^a_{\0 b} =\bar\g^c{}_b\, d\bA^a_{\0 c}$.  This is exactly
of the form of the familiar scalar Lagrangian resulting from the
dimensional reduction of pure gravity on a spacelike $n$-torus \cite{cjlp}, 
except that now
the index range is extended to include the value 0.  
Thus the usual proof of the
existence of the $SL(n,\R)$ symmetry now establishes that 
we have an $SL(n+1,\R)$ symmetry in this three-dimensional case.

          More generally, if any $N$-form potential is present in the
original $\tD$-dimensional theory, in addition to gravity, then we
expect that there would still be an $SL(n+1,\R)$ symmetry in $D=3$,
with the $N$th-degree potential yielding (after dualisation in $D=3$)
an ${n+1\choose N}$ dimensional irreducible representation of
$SL(n+1,\R)$.  For example, if we consider $D=11$ supergravity reduced
to $D=3$, we will have an $SL(9,\R)$ global symmetry, with the
Kaluza-Klein descendants of the $D=11$ 3-form potential giving an
irreducible 84-dimensional representation of $SL(9,\R)$.  Of course in
this case the symmetry actually enlarges further to $E_8$, but this
latter enlargement depends crucially on the presence (with the correct
coefficient) of the $FFA$ term in $D=11$.

     Another example is the interesting case of the dimensional
reduction of the $D=4$ Einstein-Maxwell system to $D=3$.  It has been
known that after dualising the vector potentials in $D=3$, the
resulting purely scalar theory then has an $SU(2,1)$ global symmetry
(see the first reference in \cite{cjgroups}, but this group was
actually known before; see \cite{kin}), which contains $SL(2,\R)$ as an
subalgebra.  Furthermore, if one considers $N$ Maxwell fields in $D=4$
rather than just one, then after dualising the vectors to scalars in
$D=3$ one obtains a three-dimensional purely scalar Lagrangian with an
$SU(N+1,1)$ symmetry \cite{bmg}.  Here, we present a very simple proof
of this result, by showing that the target space of the scalar sigma
model is the coset $U(N+1)\backslash SU(N+1,1)$, which is a
non-compact form of $CP^{N+1}$.

    We start from the standard Lagrangian for $N$ Maxwell fields 
$\hat F_\2^i$ coupled to gravity in $D=4$:
\be
{\cal L}_4 = \hat R\, {*\oneone} -\ft12 {*{\hat F_\2^i}}\wedge\hat F_\2^i\ .
\ee
We now make a standard Kaluza-Klein reduction to $D=3$, for which the
metric ansatz will be $d\hat s_4^2 = e^{\varphi}\, ds_3^2 + e^{-\varphi}\,
(dz + {\cal A}_\1)^2$.  Thus the $D=3$ Lagrangian will be
\be
{\cal L}_3 = R\, {*\oneone} -\ft12 {*d\varphi}\wedge d\varphi -\ft12
e^{-\varphi}\, {*F_\2^i}\wedge F_\2^i -\ft12 e^{\varphi}\, {*F_\1^i}\wedge
F_\1^i -\ft12 e^{-2\varphi}\, {*{\cal F}_\2}\wedge {\cal F}_\2\ ,
\label{d4d3lag}
\ee
where $F_\2^i = dA_\1^i - dA_\0^i\wedge {\cal A}_\1$, $F_\1^i = dA_\0^i$, and
${\cal F}_\2 = d{\cal A}_\1$.  (Here we have reduced the gauge fields
using the standard prescription $\hat A_1^i= A_\1^i + A_\0^i\, dz$.)

     Now we dualise all the 1-form potentials, to give $(N+1)$ further
axions.  Thus we add Lagrange multiplier terms $-\chi\, d{\cal F}_\2
-\psi_i\, d(F_\2^i -A_\0^i\, {\cal F}_\2)$ to (\ref{d4d3lag}), to enforce
the Bianchi identities $d{\cal F}_\2=0$ and $dF_\2^i = F_\1^i \wedge {\cal
F}_\2$, and treat ${\cal F}_\2$ and $F_\2^i$ as auxiliary fields which
we now eliminate.  We find that $e^{-2\varphi}\, {*{\cal F}_\2}
= d\chi - A_\0^i \, d\psi_i$, and $e^{-\varphi}\, {*F_\2^i} = d\psi_i$.
Substituting back into the Lagrangian, we obtain the fully-dualised
result
\be
e^{-1}\, {\cal L}_3 = R -\ft12(\del\varphi)^2 - \ft12 e^{\varphi} \,
(\del\psi_i)^2 -\ft12 e^{\varphi}\, (\del A_\0^i)^2 - \ft12 e^{2\varphi}\,
(\del\chi - A_\0^i\, \del \psi_i)^2\ .\label{d4d3duallag}
\ee

     To study the structure of the scalar manifold, we may simply
consider the metric on the $(2N+2)$-dimensional target space which, from
(\ref{d4d3duallag}), we read off to be
\be
ds^2= d\varphi^2 +e^{\varphi} \,
(d\psi_i)^2 + e^{\varphi}\, (d A_\0^i)^2 + e^{2\varphi}\,
(d\chi - A_\0^i\, d\psi_i)^2\ .\label{tsmetric}
\ee
Now define the obvious orthonormal basis,
\be
e^0=d\varphi\ ,\qquad e^i = e^{\fft12\varphi}\, d\psi_i\ ,\qquad
e^{i'} = e^{\fft12\varphi}\, d A_\0^{i'}\ ,\qquad
e^{0'} = e^{\varphi}\, (d\chi - A_\0^i\, d\psi_i)\ ,
\ee
which can be seen to satisfy
\be
de^0=0\ ,\qquad de^i = \ft12 e^0\wedge e^i\ ,\qquad de^{i'} = \ft12
e^0\wedge e^{i'}\ ,\qquad de^{0'} = e^0\wedge e^{0'} +
e^i\wedge e^{i'}\ .\label{deres}
\ee
We then easily see that the spin connection $\omega_{ab}$ is given by
\bea
&&\omega_{0i} = -\ft12 e^i\ , \qquad \omega_{0i'} = -\ft12 e^{i'}\ ,\qquad
\omega_{00'} = - e^{0'}\ ,\nn\\
&&\omega_{0'i'} = -\ft12 e^i\ , \qquad \omega_{0'i} = \ft12 e^{i'}\ ,\qquad
\omega_{ij'} = -\ft12 \delta_{ij}\, e^{0'}\ .
\eea

   It is evident from (\ref{deres}) that the 2-form
\be
J= e^0\wedge e^{0'} + e^i\wedge e^{i'} \label{kahler}
\ee
is closed, $dJ=0$, and that it is a complex structure, satisfying
$J^a{}_b\, J^b{}_c = -\delta^a_c$, where $a=(0,0',i,i')$.  In fact, 
$J$ is clearly therefore a K\"ahler form.   After
further elementary algebra, we find that the curvature 2-forms
$\Theta_{ab} = d\omega_{ab} + \omega_{ac}\wedge \omega_{cb}$ can be
written as
\be
\Theta_{ab} = -\ft14 e^a\wedge e^b - \ft14 J_{ac}\, J_{bd}\, e^c\wedge
e^d -\ft12 J_{ab}\, J\ .
\ee
This means that the components of the Riemann tensor are given by
\be
R_{abcd} = -\ft14\Big( \delta_{ac}\, \delta_{bd} - \delta_{ad}\,
\delta_{bc} + J_{ac}\, J_{bd} -J_{ad}\, J_{bc} + 2 J_{ab}\, J_{cd}
\Big) \ .
\ee
This can be recognised as the curvature tensor for a space of constant
(negative) holomorphic sectional curvature \cite{kobnom}.  Had it been
of positive curvature, it would have been $CP^{N+1}$, which is the
coset space $U(N+1)\backslash SU(N+2)$.  Since here the curvature is
negative, we can recognise it as the non-compact form of the coset,
$U(N+1)\backslash SU(N+1,1)$.  (This is related to $CP^{N+1}$ in the
same way as the hyperbolic space $H^n$ is related to the 
$n$-sphere.\footnote{It is interesting that just as $H^n$ admits a much
simpler metric than $S^n$, namely the horospherical metric $ds^2=d\rho^2
+ e^{2\rho}\, dx^i\, dx^i$, so the non-compact form of the $CP^{N+1}$
metric can be written in an analogous very simple ``horospherical'' form 
(\ref{tsmetric}).})
 Thus in particular, we see that the fully dualised scalar
Lagrangian (\ref{d4d3duallag}) in $D=3$, coming from the Einstein
Lagrangian coupled to $N$ Maxwell fields in $D=4$, has an $SU(N+1,1)$
global symmetry group.  The symmetry contains once more the group
$SL(2,\R)$ as a proper subgroup.

\section{Fully-dualised $SL(2,\R)$ coset and generalisations}

    In section 4, we established that the first-order equations
(\ref{sl2rfo}) have an $SL(2,\R)$ global symmetry for the scalars, and
two abelian gauge symmetries for the two dual fields.  It is
interesting to study the symmetries of the various second-order
equations that can be obtained by integrating out auxiliary fields.
For example, $\psi$ and $\wtd\chi$ appear in the first-order equations
only through their field strengths $P$ and $Q$, and the Bianchi
identities $dP=0=dQ$ correspond to the two second-order equations of
motion for the scalar fields $\phi$ and $\chi$.  These second-order
equations have only the $SL(2,\R)$ global symmetry.  The disappearance
of the local gauge symmetries is understandable since the scalars are
invariant under these symmetries, even in the full doubled equations.

     If, on the other hand, we integrate out the $\chi$ and $\psi$
fields, then the remaining fields $\phi$ and $\wtd \chi$ have the $\R$
global symmetry corresponding to constant shifts of the dilaton, and
the local gauge symmetry associated with $\wtd\chi$.  This case is
studied in \cite{cjlp}.  Finally, let us study the case where both of the
scalars $\phi$ and $\chi$ are integrated out.  In order to do this, we
must first make a field redefinition of the dual fields:
\be
\bar\psi = \psi + \chi \wtd \chi\ ,\qquad
\bar{\wtd\chi} = e^{-\phi}\wtd \chi\ .\label{sl2rfred}
\ee
Under this redefinition, the doubled equations (\ref{sl2rfo}) become
\be
*g = d\bar\psi -\bar{\wtd \chi}\wedge f\ ,\qquad
*f= d\bar{\wtd\chi} + g\wedge \bar{\wtd\chi}\ ,\label{fgeom}
\ee
where $f=e^{\phi} d\chi$ and $g=d\phi$.  Thus we obtain two
independent linear equations for the 1-form field strengths $f$ and $g$
for the scalars.  From these equations, we can solve for $f$ and $g$
purely in terms of the redefined dual fields $\bar\psi$ and $\bar{\wtd\chi}$.
The Bianchi identities
\be
dg = 0\ ,\qquad df - g\wedge f =0\label{fgbi}
\ee
then become equations of motion for the dual fields.

        Note that the relation between the new dual fields
$(\bar{\psi},\bar{\wtd\chi})$ and the old fields $(\psi,\wtd\chi)$
can be also expressed as
\be
\v = e^{\fft12\phi\, H}\, e^{\chi\, E_+}\, e^{\wtd\chi\, \tE_+}\,
e^{\fft12\psi\, \tH}=
e^{\bar{\wtd\chi}\, \tE_+}\, e^{\fft12\bar{\psi}\, \tH}\,
e^{\fft12\phi\, H}\, e^{\chi\, E_+}\ .\label{fredxxx}
\ee
The new dual fields are invariant under the Borel subgroup of the
global $SL(2,\R)$ symmetry group.  This has the consequence that the
completely dualised theory of the $SL(2,\R)$ scalar manifold has no
global symmetry, but it does have the abelian local gauge symmetries
of the dual potentials, which are non-diagonally realised.

         The doubled equation $*\G=\S \G$, or equivalently the
first-order equations (\ref{sl2rfo}), enable us fully to dualise the
coset, and write the equations of motion in terms of the sole dualised
pair of potentials $\bar \psi$ and $\bar{\wtd \chi}$.  The dualised
theory no longer has any global symmetry, but it does retain the local
gauge symmetry, which becomes non-linear.  Now we show that this full
dualisation of the scalar coset can also be achieved at the level of
Lagrangian, and that in particular, $\bar\psi$ and $\bar{\wtd\chi}$
are the precise Lagrangian multipliers.  To see this, we note that the
Lagrangian (\ref{sl2rlag}) can be written
\be
{\cal L} = -\ft14 (\G_0 + \G_0^{\rm T})^2\ ,\label{cosetlag}
\ee
where
\bea
\G_0&=&d\v_{0} \v_0^{-1} = \ft12 d\phi\, H + e^{\phi} d\chi E_{+}\nn\\
&=&\ft12 g\, H + f\, E_{+}=\pmatrix{\ft12 g & f\cr 0 &-\ft12 g}
\ ,
\eea
and $\v_0=\exp(\fft12 \phi H + \chi E_{+})$.  (Note that here we have
$E_{+}^{\rm T} = E_{-}$, $\tr H^2 =2$, $\tr E_{+}^2 = \tr E_{-}^2=0$
and $\tr E_{+} E_{-} =1$.  Note also that the doubled equation can
be equivalently expressed as $*(\G + \G^{\rm T}) = \S (\G + \G^{\rm
T})$.)  Thus we see that the dilaton $\phi$ and axion $\chi$ appear
in the Lagrangian only through $\G$, {\it i.e.}\ through the quantities
$f$ and $g$.  The Bianchi identities (\ref{fgbi}) for these
two fields can be expressed as $F\equiv d\G_0- \G_0\wedge \G_0 =0$.
Treating $f$ and $g$ as a new set of basic fields, we can introduce
a Lagrange multiplier $\Sigma$
\be
\Sigma = \bar\psi\, H + \bar{\wtd \chi}\, E_{-}=
\pmatrix{\bar\psi & 0\cr
                                      \bar{\wtd \chi} & -\bar\psi}\ .
\ee
The first-order Lagrangian is given by
\be
{\cal L} = -\ft14 \tr\{ (\G_0+ \G_0^{\rm T})^2 +
*F \wedge \Sigma\}\ .
\ee
Varying the Lagrangian with respect to $f$ and $g$ gives rise to the
equations of motion (\ref{fgeom}), which, as we have seen, enable us
to solve for $f$ and $g$ in terms $\bar\psi$ and $\bar{\wtd\chi}$.
Substituting the results into the above first-order Lagrangian, we
obtain the fully-dualised Lagrangian for the $SL(2,\R)$ coset (in
the Borel gauge).

        The fully-dualised Lagrangian has no global symmetry, but it
does have non-diagonally realised commuting local gauge symmetries.
It follows from (\ref{sl2rgauge}) and (\ref{sl2rfred}) that the gauge
symmetries are
\be
\delta \bar\psi = \Lambda_{\psi} + \chi \Lambda_{\wtd\chi}\ ,\qquad
\delta\bar{\wtd \chi} = e^{-\phi} \Lambda_{\wtd\chi}\ .
\ee
However, $\Lambda_{\psi}$ and $\Lambda_{\wtd\chi}$ are bad choices of 
gauge parameters for the dualised theory, in that the transformations
cannot be expressed purely in terms of the dual fields.  However, if we
define
$\Lambda_{\psi} = d\lambda_1$ and $\Lambda_{\wtd\chi}= d\lambda_2$,
the gauge transformations can now be expressed purely in terms of the dual
potentials, namely
\be
\delta\bar\psi = d\bar\lambda_1 - \bar \lambda_2\wedge f\ ,\qquad
\delta\bar\chi = d\bar\lambda_2 + \bar \lambda_2\wedge g
\ ,\label{sl2rdualgauge}
\ee
where $\bar\lambda_1 = \lambda_1 + \chi \lambda_2$ and $\bar \lambda_2
= e^{-\phi}\, \lambda_2$.   This gauge invariance of the dualised
field $\Sigma$ is a consequence of the non-abelian Bianchi identity
for $F$, namely
\be
D_{0} F \equiv dF - \G_0 \wedge F =0\ .\label{fdefintion}
\ee
Taking into account the fact that $\Sigma$ belongs to the anti-Borel
Lie algebra, we find that $\delta *\Sigma = (D\bar \lambda)^{\nabla}$,
where $\nabla$ denotes the projection onto the anti-Borel Lie algebra
along the positive root generators, and $\bar\lambda$ is parameter in
the anti-Borel algebra.  This gauge transformation rule is precisely
the same as the one given in (\ref{sl2rdualgauge}).  Note that the
relation between the fields $\bar\psi$, $\bar{\wtd\chi}$ and the
original fields $\psi$ and $\wtd\chi$, given by (\ref{sl2rfred}), can
be expressed as a double projection into the anti-Borel Lie algebra.
Denoting $\Sigma' = \psi\, H + \wtd\chi\, E_{-}$, we then have
$\Sigma' = (\v_0^{-1} \Sigma \v_0)^\nabla$, which is in fact
equivalent to the statement in (\ref{fredxxx}).  This procedure for
dualisation does not work if we choose any other parameterisation for
the coset $SL(2,\R)/O(2)$ which is not expressible as the exponentials
of a Lie algebra.  For example, the coset parameterisation in the
so-called symmetric gauge, where $\v_0$ is chosen to be a symmetric
matrix, is not dualisable.

We could, however, start with the completely covariant formulation
which admits a global $SL(2,\R)$ $\times$ local $O(2)$ invariance.
The Lagrangian is now given by
\be
{\cal L} =\tr \{ (D_\mu \v \, \v^{-1})^2 + *F\wedge \Sigma \}
\ ,\label{lagxxx}
\ee
where $D_\mu \v = (\del_\mu - h_\mu)\, \v$, the representative $\v$ is
$SL(2,\R)$-valued, and $h_\mu$ is the composite `gauge field' for the
local $O(2)$ symmetry.  Here, $F$ is defined as previously in terms of
$\G$, which is still given by $\G=d\v\v^{-1}$, but is now in the full
Lie algebra of $SL(2,\R)$.  $\Sigma$ is an arbitrary element of the
Lie algebra of $SL(2,\R)$, parameterised in terms of three fields.
The Lagrangian (\ref{lagxxx}) is invariant under the transformations
\bea
{\rm local}\,\, O(2) \times SL(2,\R):&&
\v \longrightarrow O(x)\, \v \, U^{-1}\ ,\nn\\
&& F \longrightarrow O(x) \, F \, O^{-1}(x)\ ,\nn\\
&& \Sigma \longrightarrow O(x) \, \Sigma\, O^{-1}(x)\ ,\\
{\rm gauge}:&& \delta(*\Sigma) = D_\G \lambda = d\lambda - \G\wedge
\lambda\ ,\nn
\eea
where $O(x)$ belongs to the local $O(2)$ and $U$ belongs to the global
$SL(2,\R)$.  Let us write $\G= \G^{\perp} + \G^{\Vert}$, where
$\G^{\Vert}$ belongs to Lie algebra of $O(2)$, and $\G^\perp$ is in
the orthogonal complement.  We can solve the equations for $\G^\perp$,
$\G^\Vert$ and $h$ ($h=\G^\Vert$), and obtain a highly non-linear
Lagrangian for $\Sigma$.  Although the Lagrangian still has the local
$O(2)$ invariance (as well as the gauge invariance), it is not
possible to write it in terms of only the two fields $\bar\phi$ and
$\bar{\wtd\chi}$.  However, the field equations still describe just
two degrees of freedom.

       Let us now compare the dualisations of the Borel-gauge coset,
the covariant coset, and the principal sigma model.  In all three
cases the Lagrangian can be expressed in terms of ${\cal G}$, where
${\cal G}= d{\cal V} {\cal V}^{-1}$.  In the case of the Borel-gauged
coset, ${\cal V}$ is parameterised by scalars associated with the
Cartan and the positive-roots generators. The principal sigma models
are parameterised either by the left-acting group or the group of
right-shifts.  For the covariant coset $\v$ is parameterised by the
scalars associated with the full root system.  The Lagrangians are
invariant under the following transformations
\bea
{\rm Principal\ Sigma\ Model}:&&
\v \longrightarrow \v' = {\cal O} \v \Lambda^{-1}\nn\\
{\rm Covariant\ Coset}: && \v \longrightarrow \v' = {\cal O}(x) \v
\Lambda^{-1}\nn\\
{\rm Borel-gauge\ coset}: && \v \longrightarrow
\v' = {\cal O}({\rm scalar}) \v \Lambda^{-1}\ .
\eea
In other words, the principal sigma model Lagrangian is invariant
under the global symmetry $G_{\rm L} \times G_{\rm R}$, with ${\cal
O}$ and $\Lambda$ being independent constant matrices belonging to
$G_{\rm L}$ and $G_{\rm R}$ respectively.  The covariant coset, on the
other hand, is invariant under $H({\rm local})\times G({\rm global})$
transformations.  Finally, the Borel-gauge coset is invariant only
under the global symmetry group $G$, since the local group $H$ has
been gauged.  Note that in this case the transformation ${\cal O}({\rm
scalar})$, which is a scalar-dependent transformation, is not
associated with an independent symmetry.  Rather, it is a compensating
transformation that is needed for implementing the global $G$
symmetry, and is used for restoring $\v$ to the Borel gauge after
after having performed the $G$ transformation.  In all the three cases
${\cal G}$, and hence $F\equiv d\G - \G\wedge \G$, transforms as
\be
{\cal G} \longrightarrow {\cal G}'= {\cal O} {\cal G} {\cal O}^{-1}\ ,
\qquad
F \longrightarrow F'= {\cal O} F {\cal O}^{-1}\ .\label{gftrans}
\ee
In particular, note that both ${\cal G}$ and $F$ are invariant under
the global symmetry group $G$ acting from the right.  Since all the
fields appear in the Lagrangian through ${\cal G}$, which satisfies
the Bianchi identity $F=0$, it follows that for all three cases
the first order Lagrangian can be written as
\bea
{\cal L} = {\cal L}_0({\cal G}) + {*F}\wedge \Sigma\ ,\label{fogen}
\eea
where $\Sigma$ are the dual $(D-2)$-form potentials.  In the case of
Borel-gauge coset, the number of dual potentials is $Dim(G/H)$.
In the other two cases, the number is instead equal to the
dimension of the group $G$.  Thus the Lagrangian (\ref{fogen}) is
invariant under (\ref{gftrans}), provided that $\Sigma$ transforms
as
\be
\Sigma \longrightarrow \Sigma'= {\cal O} \Sigma {\cal O}^{-1}\ .
\label{sigma}
\ee
If we solve for ${\cal G}$ in terms of $\Sigma$, and substitute back
into (\ref{fogen}), we obtain a second order Lagrangian that is
expressed purely in terms of $\Sigma$.  In the cases of the principal
sigma model and the covariant coset, the dualised Lagrangian is
expressed in terms of the $Dim(G)$ dual potentials, and the theories
are additionally invariant under global $G_{\rm L}$ or $H({\rm
local})$ symmetry groups respectively.  In the case of the Borel-gauge
coset, ${\cal O}$ is a compensating transformation that depends on the
original scalars, and it cannot be expressed locally in terms of the
dual potentials $\Sigma$. In this case there is no remaining global
symmetry that can be expressed in terms of local field
transformations.

        Note that in the coset case there are two dual Lagrangians,
one obtained by dualising the covariant coset Lagrangian, and the
other obtained by dualising the Borel-gauge Lagrangian.  Although the
two systems describe the same on-shell degrees of freedom, their
off-shell degrees of freedom differ.  To see this, we observe that the
dualised covariant-coset Lagrangian contains a number $Dim(G)$ of
$(D-2)$-form potentials, each of which has $(D-1)$ off-shell degrees
of freedom.  On the other hand the dualised Borel-gauge Lagrangian
contains only $Dim(G/H)$ $(D-2)$-form potentials.  The local $H$ gauge
symmetry of the former Lagrangian, which can be used to remove
$Dim(H)$ degrees of freedom, is not enough to remove the excess of
$Dim(H)$ $(D-2)$-form potentials, which would need to be done in order
to map it to the dualised Borel-gauge Lagrangian, since each potential
has $(D-1)$ off-shell degrees of freedom.  This phenomenon should be
contrasted with the situation for the original undualised coset
Lagrangians.  In that case, the local group $H$ is precisely enough to
fix the gauge and hence to map the covariant-coset Lagrangian to the
Borel-gauge coset Lagrangian, since each scalar has just one degree of
freedom, both off-shell as well as on-shell.


\end{document}